\documentclass[11pt]{amsart}
\usepackage{latexsym,amsmath,amssymb,amscd,eucal}
\usepackage{xypic}

\newcommand{\newsection}{\setcounter{equation}{0}\section}

\newcommand{\mbf}[1]{{\boldsymbol {#1} }}

\def\appendix#1{\addtocounter{section}{1}\setcounter{equation}{0}
\renewcommand{\thesection}{\Alph{section}}
\section*{Appendix \thesection\protect\indent \parbox[t]{11.715cm} {#1}}
\addcontentsline{toc}{section}{Appendix \thesection\ \ \ #1} }
\newcommand{\eq}{\begin{equation}}
\newcommand{\eqend}{\end{equation}}

\newenvironment{romanlist}{%

        \begin{enumerate}
        }{%
        \end{enumerate}}
\newenvironment{alphlist}{%
  \begin{enumerate}%
}{%
  \end{enumerate}}
\newbox\ncintdbox \newbox\ncinttbox
\setbox0=\hbox{$-$} \setbox2=\hbox{$\displaystyle\int$}
\setbox\ncintdbox=\hbox{\rlap{\hbox to
\wd2{\hskip-.125em\box2\relax \hfil}}\box0\kern.1em}
\setbox0=\hbox{$\vcenter{\hrule width 4pt}$}
\setbox2=\hbox{$\textstyle\int$}
\setbox\ncinttbox=\hbox{\rlap{\hbox to
\wd2{\hskip-.175em\box2\relax \hfil}}\box0\kern.1em}

\def\Dirac{{D\!\!\!\!/\,}} 

\newcommand{\complex}{{\mathbb C}} 
\newcommand{\zed}{{\mathbb Z}} 
\newcommand{\nat}{{\mathbb N}} 
\newcommand{\real}{{\mathbb R}} 
\newcommand{\rat}{{\mathbb Q}} 
\newcommand{\quat}{{\mathbb H}}
\newcommand{\oct}{{\mathbb O}}
\newcommand{\dalg}{{\mathbb K}}
\newcommand{\torus}{{\mathbb T}}
\newcommand{\id}{{1\!\!1}} 

\def\pt{{\rm pt}}
\def\sec{{\Omega}}
\def\K{{\rm K}}
\def\H{{\rm H}}
\def\C{{\rm C}}
\def\E{{\rm E}}

\def\B{{\mathbb{B}}}

\def\S{{\mathbb{S}}}
\def\P{{\mathbb{P}}}

\def\F{{\rm F}}
\def\U{{\rm U}}
\def\im{{\rm im}}

\def\BU{{\rm BU}}
\def\Tor{{\rm Tor}}
\def\tor{{\rm tor}}
\def\Ext{{\rm Ext}}
\def\Hom{{\rm Hom}}
\def\coh{{\rm coh}}
\def\Der{{\bf D}}
\def\Cl{{{\rm C}\ell}}
\def\ch{{\rm ch}}
\def\td{{\rm td}}
\def\Id{{\rm id}}

\def\Index{{\rm Index}}

\def\nn{\nonumber}

\def\e{{\,\rm e}\,}

\def\be{\begin{equation}}
\def\ee{\end{equation}}
\def\bea{\begin{eqnarray}}
\def\eea{\end{eqnarray}}
\def\bd{\begin{displaymath}}
\def\ed{\end{displaymath}}

\def\dd{{\rm d}}

\def\ii{{\,{\rm i}\,}}

\makeatletter
\newdimen\normalarrayskip              
\newdimen\minarrayskip                 
\normalarrayskip\baselineskip
\minarrayskip\jot
\newif\ifold             \oldtrue            
\def\arraymode{\ifold\relax\else\displaystyle\fi} 
\def\@arrayskip{\ifold\baselineskip\z@\lineskip\z@
     \else
     \baselineskip\minarrayskip\lineskip2\minarrayskip\fi}
\def\@arrayclassz{\ifcase \@lastchclass \@acolampacol \or
\@ampacol \or \or \or \@addamp \or
   \@acolampacol \or \@firstampfalse \@acol \fi
\edef\@preamble{\@preamble
  \ifcase \@chnum
     \hfil$\relax\arraymode\@sharp$\hfil
     \or $\relax\arraymode\@sharp$\hfil
     \or \hfil$\relax\arraymode\@sharp$\fi}}
\def\@array[#1]#2{\setbox\@arstrutbox=\hbox{\vrule
     height\arraystretch \ht\strutbox
     depth\arraystretch \dp\strutbox
     width\z@}\@mkpream{#2}\edef\@preamble{\halign \noexpand\@halignto
\bgroup \tabskip\z@ \@arstrut \@preamble \tabskip\z@ \cr}%
\let\@startpbox\@@startpbox \let\@endpbox\@@endpbox
  \if #1t\vtop \else \if#1b\vbox \else \vcenter \fi\fi
  \bgroup \let\par\relax
  \let\@sharp##\let\protect\relax
  \@arrayskip\@preamble}
\makeatother

\newcommand{\beq}{\begin{eqnarray}}
\newcommand{\eeq}{\end{eqnarray}}

\newcommand{\rh}{\rho}

\def\appendix#1{\addtocounter{section}{1}\setcounter{equation}{0}
\renewcommand{\thesection}{\Alph{section}}
\section*{Appendix \thesection. #1}
\addcontentsline{toc}{section}{Appendix \thesection\ \ \ #1} }

\def\ii{{\,{\rm i}\,}}

\newtheorem{theorem}{Theorem}[section]
\newtheorem{lemma}{Lemma}[section]
\newtheorem{cor}{Corollary}[section]
\newtheorem{proposition}{Proposition}[section]
\theoremstyle{definition}
\newtheorem{definition}{Definition}[section]
\theoremstyle{remark}
\newtheorem{example}{Example}[section]
\newtheorem{remark}{Remark}[section]

\numberwithin{equation}{section}

\begin{document}

\hfill{\small HWM--05--16 \ \ EMPG--05--13 \ \ hep-th/0507043}

\hfill{\small July 2005}

\vskip 1cm

\title[Geometric K-Homology of Flat D-Branes]
{Geometric K-Homology of Flat D-Branes}

\author{Rui M.G. Reis and Richard J. Szabo}

\address{Department of Mathematics, School of Mathematical and
  Computer Sciences, Heriot-Watt University, Colin Maclaurin
  Building, Riccarton, Edinburgh EH14 4AS, U.K.}

\email{Reis@ma.hw.ac.uk}

\email{R.J.Szabo@ma.hw.ac.uk}

\begin{abstract}
We use the Baum-Douglas construction of K-homology to explicitly
describe various aspects of D-branes in Type~II superstring theory in
the absence of background supergravity form fields. We rigorously
derive various stability criteria for states of D-branes and show how
standard bound state constructions are naturally realized directly in
terms of topological K-cycles. We formulate the mechanism of flux
stabilization in terms of the K-homology of non-trivial fibre
bundles. Along the way we derive a number of new mathematical results
in topological K-homology of independent interest.
\end{abstract}

\maketitle

\newsection*{Introduction}

One of the most exciting recent interactions between physics and mathematics
has been through the realization that D-branes in string theory are classified
by generalized cohomology theories such as K-theory. The charges of
D-branes in Type~II superstring theory are classified by the K-theory
groups of the spacetime in which they
live~\cite{Horava1,MM1,16,Witten1}. In Type~I superstring theory one
uses instead KO-theory, while the charges of branes in orbifolds and
orientifolds are classified by various equivariant K-theories,
KR-theories, and extensions
thereof~\cite{Berg1,Berg2,G-C1,Gukov1,16,Witten1}. In curved
backgrounds and in the presence of a non-trivial $B$-field, D-brane
charge takes values in a twisted K-theory
group~\cite{BM1,Kapustin1,Witten1}. In addition, the Ramond-Ramond
fields which are typically supported on D-branes similarly take values
in appropriate K-theory groups~\cite{Berg2,MW1,FH1}. These
realizations have prompted intensive investigations in both the
mathematics and physics literature into the properties and definitions
of various K-theory groups. At the heart of the excitement in these
investigations is the fact that the correct physical picture of
D-brane charges (and Ramond-Ramond fields) cannot be properly captured
in general by ordinary cohomology but rather requires
K-theory~\cite{FW1,DMW1}, and conversely that the known physical properties
of D-branes in string theory give insights into the rigorous
characterizations of certain less widely explored generalized
cohomology functors such as those of twisted K-theory~\cite{MMS1}.

In this paper we will elucidate in detail the observation that
K-homology, the homological version of K-theory, is really the more
appropriate arena in which to classify
D-branes~\cite{KHomM,13,Matsuo1,Periwal1,Sz1,Sz2}. We treat only the
case of Type~II D-branes in the absence of non-trivial $B$-fields. We
will build on the classic Baum-Douglas
construction of K-homology~\cite{1,2} which is called
{\it topological} K-homology in order to distinguish it from {\it
  analytic} K-homology, another homological realization of
K-theory. In~\cite{1,2} it is proven that this is indeed the
homology theory dual to K-theory. For a unified treatment which
works for a generic cohomology theory, see~\cite{3}. The main
advantage of this geometric formulation is that the K-homology
cycles encode the most primitive requisite objects that must be
carried by any D-brane, such as a spin$^c$ structure and a complex
vector bundle.

Generally, D-branes are much more complicated objects than just
subspaces in an ambient spacetime and require a more abstract
mathematical notion, such as that of a derived
category~\cite{Douglas1,Witten3}. Nevertheless, the realization of
D-branes in topological K-homology gives them a very natural robust
definition in fairly general spacetime backgrounds and reveals
various important properties of their (quantum) dynamics that could
not be otherwise detected if one classified the brane worldvolumes
using only ordinary singular homology. As we will describe in detail
in the following, such effects include important stability
properties as well as the fact that D-branes do not always wrap
submanifolds of the spacetime. Moreover, there is a natural
relationship between the Baum-Douglas construction and the
realization of certain D-branes as objects in a particular
triangulated category.

The requisite mathematical material used for this investigation is
surveyed in Section~1. We elaborate on various aspects of
K-homology, giving all the relevant definitions (in terms of the
Baum-Douglas approach) and describing some new results which to the
best of our knowledge have not previously appeared in the
literature. We also present different perspectives on K-homology:
the spectrum based definition, which gives a more algebraic
topological setting, allowing us to use general results in topology
to investigate the structure of D-branes; and the analytic
definition (based on the Baum-Douglas approach and the
Brown-Douglas-Fillmore construction, although one can also use
Kasparov's approach), which relates the study of D-branes with the
study of C$^*$-algebras and their K-homology. This second approach
makes way to the study of operator algebras, as well as giving a
natural setting in which to express the famous index results. Of
importance in this section is the set up of several methods to
calculate the K-homology groups such as Poincar$\acute{\rm{e}}$
Duality, the Universal Coefficient Theorem, and the Chern Character.

We then undertake the task of attempting to describe D-branes within
this rigorous formalism in Section~2. Our goal throughout is
two-fold. Firstly, we define and describe the physics of D-branes in
a rigorously precise K-homological setting which we hope is
accessible to mathematicians with little or no prior knowledge of
string theory, while at the same time describing another
mathematical framework in which to set and do string theory.
Secondly, we emphasize the fact that the (sometimes surprising)
physical properties of D-branes are completely transparent when the
branes are defined and analysed within the mathematical framework of
topological K-homology. Our basic aim will be to find generators for
the pertinent K-homology groups which will in turn be identified
geometrically with the D-branes of the spacetime. We give some
general results on how to do this and some of our results on the
description of the K-homology generators (Th. 2.1, 2.3) deal with
the mathematical problem of representation of cycles in a homology
theory. These results also relate K-homology with homology and
spin$^c$ bordism. Many non-trivial dynamical aspects of D-branes are
then reformulated as the problem of finding appropriate changes of
bases for the generators of the K-homology groups. Included in this
list are the constructions of bound states of D-branes from both the
``branes within branes'' mechanism~\cite{Douglas2} and the
dielectric effect~\cite{Myers1}, as well as the decay of unstable
systems of D-branes into stable bound states via tachyon
condensation on their worldvolumes~\cite{Horava1,16,Sen1,Witten1}.
While our constructions find their most natural interpretation in
the physics of D-branes, the results may also be of independent
mathematical interest. In this section we also relate our
K-homological approach with the one using derived categories.

In Section~3 we start turning our attention to explicit examples,
including a simple analysis of D-branes in spheres and projective
spaces as well as a homological treatment of T-duality. We apply the
structure results from Section~2 and introduce the Hurewicz
homomorphism. This homomorphism is important in the problem of
representation of cycles, as well as that of finding the K-homology
generators.

In Section~4 we look at some more complicated examples of D-branes
which carry torsion charges. In both the torsion and torsion-free
cases, we study in detail the phenomenon of brane
instability~\cite{DMW1,MMS1}, i.e. that some D-branes may be
unstable even though they wrap non-trivial spin$^c$ homology cycles
of the spacetime. This problem becomes particularly transparent in
topological K-homology. A number of explicit examples of torsion
D-branes are presented, including those in Lens spaces, in all (even
and odd dimensional) real projective spaces and some products
thereof, and in the basic Fermat quintic threefold and its mirror
Calabi-Yau threefold. We introduce another method of calculating the
K-homology groups, namely the Atiyah-Hirzebruch-Whitehead (AHW)
spectral sequence, and apply it to Lens spaces. We also deal with
the problem of factorizing the Hurewicz homomorphism through
K-homology. This last issue is non-trivial, since there is no map
between the K-homology and the ordinary homology spectra. We give
conditions and particular instances in which this occurs. We also
calculate and give a geometric interpretation (to the best of our
knowledge, for the first time) of the K-homology of the even real
projective spaces, which is a good example of a non spin$^c$ space,
therefore of one that doesn't allow for the application of the usual
tools (i.e. Poincar$\acute{\rm{e}}$ Duality).

 Finally, in Section~5 we examine the problem of
stabilizing certain D-branes even when they wrap homologically
trivial worldvolumes. This is achieved by regarding the ambient
spacetime as the total space of a non-trivial fibration. The
characteristic class of the fibre bundle then acts as a source of
stabilization and effectively renders the D-brane stable. We present
a number of classes of examples of this type, and in each instance
explicitly determine the topological K-homology groups. Our analysis
includes as a special case the well-known example of spherical
D-branes wrapping $\S^2\subset\S^3$~\cite{BDS1}, which in our
construction are rendered stable by virtue of the Hopf fibration
over $\complex\P^1$. The methods of this section are based in the
Leray-Serre spectral sequence, of which the AHW spectral sequence is
a special case. By the use of this spectral sequence we give general
results in the structure of the K-homology groups of spaces which
are total spaces of certain types of fibrations (coverings,
spherical fibrations, and fibrations with fibre a sphere).

In summary, in this paper we present another way of viewing
D-branes. In this setting a D-brane on a background spacetime $X$ is
a triple $(M,E,\phi)$, where $M$ is a spin$^c$ submanifold of $X$
(usually interpreted as a D-brane in string theory), $E$ is the
Chan-Paton bundle on $M$, and $\phi : M \rightarrow X$ is the
embedding. This contains more information than the usual picture of
a D-brane as a submanifold of spacetime, since it also considers the
specific embedding of $M$ in $X$ and its Chan-Paton bundle. D-brane
charges are then classified by the K-homology group of $X$, instead
of its K-theory. The relations used to define the equivalence
relation defining the group are physically meaningful and therefore
natural. In general, the K-homology group may contain elements in
which the generators aren't as above (e.g., $\phi$ may only be a
continuous map instead of an embedding of manifolds), and the
question of whether this is always the case relates to the
mathematical problem of representation of cycles for which we have
given some answers (for instance in Theorems 2.1, 2.3, the numerous
examples given). In terms of string theory, one of the main
advantages of this method is that it allows, through calculation, to
obtain a complete description, up to equivalence, of all the
D-branes that may appear in a specific spacetime. Several methods
are given to calculate these, and abstract as well as specific
examples are worked out in detail. Also, this paper recasts
interpretations of D-brane phenomena done in the setting of
K-homology of C$^*$-algebras into an algebraic topological
framework, while supplying the necessary tools to work in this new
topological framework. Although we lose the more rich structure of a
cohomology theory, we think that the ability to obtain a full
geometric characterization of D-branes more than compensates for
that.

Another task we tried to fulfil with this paper was to create, as
far as possible, a dictionary between string theory (and
particularly D-brane theory) and mathematics, trying with this to
make it easier for mathematicians not necessarily connected with
mathematical physics to acquaint themselves with some of the
mathematical problems that lye at the heart of string theory. Also,
as an effort to put the physical notions in a mathematically
rigorous setting as much as possible.

\subsection*{Acknowledgments}

We thank P.~Baum, D.~Calderbank, R.~Hepworth, J.J.~Manjar\'{\i}n,
V.~Mathai, R.~Minasian, A.~Ranicki, and E.~Rees  for helpful
discussions. We also thank the referee for various comments and
corrections which have helped to improve the material presented
herein. The work of R.M.G.R. is supported in part by FCT grant
SFRH/BD/12268/2003. The work of R.J.S. is supported in part by a
PPARC Advanced Fellowship, by PPARC Grant PPA/G/S/2002/00478, and by
the EU-RTN Network Grant MRTN-CT-2004-005104.

\newsection{K-Homology}\label{TKH}

In this section we shall develop the requisite mathematical material that
will be required throughout this paper, postponing the start of our
string theory considerations until the next section. We define
geometric K-homology and describe some basic properties of the
topological K-homology groups of a topological space. We also compare
this homology theory with other formulations of K-homology as the dual
theory to K-theory.

Unless otherwise stated, in this paper $X$ will denote a finite
CW-complex, i.e. a compact CW-complex (see \cite{8} for a rigorous
mathematical definition). For instance, compact manifolds are finite
CW-complexes.

\subsection{K-Cycles\label{ss1}}

\begin{definition}
A {\it K-cycle} on $X$ is a triple $(M,E,\phi)$ where
\begin{romanlist}
\item $M$ is a compact spin$^c$ manifold without boundary;
\item $E$ is a complex vector bundle over $M$; and
\item $\phi : M \rightarrow X$ is a continuous map.
\end{romanlist}
\label{kcycle}\end{definition}
\noindent
There are no connectedness requirements made upon $M$, and hence the
bundle $E$ can have different fibre dimensions on the different
connected components of $M$. It follows that disjoint union
$$(M_1,E_1, \phi_1) \amalg (M_2,E_2,\phi_2):=(M_1 \amalg M_2,E_1 \amalg E_2,
\phi_1 \amalg \phi_2)$$
is a well-defined operation on the set of K-cycles on $X$.

\begin{definition}
Two K-cycles $(M_1,E_1,\phi_1)$ and $(M_2,E_2,\phi_2)$ on $X$ are
  {\it isomorphic} if there exists a diffeomorphism $h : M_1 \rightarrow M_2$
  such that
\begin{romanlist}
\item $h$ preserves the spin$^c$ structures;
\item $h^{*}(E_2) \cong E_1$; and
\item The diagram
  $$\xymatrix{ M_1 \ar[r]^h  \ar[rd]_{\phi_1} & M_2 \ar[d]^{\phi_2}\\
     & X }$$
commutes.
\end{romanlist}
The set of isomorphism classes of K-cycles on $X$ is denoted $\Gamma(X)$.
\label{isoK-cycles} \end{definition}

\begin{definition}[{\bf Bordism}]
Two K-cycles $(M_1,E_1,\phi_1)$ and $(M_2,E_2,\phi_2)$ on $X$ are
{\it bordant} if there exist a compact spin$^c$ manifold $W$
with boundary, a complex vector bundle $E \rightarrow W$, and a
continuous map $\phi : W \rightarrow X$ such that the two K-cycles
$(\partial W, E|_{\partial W},\phi|_{\partial W})$ and $(M_1 \amalg
(-M_2), E_1 \amalg E_2, \phi_1 \amalg \phi_2)$ are isomorphic. Here
$-M_2$ denotes $M_2$ with the spin$^c$ structure on its tangent bundle
$TM_2$ reversed.
\label{bord} \end{definition}

\subsection{Clutching Construction\label{ss2}}

Before proceeding with further definitions, we need a construction
that will be instrumental in defining the topological K-homology groups.
Let $M$ be a compact spin$^c$ manifold and $F\to M$ a $\C^\infty$ real
spin$^c$ vector bundle with even-dimensional fibres and projection map
$\rho$. Let $\id_M^\real:=M \times \mathbb{R}$ denote the trivial real line
bundle over $M$. Then $F \oplus \id_M^\real$ is a real vector bundle
over $M$ with odd-dimensional fibres. By choosing a $\C^{\infty}$ metric
on it, we may define the unit sphere bundle
\beq
\widehat{M}=\S\bigl(F \oplus \id_M^\real\bigr)
\label{hatMdef}\eeq
by restricting the set of fibre vectors of $F \oplus \id_M^\real$ to
those which have unit norm. The spin$^c$ structures on $TM$
and $F$ induce a spin$^c$ structure on $T\widehat{M}$ by the exact
sequence lemma~\cite{1}, and hence $\widehat{M}$ is a spin$^c$
manifold. By construction, $\widehat{M}$ is a sphere bundle over $M$
with even-dimensional spheres as fibres. We denote the bundle projection
by
\beq
\pi \,:\, \widehat{M} ~\longrightarrow~ M \ .
\label{pidef}\eeq
Alternatively, we may regard the total space $\widehat{M}$ as consisting of two
  copies $\B^\pm(F)$ of the unit ball bundle $\B(F)$ of $F$ (carrying
  opposite spin$^c$ structures) glued
  together by the identity map $\Id_{\S(F)}$ on its boundary, so that
\beq
\widehat{M} = \B^+(F) \cup_{\S(F)} \B^-(F) \ .
\label{hatMBF}\eeq

For $p\in M$, let $2n=\dim_\real F_p$ with $n\in\nat$. The group
Spin$^{c}(2n)$ has two irreducible half-spin representations. The
spin$^c$ structure on $F$ associates to these representations complex
vector bundles $S_0(F)$ and $S_1(F)$ of equal rank $2^{n-1}$ over
  $M$. Their Whitney sum $S(F)=S_0(F)\oplus S_1(F)$ is a bundle of
  Clifford modules over $M$ such that $\Cl(F)\otimes\complex\cong{\rm
    End}\,S(F)$, where $\Cl(F)$ is the real Clifford algebra bundle of
  $F$. Let $S_+(F)$ and
  $S_-(F)$ be the spinor bundles over $F$ obtained from pullbacks to $F$ by the
  bundle projection $\rho :F \rightarrow M$ of $S_0(F)$ and $S_1(F)$,
  respectively. Clifford multiplication induces a bundle map $F
  \otimes S_0(F)\rightarrow S_1(F)$ that defines a vector bundle map
  $\sigma :S_+(F)\rightarrow S_-(F)$ covering $\Id_F$ which is an
  isomorphism outside the zero section of $F$. Since the ball bundle
  $\B(F)$ is a sub-bundle of $F$, we may form spinor bundles over
$\B^\pm(F)$ as the restriction bundles $\Delta^\pm(F)
=S_\pm(F)|_{\B^\pm(F)}$. We can then glue $\Delta^{+}(F)$ and $\Delta^{-}(F)$
along
  $\S(F)=\partial\B(F)$ by the Clifford multiplication map $\sigma$
  giving a vector bundle over $\widehat{M}$ defined by
\beq
H(F) = \Delta^{+}(F)\cup_{\sigma}\Delta^{-}(F) \ .
\label{HFdef}\eeq
For each $p\in M$, $H(F)|_{ \pi^{-1}(p)}$ is the
  Bott generator vector bundle over the even-dimensional sphere
  $\pi^{-1}(p)$~\cite{1}. Thus, starting from the triple $(M,F,\rho)$
  we have constructed another triple $(\widehat{M},H(F),\pi)$.

  \begin{definition}
If $(M,E,\phi)$ is a K-cycle on $X$ and $F$ is a $\C^\infty$ real spin$^c$
vector
bundle over $M$ with even-dimensional fibres, then the process of
obtaining the K-cycle $(\widehat{M},H(F) \otimes \pi^{*}(E),\phi \circ
\pi)$ from $(M,E,\phi)$ is called \it{vector bundle modification}.
    \label{VBM} \end{definition}

\subsection{Topological K-Homology\label{KHom}}

We are now ready to define the topological K-homology groups of the
space $X$.

\begin{definition}
  The \textit{topological K-homology group of
  $X$} is the group obtained from quotienting $\Gamma(X)$ by the
  equivalence relation $\sim$ generated by the relations of
\begin{romanlist}
\item bordism;
\item direct sum: if $E= E_1 \oplus E_2$, then $(M,E,\phi) \sim (M,E_1,\phi)
  \amalg (M,E_2,\phi)$; and
\item vector bundle modification.
\end{romanlist}
  The group operation is induced by disjoint union of K-cycles. We denote this
  group by $\K_\sharp^{\rm t}(X):=\Gamma(X)\,/\sim$, and the homology class of
  the K-cycle $(M,E,\phi)$ by $[M,E,\phi]\in\K^{\rm t}_\sharp(X)$.
\label{Tgroup} \end{definition} \noindent The manifolds $M \amalg N$
and $N \amalg M$ are bordant through a bordism which clearly induces
a bordism of the respective K-cycles. It follows that
$[\emptyset,\emptyset,\emptyset]$ is the identity for the operation
induced by disjoint union of K-cycles and $[-M,E,\phi]$ is the
inverse of $[M,E,\phi]$, where $-M$ denotes $M$ with its spin$^c$
structure reversed.The operation is also clearly associative and
commutative. Thus $\K_\sharp^{\rm t}(X)$ is an abelian group. Since
the equivalence relation on $\Gamma(X)$ preserves the parity of the
dimension of $M$ in K-cycles $(M,E,\phi)$, one can define the
subgroup $\K_{0}^{\rm t}(X)$ (resp. $\K_{1}^{\rm t}(X)$) consisting
of classes of K-cycles $(M,E,\phi)$ for which all connected
components $M_i$ of $M$ are of even (resp. odd) dimension. Then
$\K_\sharp^{\rm t}(X) = \K_{0}^{\rm t}(X) \oplus \K_{1}^{\rm t}(X)$
has a natural $\zed_2$-grading.

The geometric construction of K-homology is functorial. If $f:X
\rightarrow Y$ is a continuous map, then the induced
  homomorphism $$f_{*} \,:\, \K_\sharp^{\rm t}(X) ~\longrightarrow~
  \K_\sharp^{\rm t}(Y)$$
of $\zed_2$-graded abelian groups is given on classes of K-cycles
$[M,E,\phi]\in\K^{\rm
  t}_\sharp(X)$ by $$f_{*}[M,E,\phi] := [M,E,f\circ\phi] \ . $$
One has $(\Id_X)_*=\Id_{\K^{\rm t}_\sharp(X)}$ and $(f\circ g)_*=f_*\circ
g_*$. Since vector bundles over $M$ extend to vector bundles over
$M\times[0,1]$, it follows by bordism that induced homomorphisms
depend only on their homotopy classes.

If $\pt$ denotes a one-point topological space, then the only
K-cycles on $\pt$ are $(\pt,\pt\times\complex^k,\Id_\pt)$ with
$k\in\nat$. Thus $\K_{0}^{\rm t}(\textrm{pt}) \cong \mathbb{Z}$ and
$\K_{1}^{\rm t}(\textrm{pt}) \cong 0$. The collapsing map
$\varepsilon:X \rightarrow \textrm{pt}$ then induces an epimorphism
\beq \varepsilon_{*} \,:\, \K_\sharp^{\rm t}(X)  ~\longrightarrow ~
\K_\sharp^{\rm t}(\textrm{pt}) \cong \mathbb{Z} \ .
\label{collapseepi}\eeq The {\it reduced} topological K-homology
group of $X$ is \beq \widetilde{\K}_\sharp^{\rm t}(X):=
\ker\varepsilon_{*} \ . \label{redKhom}\eeq Since the map
(\ref{collapseepi}) is an epimorphism with left inverse induced by
the inclusion of a point $\iota:\pt\hookrightarrow X$, one has
$\K_\sharp^{\rm t}(X)\cong\zed\oplus\widetilde{\K}_\sharp^{\rm
t}(X)$ for any space $X$.

\subsection{Computational Tools\label{CompTools}}

Before adding further structure to this K-homology theory, we pause to
describe some basic technical results which will aid in calculating
the groups $\K_\sharp^{\rm t}(X)$, particularly in the subsequent sections
when we shall seek explicit K-cycle representatives for their
generators. In what follows we shall use the notation
$[n]:=\{1,2,\dots,n\}$.

\begin{lemma} \label{l1}
  $\K_\sharp^{\rm t}(X)$ is generated by classes of K-cycles $[M,E,\phi]$ where
  $M$ is connected.
\end{lemma}
\begin{proof}
Let $\{ M_{i}\}_{i \in I}$  be the set of connected components of $M$.
Since $M$ is compact, $I$ is a finite set. Defining $E_{i} := E|_{
  M_{i}}$ and $\phi_{i} := \phi|_{M_{i}}$, we have $E \cong \coprod_{i \in I}
\,E_{i}$ and $\phi = \coprod_{i \in I}\,\phi_{i}$ so that $[M,E,\phi]
= \sum_{i \in I}\,[M_{i},E_{i},\phi_{i}]$.
\end{proof}

\begin{lemma}\label{l2}
If $\{X_{j}\}_{j \in J}$ is the set of connected components of $X$
then $\K_\sharp^{\rm t}(X)= \bigoplus _{j \in J}\,\K_\sharp^{\rm t}(X_{j})$.
\end{lemma}
\begin{proof}
Let $[M,E,\phi] \in\K_\sharp^{\rm t}(X)$ with $\{ M_{i}\}_{i \in I}$ the
set of connected components of $M$. As in the proof of Lemma~\ref{l1},
one has $[M,E,\phi] = \sum_{i \in I}\,
[M_{i},E_{i},\phi_{i}]$. For any $i \in I$, $M_{i}$ is connected
and $\phi_{i}$ is continuous, and so there exists $j_{i} \in J$ such that
$\phi_{i}(M_{i}) \subset X_{j_{i}}$. Thus $[M_{i},E_{i},\phi_{i}]\in
\K_\sharp^{\rm t}(X_{j_{i}})$ and so $[M,E,\phi] \in  \bigoplus _{j \in
  J}\,\K_\sharp^{\rm t}(X_{j})$. Conversely, let  $[M_{i},E_{i},\phi_{i}]
\in\K_\sharp^{\rm t}(X_{j_{i}})$ for some $i \in [n]$ and $j_{i} \in J$.
Defining $M:= \coprod_{i \in [n]}\,M_{i}$, $E:= \coprod_{i \in
  [n]}\,E_{i}$ and $\phi := \coprod_{i \in [n]}\,\phi_{i}$, one has
$[M,E,\phi ] \in\K_\sharp^{\rm t}(X)$. The conclusion now follows by
considering the image of the class $\sum_{i \in
  [n]}\,[M_{i},E_{i},\phi_{i}]$ in $\K_\sharp^{\rm t}(X)$ under the
homomorphism induced by the continuous map $\coprod_{i \in [n]}\,
\iota_{j_{i}}$, where $\iota_{j_{i}} : X_{j_{i}} \hookrightarrow X$
are the canonical inclusions.
\end{proof}

\begin{lemma} \label{l3}
  Let $(M,E,\phi)$ be a K-cycle on $X$. Suppose that the degree~$0$
  topological K-theory group $\K_{\rm t}^{0}(M)$ of $M$ is
  generated as a $\mathbb{Z}$-module by classes $[F_1],
  \ldots,[F_p]$ of complex vector bundles over $M$. Then $[M,E,\phi]$
  belongs to the $\mathbb{Z}$-submodule of $\K_\sharp^{\rm t}(X)$ generated by
$\{
  [M,F_{i},\phi] \}_{i \in [p]}$.
\end{lemma}
\begin{proof}
By hypothesis there exist integers $n_1,\ldots ,n_p$ such that $[E] =
\sum_{i\in[p]}\,n_{i}~[F_{i}]$. Without loss of generality we may
suppose that $n_{j} \geq 0$ for all $1 \leq j \leq m$ while $n_{j}<
0$ for all $m+1 \leq j \leq p$, for some $m$ with $1 \leq m \leq p$. Then
$$ [E] + \sum_{i=m+1}^p (-n_{i})~[F_{i}] = \sum_{i=1}^m\,n_{i}~[F_{i}]
\ , $$ which implies that there exists an integer $k \geq 0$ such that
$$E\oplus \Big(\,\bigoplus _{i=m+1}^p~\bigoplus_{j=1}^{-n_{i}}\,F_{i}\Big)
\oplus\id_M^{\complex^{k}} = \Big(\,\bigoplus _{i=1}^m
{}~\bigoplus_{j=1}^{n_{i}}\,F_{i}\Big) \oplus\id_M^{\complex^{k}} \ . $$
Going down to classes in
$\K_\sharp^{\rm t}(X)$ using the direct sum relation, we then have
$$\big[M,E,\phi\big] + \sum_{i=m+1}^p
(-n_{i})~\big[M,F_{i},\phi] +
\big[M,\id_M^{\complex^k},\phi\big] = \sum_{i=1}^m\,n_{i}~
\big[M,F_{i},\phi\big] +\big[M,\id_M^{\complex^k},\phi\big]$$
which implies that $[M,E,\phi] =
\sum_{i\in[p]}\,n_{i}~[M,F_{i},\phi]$.
\end{proof}

\begin{cor}\label{corKclass}
The homology class of a K-cycle $(M,E,\phi)$ on $X$ depends only
  on the K-theory class of $E$ in $\K_{\rm t}^{0}(M)$.
\end{cor}

\begin{lemma}\label{hom}
The homology class of a K-cycle $(M,E,\phi)$ on $X$ depends only on
the homotopy class of $\phi$ in $[M,X]$.
\end{lemma}
\begin{proof}
This follows from $[M,E,\phi]=[M,E, \phi \circ \textrm{id}_M]=
\phi_{*}[M,E,\textrm{id}_M]$ and the fact that induced homomorphisms
depend only on their homotopy classes.
\end{proof}

\begin{cor}\label{l5}
If $X$ is a compact spin$^{c}$ manifold without boundary, $E\to X$ is
a complex vector bundle and $\phi : X
\rightarrow X$ is a continuous map, then $[X,E,\phi]$
  depends only on the homotopy class of $\phi$ in $[X,X]$.
\end{cor}

\subsection{Cap Product\label{CapProd}}

The {\it cap product} is the $\zed_2$-degree preserving bilinear
pairing $$\cap\,:\,\K_{\rm t}^{0}(X)\otimes
\K_\sharp^{\rm t}(X)~\longrightarrow~\K_\sharp^{\rm t}(X)$$ given for any
complex
vector bundle $F\to X$ and K-cycle class $[M,E,\phi]\in\K^{\rm
  t}_\sharp(X)$ by $$[F] \cap [M,E,\phi]:=[M,\phi^{*}F\otimes E ,\phi] $$
and extended linearly. It makes $\K_\sharp^{\rm t}(X)$ into a module over the
ring $\K^0_{\rm t}(X)$. Later on (see Sections~\ref{IIArem}
and~\ref{suspension}) we will see that this product can be extended to
a bilinear form
\beq
\cap\,:\,\K_{\rm t}^i(X)\otimes\K^{\rm t}_j(X)~\longrightarrow~
\K^{\rm t}_{e(i+j)}(X) \ ,
\label{capprodext}\eeq
where we have denoted the mod~2 congruence class of an integer
$n\in\zed$ by $$e(n):=\left\{\begin{array}{ll}0 &
      , \quad n~\textrm{even}\\
      1 & , \quad n~\textrm{odd}
\end{array} \right. \ . $$
The construction utilizes Bott periodicity and the isomorphism $\K^1_{\rm
  t}(X)\cong\K^0_{\rm t}(\Sigma X)$, where $\Sigma X$ is the reduced
suspension of the topological space $X$. The product $ \cap : \K_{\rm
  t}^{1}(X)\otimes\K_i^{\rm t}(X) \rightarrow\K_{e(i+1)}^{\rm t}(X)$
is given by the pairing $
\cap :\K_{\rm t}^{0}(\Sigma X)\otimes\K_{e(i-1)}^{\rm t}(\Sigma X) \rightarrow
\K_{e(i-1)}^{\rm t}(\Sigma X).$

\subsection{Exterior Product\label{ExtProd}}

If $X$ and $Y$ are spaces, then the {\it exterior product}
$$ \times \,:\, \K^{\rm t}_{i}(X) \otimes \K^{\rm t}_{j}(Y) ~\longrightarrow~
\K^{\rm t}_{e(i+j)}(X \times Y)$$ is given for classes of K-cycles
$[M,E,\phi]\in\K^{\rm t}_i(X)$ and $[N,F,\psi]\in\K^{\rm t}_j(Y)$ by
$$\big[M,E,\phi\big]\times \big[N,F,\psi\big]:=\big[M \times N,E
\boxtimes F, (\phi,\psi)\big] \ , $$ where $M \times N$ has the
spin$^c$ product structure uniquely induced by the spin$^c$
structures on $M$ and $N$, and $E \boxtimes F$ is the vector bundle
over $M \times N$ with fibres $(E \boxtimes F)_{(p,q)}=E_p \otimes
F_q$ for $(p,q)\in M\times N$. This product is natural with respect
to continuous maps and there is the following version (due to
Atiyah) of the K\"{u}nneth theorem in K-homology~\cite{17}.

\begin{theorem} If $X$ is a CW-complex and  $Y$ is a compact topological space, then for
 each $i=0,1$ there is a natural short exact sequence
$$0 ~\longrightarrow~ \bigoplus_{e(j+l)=i}
\K^{\rm t}_{j}(X) \otimes\K^{\rm t}_{l}(Y)
{}~\longrightarrow~\K^{\rm t}_i(X \times
Y)~\longrightarrow~\bigoplus_{e(j+l)=e(i+1)} \Tor\big(\K^{\rm
t}_{j}(X)\,,\,\K^{\rm t}_l(Y)\big)~\longrightarrow~ 0 \ . $$
\label{Kunneththm}\end{theorem} \noindent

This sequence always splits, although unnaturally (see \cite{Bod},
\cite{Deu} where a generalization to the case of C$^*$-algebras is
given, and \cite{RosSch}).

\subsection{Spectral K-Homology\label{RelK}}

Let $X$ be a general (not necessarily compact) CW-complex. We then
define
$$\K_{i}^{\rm t}(X) :=
\lim_{\stackrel{\scriptstyle\longrightarrow}{Y}}\K_{i}^{\rm t}(Y),$$
for $i=0,1$, where the limit runs over the finite CW-subcomplexes
$Y$ of $X$. By defining $\K_{2k}^{\rm t}(X) := \K_{0}^{\rm t}(X)$
and $\K_{2k+1}^{\rm t}(X) := \K_{1}^{\rm t}(X)$ for all $k \in
\mathbb{Z}$, one has that $\K_\sharp^{\rm t}(X)$ is a 2-periodic
unreduced homology theory on the category of CW-complexes. On the
other hand, K-theory is a 2-periodic cohomology theory which can be
defined in terms of its spectrum $\K^{\U}= \{ \K^{\U}_{n}\}^{~}_{n
  \in \mathbb{Z}} $, where $\K^{\U}_{2k} := \mathbb{Z}
\times \BU(\infty)$ and $\K^{\U}_{2k+1} := \U(\infty)$ are the
classifying spaces for $\K_{\rm t}^0$ and $\K_{\rm t}^1$, respectively. Thus we
can define~\cite{8} a homology theory related to $\K_{\rm t}^\sharp$ by
the inductive limit
$$\K^{\rm s}_i(X,Y) := \lim_{\stackrel{\scriptstyle\longrightarrow}
{\scriptstyle n}} \, \pi^{~}_{n+i} \big((X/Y) \wedge \K^{\U}_{n}\,\big)$$
for all $i\in \mathbb{Z}$, where $Y$ is a closed subspace of the
topological space $X$ and $\wedge$ denotes the smash product. Bott
periodicity then implies that this is a 2-periodic homology theory.

For any finite CW-complex $X$, we can construct a map $$\mu^{\rm s} \,:\,
\K_{i}^{\rm t}(X)~\longrightarrow~\K^{\rm s}_{i}(X):=\K_i^{\rm
  s}(X,\emptyset)$$ given by
$$\mu^{\rm s}\big([M,E,\phi]\big) := \phi_{*}\big([E] \cap [M]_{\rm s}\big)$$
on
classes of K-cycles and extended by linearity. Here $\cap :\K_{\rm t}^{0}(X)
\otimes\K^{\rm s}_{i}(X) \rightarrow\K^{\rm s}_{i}(X)$ for $i=0,1$ is the
spectrally
defined cap product with $[M]_{\rm s}$ the fundamental class of the manifold
$M$ in $\K^{\rm s}_{i}(X)$~\cite{8}. The transformation $\mu^{\rm s}$ is an
isomorphism which is natural in $X$, and so it defines a natural
equivalence between the functors $\K^{\rm t}_\sharp$ and $\K^{\rm
  s}_\sharp$~\cite{3}. It follows that $\K^{\rm t}_\sharp(X)$ is a
realization of $\K^{\rm s}_\sharp(X)$. The map $\mu^{\rm s}$ is also
compatible with cap products, i.e. $\mu^{\rm s} (\xi\cap \alpha) = \xi\cap
\mu^{\rm s}(\alpha)$ for all $\xi\in\K_{\rm t}^\sharp(X)$ and $\alpha \in
\K^{\rm t}_\sharp(X)$, or equivalently there is a commutative diagram
 $$\xymatrix{ \K_{\rm t}^i(X)\otimes\K_j^{\rm t}(X) \ar[r]^{\quad\cap}
   \ar[d]_{\textrm{id}_{\K_{\rm t}^i(X)}\otimes \mu^{\rm s}}
   &\K_{e(i+j)}^{\rm t}(X) \ar[d]^{\mu^{\rm s}}\\
      \K_{\rm t}^i(X)\otimes\K_j^{\rm s}(X) \ar[r]^{\quad\cap} &
\K^{\rm s}_{e(i+j)}(X) \ . } $$
In particular, if $X$ is a compact connected spin$^c$ manifold
without boundary, then $$\mu^{\rm s}\big([X,\id_X^\complex,{\rm
id}^{~}_{X}]\big)
 ~=~ ({\rm id}^{~}_{X})_{*}\big([\id_X^\complex] \cap
  [X]_{\rm s}\big) ~=~({\rm id}^{~}_{X})_{*}\big([X]_{\rm
    s}\big)~=~[X]_{\rm s}$$
in $\K^{\rm s}_\sharp(X)$, with $\id_X^\complex$ the
trivial complex line bundle over $X$. Since $\mu^{\rm s}$ is a natural
equivalence between $\K^{\rm t}_\sharp$ and $\K^{\rm s}_\sharp$ it follows
that,
within the framework of topological K-homology as the dual theory
to K-theory, $[X,\id_X^\complex,\textrm{id}^{~}_{X}]$ is the fundamental
class of $X$ in $\K^{\rm t}_\sharp(X)$.

One can give a definition of relative K-homology groups $\K_i^{\rm
  t}(X,Y)$ in such a way that there is also a map $\mu^{\rm s} :\K_i^{\rm
  t}(X,Y) \rightarrow\K^{\rm s}_i(X,Y)$ which defines a natural equivalence
between functors on the category of topological spaces having the
homotopy type of finite CW-pairs $(X,Y)$~\cite{3}. One can also give
a bordism description of $\K^{\rm t}_\sharp(X,Y)$ as follows. We
consider the set of all triples $(M,E,\phi)$ where
\begin{romanlist}
\item $M$ is a compact spin$^c$ manifold with boundary;
\item $E$ is a complex vector bundle over $M$; and
\item $\phi : M \rightarrow X$ is a continuous map with $\phi(\partial
  M)\subset Y$.
\end{romanlist}
This set is quotiented by relations of bordism (modified from
Definition~\ref{bord} by the requirement that
$M_1\amalg(-M_2)\subset\partial W$ is a regularly embedded submanifold
of codimension~$0$ with
$\phi(\partial W\setminus M_1\amalg(-M_2))\subset Y$), direct sum and
vector bundle modification. The collection of equivalence classes is
a $\zed_2$-graded abelian group with operation induced by disjoint
union of relative K-cycles~\cite{3}.

Since K-homology is a generalized homology theory, there is a long
exact homology sequence for any pair $(X,Y)$. Because $\K^{\rm t}_\sharp$
is a 2-periodic theory, this sequence truncates to the six-term exact
sequence
$$\xymatrix{\K_{0}^{\rm t}(Y) \ar[r]^{\iota_*} &\K_{0}^{\rm
    t}(X) \ar[r]^{\varsigma_*} &\K_{0}^{\rm t}(X,Y) \ar[d]^{\partial} \\
\K_{1}^{\rm t}(X,Y)\ar[u]^{\partial} &\K_{1}^{\rm t}(X)
\ar[l]^{\quad\varsigma_*}
&\K_{1}^{\rm t}(Y) \ar[l]^{\iota_*} } $$
where the horizontal arrows are induced by the canonical inclusion
maps $\iota:Y\hookrightarrow X$ and
$\varsigma:(X,\emptyset)\hookrightarrow(X,Y)$. In the bordism
description, the connecting homomorphism is given by the boundary map
$$\partial[M,E,\phi]:=[\partial M,E|_{\partial M},\phi|_{\partial
M}]$$ on classes of K-cycles and extended by linearity. One also has
the usual excision property. If $U\subset Y$ is a subspace whose
closure lies in the interior of $Y$, then the inclusion
$\varsigma^U:(X\setminus U,Y\setminus U)\hookrightarrow(X,Y)$ induces an
isomorphism $$\varsigma_*^U\,:\,\K^{\rm t}_\sharp(X\setminus U,Y\setminus
U)~\stackrel{\approx}{\longrightarrow}~\K^{\rm t}_\sharp(X,Y)$$ of
$\zed_2$-graded
abelian groups.

\subsection{Analytic K-Homology\label{AnK}}

We will now briefly describe the relationship between $\K_\sharp^{\rm
  t}(X)$ and the analytic K-homology groups of a finite CW-complex
  $X$ (the construction of these groups follows verbatim to the case of a compact metrizable
topological space -- see \cite{1}).

\subsubsection{The Group $\K^{\rm a}_{0}(X)$}\label{ak0}

Let $\Omega_0(X)$ be the set of all quintuples $(\mathcal{H}_{0},\psi_{0},
\mathcal{H}_{1},\psi_{1},T)$ where
\begin{romanlist}
\item for each $i=0,1$, $\mathcal{H}_{i}$ is a separable Hilbert space;
\item for each $i=0,1$, $\psi_{i} : \C(X) \rightarrow
  \mathcal{L}(\mathcal{H}_{i})$ is a unital algebra $*$-homomorphism, where
  $\C(X)$ is the $C^*$-algebra of continuous complex-valued functions
  on $X$ and $\mathcal{L}(\mathcal{H}_{i})$ is the $C^*$-algebra of
  bounded linear operators on $\mathcal{H}_{i}$; and
\item $T : \mathcal{H}_{0} \rightarrow \mathcal{H}_{1}$ is a bounded
  Fredholm operator such that the operator $T \circ
\psi_{0} (f) - \psi_{1}(f) \circ T$ is compact for
all $f \in \C(X)$.
\end{romanlist}
We can define on $\Omega_0(X)$ a direct sum operation and an
equivalence relation generated by isomorphism,
direct sum with a trivial object, and compact perturbation of Fredholm
operators. The quotient set is, with direct sum, an abelian group
$\K^{\rm a}_{0}(X)$ called the {\it degree~$0$ analytic} K-homology group of
$X$. There is an epimorphism
\beq
\Index\,:\,\K^{\rm a}_{0}(X)~\longrightarrow~ \mathbb{Z}
\label{Indexepi}\eeq
given by
$$\Index[\mathcal{H}_{0},\psi_{0},\mathcal{H}_{1},\psi_{1},T]:=
\Index~T \ . $$

Suppose that $X$ is a closed C$^{\infty}$ manifold, $E_{0}$, $E_{1}$ are
complex
C$^\infty$ vector bundles over $X$ and $D :\C^{\infty}(E_{0})
\rightarrow \C^{\infty}(E_{1})$ is an elliptic pseudo-differential
operator on $X$. Then one can construct an element $[D] \in\K^{\rm
  a}_{0}(X)$ which depends only on $D$. All elements of $\K^{\rm
  a}_{0}(X)$ arise in this way, and in this case we have that
$$\Index[D]=\Index~D$$ is the analytic index of $D$ regarded as a
Fredholm operator~\cite{1}.

\subsubsection{The Group $\K^{\rm a}_{1}(X)$}\label{ak1}

Let $\Omega_1(X)$ be the set of all pairs $(\mathcal{H},\tau)$ where
\begin{romanlist}
\item $\mathcal{H}$ is a separable Hilbert space; and
\item $\tau : \C(X) \rightarrow \mathcal{Q}(\mathcal{H})$ is a unital algebra
$*$-homomorphism, where $\mathcal{Q}(\mathcal{H})=\mathcal{L}(\mathcal{H})/
\mathcal{K}(\mathcal{H})$ is the Calkin algebra with
$\mathcal{K}(\mathcal{H})$ the closed ideal in
$\mathcal{L}(\mathcal{H})$ consisting of compact operators on
$\mathcal{H}$.
\end{romanlist}
On $\Omega_1(X)$ we can define a direct sum operation and an
equivalence relation using unitary equivalence and triviality. The
quotient set is, with direct sum, an abelian group $\K^{\rm a}_{1}(X)$
called the {\it degree~$1$ analytic} K-homology group of $X$. It
coincides with the Brown-Douglas-Fillmore group
$\Ext(X):=\Ext(\C(X),\mathcal{K})$ of equivalence classes of
extensions of the $C^*$-algebra $\C(X)$ by compact operators
$\mathcal{K}$~\cite{BDF1}, defined by $C^*$-algebras
$\mathcal{A}$ which fit into the short exact sequence
\beq
0~\longrightarrow~\mathcal{K}~\longrightarrow~\mathcal{A}~
\longrightarrow~\C(X)~\longrightarrow~0 \ .
\label{Extseq}\eeq

Suppose that $X$ is a closed C$^{\infty}$ manifold, $E$ is a complex
C$^\infty$ vector bundle over $X$ and $A : \C^{\infty}(E) \rightarrow
\C^{\infty}(E)$ is a self-adjoint elliptic pseudo-differential
operator on $X$. Then one can construct an element $[A] \in\K^{\rm
  a}_{1}(X)$ which depends only on $A$. All elements of $\K^{\rm
  a}_{1}(X)$ arise in this way~\cite{1}.

\subsubsection{The Group $\K^{\rm a}_\sharp(X)$}\label{ak}

We define $\K^{\rm a}_\sharp(X):=\K^{\rm a}_{0}(X) \oplus\K^{\rm
  a}_{1}(X)$ to be the {\it analytic
K-homology group of $X$}. There is a natural notion of induced homomorphism
$f_*:\K^{\rm a}_\sharp(X)\to\K^{\rm a}_\sharp(Y)$ for continuous maps $f: X
\rightarrow Y$ such that $\K^{\rm a}_\sharp$ is a 2-periodic homology
theory. Let us now describe its explicit relation to the topological
K-homology theory $\K_\sharp^{\rm t}$.

Let $(M,E,\phi)$ be a K-cycle on the finite CW-complex $X$ and
$\Dirac$ the Dirac operator on the spin$^c$ manifold $M$. Then the
twisted Dirac operator $\Dirac_{E}$ is an elliptic first order
differential operator on $M$ (self-adjoint if $\dim_\real M$ is odd;
the "corner operator" that maps odd spinors to even spinors, and
vice-versa, if $\dim_\real M$ is even). Hence it determines an
element $[\Dirac_{E}]=[E]\cap[\Dirac]\in\K^{\rm a}_\sharp(M)$ with
the degree preserved, and $\phi_{*}[\Dirac_{E}]\in\K^{\rm
a}_\sharp(X)$. This element depends only on the K-homology class
$[M,E,\phi]\in\K^{\rm t}_\sharp(X)$, and so we get a well-defined
map of $\zed_2$-graded abelian groups
$$\mu^{\rm
  a}\,:\,\K^{\rm t}_\sharp(X)~\longrightarrow~\K^{\rm
  a}_\sharp(X)$$ given by $$\mu^{\rm a}[M,E,\phi]:= \phi_{*}[\Dirac_{E}]$$
on classes of K-cycles. This map is an isomorphism which is
natural~\cite{1}. The index epimorphism (\ref{Indexepi}) and the
epimorphism (\ref{collapseepi}) induced by the collapsing map
together with the isomorphism $\mu^{\rm a}$ generate a commutative
diagram \beq \xymatrix{\K^{\rm t}_\sharp(X)
\ar[rd]^{\varepsilon_{*}}
   \ar[d]_{\mu^{\rm a}} & \\ \K^{\rm a}_\sharp(X)
   \ar[r]_{\quad\Index} & \mathbb{Z} \ . }
\label{Indexcommdiag}\eeq

As a consequence of what was said in the previous and this
subsections one has the following.

\begin{theorem}
Let $X$ be a finite CW-complex. Then all three forms of K-homology
are isomorphic, i.e.
$$\K^{\rm t}_\sharp(X) \, \cong \, \K^{\rm a}_\sharp(X) \, \cong \K^{\rm s}_i(X)$$
are isomorphic as graded groups.
\end{theorem}

\subsection{Poincar\'{e} Duality}\label{ss7}

Let $X$ be an $n$-dimensional compact manifold with (possibly empty)
boundary, and $\B(TX) \rightarrow X$ and $\S(TX) \rightarrow X$ the
unit ball and sphere bundles of $X$. An element
$\tau\in\K_{\rm t}^{e(n)}(\B(TX),\S(TX))$ is called a
\textit{Thom class} or an \textit{orientation} for $X$ if
$\tau|_{(\B(TX)_x,\S(TX)_x)} \in \K_{\rm
  t}^{e(n)}(\B(TX)_x,\S(TX)_x)\cong\K_{\rm t}^{0}(\textrm{pt})$ is a
generator for all $x\in X$. The manifold $X$ is said to be {\it
  $\K_{\rm t}^\sharp$-orientable} if it has a Thom class. In that case the
usual cup product on the topological K-theory ring yields the Thom
isomorphism
$$\mathfrak{T}_X^{~} \,:\,\K_{\rm t}^i\big(X\big)~
\stackrel{\approx}{\longrightarrow}~\K_{\rm
  t}^{e(i+n)}\big(\B(TX)\,,\,\S(TX)\big)$$ given for $i=0,1$ and
$\xi\in\K_{\rm t}^i(X)$ by
$$\mathfrak{T}_X^{~}(\xi):=\pi_{\B(TX)}^{*}(\xi)\cup\tau \ , $$
where $\pi^{~}_{\B(TX)}:\B(TX) \rightarrow X$ is the bundle projection. This
construction also works by replacing the tangent bundle of $X$ with any ${\rm
  O}(r)$ vector bundle $V\to X$, defining a Thom isomorphism
$$\mathfrak{T}_{X,V}^{~}\,:\,\K^i_{\rm t}\big(X\big) ~
\stackrel{\approx}{\longrightarrow}~\K_{\rm
  t}^{e(i+r)}\big(\B(V)\,,\,\S(V)\big)$$ given by
\beq
\mathfrak{T}^{~}_{X,V}(\xi):=\pi_{\B(V)}^{*}(\xi)\cup\tau^{~}_{V} \ ,
\label{ThomisodefV}\eeq
where the element $\tau^{~}_{V}\in\K_{\rm t}^{e(r)}(\B(V),\S(V))$ is called
the \textit{Thom class of $V$}.

Any $\K^\sharp_{\rm t}$-oriented manifold $X$ of dimension $n$ has a
uniquely determined fundamental class $[X]_{\rm s}
\in\K^{\rm s}_{e(n)}(X,\partial X)$. One then has the
Poincar\'{e} duality isomorphism
$$\Phi_X^{~}\,:\,\K^i_{\rm
  t}(X)~\stackrel{\approx}{\longrightarrow}~\K^{\rm s}_{e(i+n)}(X,\partial X)$$
given for $i=0,1$ and $\xi\in\K_{\rm t}^i(X)$ by taking the cap
product
\beq
\Phi^{~}_X(\xi):=\xi\cap [X]_{\rm s} \ .
\label{PhiXdef}\eeq
In particular, if $X$ is a compact spin$^c$ manifold of dimension~$n$
without boundary, then $X$ is $\K^\sharp_{\rm t}$-oriented and so in this
case we also have a Poincar\'{e} isomorphism as above~\cite{3,8}
giving
\begin{displaymath}
\K_{0}^{\rm t}(X) ~\cong~ \K_{\rm t}^{e(n)}(X) \ , \qquad
\K_{1}^{\rm t}(X) ~\cong~ \K_{\rm t}^{e(n+1)}(X) \ .
\end{displaymath}

\subsection{Universal Coefficient Theorem\label{UCT}}

Let $X$ be a compact $n$-dimensional spin$^c$ manifold without
boundary. In the framework of analytic K-homology, the six-term
exact sequence in K-theory corresponding to an extension
(\ref{Extseq}) reduces to the short exact sequence
$$\begin{matrix}0&\longrightarrow&\K^{~}_0(\mathcal{K})&\longrightarrow&
\K^{~}_0(\mathcal{A})&\longrightarrow&
\K_{\rm t}^0(X)&\longrightarrow&0 \\[2pt]
 & & \parallel & & & & & & \\[2pt] & & \zed & &
 & & & & \end{matrix} $$
and therefore defines an element of $\Ext(\K_{\rm t}^0(X),\zed)$ in
homological algebra. Conversely, there is a universal coefficient
theorem given by the short exact sequence~\cite{RosSch,12,13} (here
we're assuming that this extension is in the kernel of $\gamma_{00}
: Ext(X) \rightarrow Hom(K^1_t(X),\mathbb{Z})$ -- see \cite{RDou}
for a definition of this homomorphism)
$$\begin{matrix}0&\longrightarrow&\Ext\big(\K_{\rm
    t}^{0}(X)\,,\,\mathbb{Z}\big)&\longrightarrow&
\Ext\big(\C(X)\,,\,\mathcal{K}\big)&\longrightarrow&
\Hom\big(\K_{\rm t}^{1}(X)\,,\,\mathbb{Z})&\longrightarrow&0 \ . \\[2pt]
 & & & & \parallel & & & & \\[2pt] & & & & \K^{\rm a}_{1}(X)
 & & & & \end{matrix} $$
This sequence splits, although not naturally. For definiteness, suppose
that the degree~$0$ K-theory of $X$ can be split as $\K^0_{\rm
  t}(X)=\Lambda_{\K^0_{\rm t}(X)}\oplus\tor_{\K^0_{\rm t}(X)}$, where
the lattice $\Lambda_{\K^0_{\rm t}(X)}=\K^0_{\rm
t}(X)\,/\,\tor_{\K^0_{\rm
  t}(X)}$ is the free part of the K-theory group and $\tor_{\K^0_{\rm
  t}(X)}=\bigoplus_{i=1}^m\mathbb{Z}_{n_{i}}$ is its torsion
subgroup (such a split is neither unique nor natural). Since $X$ is
a finite CW-complex, the abelian group $\K^{0}_{\rm t}(X)$ is
finitely generated and we have
$$\Ext\big(\K_{\rm t}^{0}(X)\,,\,\mathbb{Z}\big)~=~\Ext\big(\tor_{\K^0_{\rm
    t}(X)}\,,\,\mathbb{Z}\big)
{}~\cong~\bigoplus_{i=1}^m\,\Ext\big(\zed_{n_i}\,,\,\zed\big)~\cong~
\bigoplus_{i=1}^m\,\zed_{n_i}~=~\tor_{\K^0_{\rm t}(X)}$$ from which
it follows that $$\K^{\rm a}_{1}\big(X\big)\cong\Hom\big(\K_{\rm
  t}^{1}(X)\,,\,\mathbb{Z}\big)
\oplus\tor_{\K_{\rm t}^{0}(X)} \ . $$ Although the torsion part of
the dual homology group to the topological K-theory group $\K_{\rm
t}^{1}(X)$ can differ from that of the analytic K-homology $\K^{\rm
a}_1(X)$, Poincar\'e duality always asserts an isomorphism between
the full groups $\K^{\rm
  a}_\sharp(X)\cong\K^{\rm t}_\sharp(X)$ and $\K_{\rm t}^\sharp(X)$. Note that
if
$\dim_\real X$ is even and $\K_{\rm t}^{0}(X)$ is a free
abelian group, then $\K^{1}_{\rm t}(X) \cong\K^{\rm t}_{1}(X) \cong
\K^{\rm a}_{1}(X) \cong \Hom(\K_{\rm t}^{1}(X),\mathbb{Z})$.

One can make a stronger statement (due to D. W. Anderson) which
works for any CW-complex $X$ with finitely generated K-theory. Since
$\K^\U$ is a CW-spectrum and a ring-spectrum, there is a universal
coefficient theorem expressed by the (split) exact
sequence~\cite{UCT3,UCT4,UCT1}
$$0 ~\longrightarrow~ {\rm Ext}\big(\K^{\rm
  t}_{e(i-1)}(X)\,,\,\zed\big) ~\longrightarrow~ \K_{\rm
  t}^i\big(X\big) ~\longrightarrow~ {\rm Hom}\big(\K^{\rm
  t}_{i}(X)\,,\,\zed\big) ~\longrightarrow~ 0$$
for $i=0,1$. The epimorphism is given by the index map. As above, let
us consider the splits $\K_{\rm t}^i(X)=\Lambda_{\K_{\rm
    t}^i(X)}\oplus\tor_{\K_{\rm t}^i(X)}$ and $\K^{\rm t}_i(X)=\Lambda_{\K^{\rm
    t}_i(X)}\oplus\tor_{\K^{\rm t}_i(X)}$. One then easily concludes that
\bea
{\rm Ext}\big(\K^{\rm t}_{e(i-1)}(X)\,,\,\zed\big) &\cong& {\rm
  Ext}\big(\tor_{\K^{\rm t}_{e(i-1)}(X)}\,,\,\zed\big) ~\cong~
\tor_{\K^{\rm t}_{e(i-1)}}(X) \ , \nonumber\\
{\rm Hom}\big(\K^{\rm t}_{i}(X)\,,\,\zed\big) &\cong& {\rm
  Hom}\big(\Lambda_{\K^{\rm t}_i(X)}\,,\,\zed\big) ~\cong~ \Lambda_{\K^{\rm
    t}_i(X)} \ . \nonumber
\eea By the universal coefficient theorem it follows that
$\tor_{\K_{\rm
    t}^{e(i-1)}(X)} \cong \tor_{\K^{\rm t}_i(X)}$ and
$\Lambda_{\K_{\rm t}^i(X)} \cong \Lambda_{\K^{\rm t}_i(X)}$. We can
also conclude the following from these general calculations.

\begin{proposition}\label{fingen}
Let $X$ be a finite CW-complex. Then its K-homology group is
finitely generated.
\end{proposition}
\begin{proof}
Since $X$ is a finite CW-complex, its K-theory groups are finitely
generated. The isomorphisms above then imply the conclusion.
\end{proof}

\subsection{Chern Character }\label{ss5}

There is a natural transformation $\ch_{\bullet} :\K_\sharp^{\rm t}(X)
\rightarrow\H^{~}_\sharp(X;\mathbb{Q})$ of $\zed_2$-graded homology theories
called the (homology) Chern character which is defined in the
following way. Recall the $\zed_2$-grading on
singular homology given by $\H_\sharp(X;\rat)=\H_{\rm
  even}(X;\rat)\oplus\H_{\rm odd}(X;\rat)$ with $\H_{\rm
  even}(X;\rat)=\bigoplus_{e(k)=0}\,\H_k(X;\rat)$ and $\H_{\rm
  odd}(X;\rat)=\bigoplus_{e(k)=1}\,\H_k(X;\rat)$.
Given a K-cycle $(M,E,\phi)$ on $X$, let $\phi_{*}
:\H_\sharp(M;\mathbb{Q})\rightarrow\H_\sharp(X;\mathbb{Q})$ be the
homomorphism induced on rational homology by $\phi$. Then
$\ch^\bullet(E)\cup\td(TM) \cap [M]$ is the Poincar\'{e} dual on $M$
of the even degree cohomology class $\ch^\bullet(E)\cup\td(TM)$, where
$\ch^\bullet:\K^0_{\rm t}(-)\to\H^{\rm even}(-;\mathbb{Q})$ is the
(cohomology) Chern character in K-theory, $\td$ denotes the Todd class
of a spin$^{c}$ vector bundle and $[M]$ is the orientation cycle of
$M$ in $\H_\sharp(M;\rat)$ induced by the spin$^c$ structure on $TM$. Then
\beq
\ch_\bullet(M,E,\phi):= \phi_{*}\big(\ch^\bullet(E)\cup\td(TM) \cap
[M]\big)
\label{Chernchar}\eeq
is an element of $\H_\sharp(X;\mathbb{Q})$ which depends only on
the K-homology class $[M,E,\phi]\in\K_\sharp^{\rm t}(X)$. This map preserves
the $\zed_2$-grading. The Chern
characters $\ch^\bullet$ and $\ch_{\bullet}$ preserve the cap product,
i.e.~for every topological space $X$ there is a $\zed_2$-degree
preserving commutative diagram
 $$\xymatrix{\K_{\rm t}^\sharp(X)\otimes\K_\sharp^{\rm t}(X) \ar[r]^{\qquad
     \cap}\ar[d]_{\ch^\bullet  \otimes\ch_{\bullet} } & \K_\sharp^{\rm
     t}(X) \ar[d]^{\ch_{\bullet}}\\
      \H^\sharp(X;\mathbb{Q})\otimes\H_\sharp(X;\mathbb{Q})
      \ar[r]^{\qquad\cap} &\H_\sharp(X;\mathbb{Q}) \ . } $$
Since $X$ is a finite CW-complex, $\K_\sharp^{\rm t}(X)$ is a
finitely generated abelian group and $\ch_{\bullet}$ induces an
isomorphism $\K_\sharp^{\rm t}(X) \otimes_\zed \mathbb{Q}
\cong\H^{~}_\sharp(X;\mathbb{Q})$ of $\zed_2$-graded vector spaces
over $\rat$.

The Chern character can be used to give an explicit formula for the
epimorphism (\ref{collapseepi}) in terms of characteristic classes as
$\varepsilon_{*}[M,E,\phi]=\ch^\bullet(E) \cup\td(TM)[M]$.
Then the commutative diagram (\ref{Indexcommdiag}) can be recast as
the equality
\beq
\qquad \Index~\phi_{*}[\Dirac_{E}]~=~\Index~\mu^{\rm a}[M,E,\phi]~=~
\varepsilon_{*}[M,E,\phi]~=~\ch^\bullet(E) \cup\td(TM)[M] \ .
\label{IndexphiE}\eeq
 In the special case where $X$ is a point this becomes
 $\Index~\Dirac_{E}=\ch^\bullet(E) \cup \td(TM)[M]$, which is a
 particular instance of the Atiyah-Singer index theorem.

\newsection{K-Cycles and D-Brane Constructions\label{BraneConstr}}

We will now bring string theory into the story. Many of our
subsequent results find their most natural interpretations in terms
of D-branes, with the mathematical formalism of K-homology leading to
new insights into the properties of D-branes wrapping
cycles in non-trivial spacetimes. We begin with some heuristic physical
discussion aimed at motivating the interpretation of D-branes as K-cycles in
topological K-homology. Then we move on to more mathematical
computations explaining the interplay between K-homology and the properties of
D-branes. The analysis will center around finding explicit K-cycle
representatives for the generators of the K-homology groups, which
will be interpreted as D-branes in the pertinent spacetime. For
physical definitions and descriptions of D-branes in string theory,
see~\cite{Johnson1,Polchinski1}.

\subsection{D-Branes\label{DBranes}}

Consider Type~II superstring theory on a spacetime $X$ with all
background supergravity form fields turned off. $X$ is an
oriented ten-dimensional spin manifold. A {\it D-brane} in $X$ is an oriented
spin$^c$ submanifold $M\subset X$ together with a complex vector bundle
$E\to M$ called the {\it Chan-Paton bundle}. $M$ itself is refered to
as the {\it worldvolume} of the D-brane and when $\dim_\real M=p+1$ we will
sometimes refer to the brane as a D$p$-brane to emphasize its
dimensionality. The presence of non-vanishing background form fields would
mean that the classification of D-branes requires algebraic and/or
twisted (co)homology tools. Their absence means that we can resort to
topological methods, which thereby classify {\it flat} D-branes. While
this works whenever the tangent bundle $TX$
over spacetime is stably trivial, in more general cases the set-up
would not describe a true background of string theory since the
terminology `flat' used here is not meant to imply that we consider a
flat spacetime geometry. Nonetheless,
the topological description will provide geometric insight into the
nature of D-branes in curved backgrounds, even in this simplified
setting. Thus a very crude definition of a D-brane is
as a K-cycle $(M,E,\phi)$ on the spacetime $X$, with
$\phi=\iota:M\hookrightarrow X$ the natural inclusion (The remaining
elements of $\Gamma(X)$ then ensure that the quotient $\Gamma(X)\,/\sim$
really is $\K^{\rm t}_\sharp(X)$). Sometimes we regard D-branes as
sitting in an ambient space which is a proper subspace of spacetime,
for instance when $X$ is a product $X=Q\times Y$ we may be interested in
worldvolumes $M\subset Q$. When there is no danger of confusion we will
also use the symbol $X$ for this ambient space. In either case, $X$
is also customarily called the {\it target space}.

D-branes are generally more complicated objects than just submanifolds
carrying vector bundles, because in string theory they are realized as
(Dirichlet) boundary conditions for a two-dimensional superconformal
quantum field theory with target space $X$. Any
classification based solely on K-theory is expected to capture only
those properties that depend on D-brane {\it charge}. Nevertheless,
the primitive definition of a D-brane as a K-cycle in topological
K-homology is very natural and carries much more information than its
realization in the dual K-theory framework. We shall see that the
geometric description of K-homology is surprisingly rich and provides
a simple context in which non-trivial D-brane effects are
exhibited in a clear geometrical fashion.

\begin{example}[{\it B-Branes}]\label{BBranes}
Let $X$ be a (possibly singular) complex $n$-dimensional projective
algebraic variety, and let $\mathcal{O}_X$ be the
structure sheaf of regular functions on $X$. Recall that a coherent
sheaf on $X$ is a sheaf of $\mathcal{O}_X$-modules which is locally
the cokernel of a morphism of holomorphic vector bundles over $X$. The coherent
algebraic sheaves on $X$ form an abelian category denoted
$\coh(X)$. The bounded derived category of coherent sheaves on $X$, denoted
${\bf D}(X):=\Der(\coh(X))$, is the triangulated category of topological
B-model D-branes (or {\it B-branes} for short) in
$X$~\cite{Douglas1,Witten3}. An object of
this category is a bounded differential complex of coherent sheaves.
It contains $\coh(X)$ as a full subcategory by identifying a
coherent sheaf $\mathcal{F}$ with the trivial complex
$$\mathcal{F}^\bullet=\big(0~\stackrel{0}{\longrightarrow}~\dots~
\stackrel{0}{\longrightarrow}~0~\stackrel{0}{\longrightarrow}~\mathcal{F}
{}~\stackrel{0}{\longrightarrow}~0~\stackrel{0}{\longrightarrow}~
\dots~\stackrel{0}{\longrightarrow}~0\big)$$ having $\mathcal{F}$ in
degree~$0$. For details of these and other constructions,
see~\cite{Aspinwall1}.

The relevant K-homology group, denoted $\K^\omega_0(X)$, is the
Grothendieck group of coherent algebraic sheaves on $X$ obtained by
applying the usual Grothendieck completion functor to the abelian category
$\coh(X)$. There is a natural transformation $\alpha^\omega:{\bf
  D}(X)\to\K^\omega_0(X)$ which may be described as follows. Let
$\mathcal{F}^\bullet\in{\bf D}(X)$ be a complex. Using a
locally-free resolution we may replace $\mathcal{F}^\bullet$ by a
quasi-isomorphic complex of locally-free sheaves, each of which has
associated to it a holomorphic vector bundle. The virtual Euler class
of the complex obtained by replacing locally-free sheaves with
their corresponding vector bundles is then the K-homology class
$\alpha^\omega(\mathcal{F}^\bullet)$ that we are looking for. On the
other hand, the underlying topological space of $X$ is a finite
CW-complex and has topological K-homology group $\K_\sharp^{\rm
  t}(X)$. We will now construct a natural map
$\mu^\omega:\K_0^\omega(X)\to\K_0^{\rm t}(X)$, which by composition
gives a natural map from the derived category
$\mu^\omega\circ\alpha^\omega:\Der(X)\to\K_0^{\rm t}(X)$ and gives an
intrinsic description of B-branes in terms of K-cycles.

Consider triples $(M,E,\phi)$ where
\begin{romanlist}
\item $M$ is a non-singular complex projective algebraic variety;
\item $E$ is a complex algebraic vector bundle over $M$; and
\item $\phi:M\to X$ is a morphism of algebraic varieties.
\end{romanlist}
Two triples $(M_1,E_1,\phi_1)$ and $(M_2,E_2,\phi_2)$ are said to be
isomorphic if there exists an isomorphism $h:M_1\to M_2$ of complex
projective algebraic varieties such that $h^*(E_2)\cong E_1$ as
complex algebraic vector bundles over $M_1$ and $\phi_1=\phi_2\circ
h$. The set of isomorphism classes of triples is denoted
$\Gamma^\omega(X)$. Given such a triple $(M,E,\phi)$, the morphism
$\phi:M\to X$ induces the direct image functor
$\coh(\phi):\coh(M)\to\coh(X)$, defined by
$\coh(\phi)(\mathcal{F})=\phi_*(\mathcal{F})$ for
$\mathcal{F}\in\coh(M)$, which is left exact and induces the $i$-th
right derived functor ${\mbf R}^i\,\coh(\phi):\Der(M)\to\Der(X)$ for
$i=0,1,\dots,n$ as follows. We include the category of coherent
sheaves into the category of quasi-coherent sheaves, replace a
complex by a quasi-isomorphic complex of injectives, and apply the
functor $\coh(\phi)$ componentwise to the complex of injectives (If
$X$ is singular this requires a resolution of its singularities). Then
the induced map
$$\phi_*\,:\,\K_0^\omega(M)~\longrightarrow~\K_0^\omega(X)$$ is given
for coherent algebraic sheaves $\mathcal{F}$ on $M$ by
\beq
\phi_*[\mathcal{F}]
:=\sum_{i=0}^n\,(-1)^i~\alpha^\omega\circ{\mbf R}^i\,\coh(\phi)
[\mathcal{F}^\bullet] \ .
\label{phicoh}\eeq
In particular, if $\underline{E}$ denotes
the sheaf of germs of algebraic sections of $E\to M$, then
$\phi_*[\,\underline{E}\,]\in\K_0^\omega(X)$. By using a resolution of the
singularities of $X$ if necessary and the fact that any coherent sheaf
on a non-singular variety admits a resolution by locally free sheaves,
one can show that the $\phi_*[\,\underline{E}\,]$ obtained from
triples in $\Gamma^\omega(X)$ generate the abelian group
$\K_0^\omega(X)$~\cite{1}. By forgetting some structure a triple
$(M,E,\phi)$ becomes a K-cycle on $X$ and hence determines an element
$[M,E,\phi]\in\K_0^{\rm t}(X)$.

Thus we get a well-defined map
$$\mu^\omega\,:\,\K_0^\omega(X)~\longrightarrow~\K^{\rm t}_0(X)$$
given on generators by $$\mu^\omega\phi_*\big[\,\underline{E}\,\big]:=
\big[M,E,\phi\big] \ . $$ This map is a natural transformation of the
covariant functors $\K_0^\omega$ and $\K_0^{\rm t}$, thus providing
an extension of the Grothendieck-Riemann-Roch theorem. However, in
contrast to the transformations $\mu^{\rm s}$ and $\mu^{\rm a}$ of the
previous section, $\mu^\omega$ is {\it not} an isomorphism~\cite{1}. This
suggests that the topological K-homology group $\K_\sharp^{\rm
  t}(X)$ carries more information about the category of B-branes than
the Grothendieck group $\K_\sharp^\omega(X)$.

One of the virtues of this mapping of elements in the derived category
$\Der(X)$ to classes of K-cycles
in $\K^{\rm t}_0(X)$ is that it allows one to compute B-brane charges
even when the variety $X$ is singular. The collapsing map
$\varepsilon:X\to\pt$ induces, as before, an epimorphism
$\chi:\K_0^\omega(X)\to\K_0^\omega(\pt)$. Since a coherent algebraic sheaf over
a point is just a finite dimensional complex vector space, which can be
characterized by its dimension, one has
$\K_0^\omega(\pt)\cong\zed$. The {\it charge} of a B-brane represented by a
coherent algebraic sheaf $\mathcal{F}$ on $X$ is the image of
$[\mathcal{F}]\in\K_0^\omega(X)$ in $\zed$ under the epimorphism
$\chi$, which using (\ref{phicoh}) is given explicitly by the Euler number
$$\chi(X;\mathcal{F})~:=~\chi[\mathcal{F}]~=~
\sum_{i=0}^n\,(-1)^i~\dim_\complex\H^i(X;\mathcal{F})
\ , $$ where $\H^i(X;\mathcal{F})$ is the $i$-th cohomology group of
$X$ with coefficients in $\mathcal{F}$. Together with the epimorphism
(\ref{collapseepi}) and the transformation $\mu^\omega$, there is a
commutative diagram similar to (\ref{Indexcommdiag}) given by
$$\xymatrix{\K^{\rm t}_{0}(X) \ar[rd]^{\varepsilon_{*}}
   & \\ \K^\omega_{0}(X)\ar[u]^{\mu^\omega}
   \ar[r]_{\quad\chi} & \mathbb{Z} \ . } $$
For a B-brane represented by a K-cycle $(M,E,\phi)$, the
characteristic class formula for $\varepsilon_*$ in
(\ref{IndexphiE}) then gives the charge
$$\chi\big(X\,;\,\phi_*\underline{E}\,\big)=\ch^\bullet(E)\cup\td(TM)[M]
\ . $$ If $X$ is non-singular and $E$ is a complex vector bundle over
$X$, then this charge formula applied to the K-cycle $(X,E,\Id_X)$ is just the
Hirzebruch-Riemann-Roch theorem which computes the analytic index of
the twisted Dolbeault operator $\overline{\partial}_E$ on
$X$. See~\cite{1} for more details.~\hfill{$\lozenge$}
\end{example}

\subsection{Qualitative Description of K-Cycle Classes\label{Qual}}

We will now explain how the equivalence relations of topological
K-homology, as spelled out in Definition~\ref{Tgroup}, translate
into physical statements about D-branes. Let us begin with bordism.
Suppose that $X$ is a locally compact topological space and
$\pt\nsubseteq X$ contains a distinguished point called
``infinity''. Let $X^\infty:=X\amalg\pt$ be the one-point
compactification of $X$. We are interested in configurations of
D-branes in $X$ which have finite energy. This means that they
should be regarded as equivalent to the closed string vacuum
asymptotically in $X^\infty$. The {\it charge} of a D-brane
$(M,E,\phi)$ is given through the index formula (\ref{IndexphiE})
(In this paper we do not deal with the square root of the
$\widehat{A}$-genus which usually defines D-brane
charges~\cite{MM1}). The condition that the D-brane should have
vanishing charge at infinity is tantamount to requiring that its
K-homology class have a trivialization at infinity, i.e. that there
is a compact subset $U$ of $X$ such that $E|_{X-U}$ is a trivial
bundle and $\phi|_{X-U}$ is homotopically trivial, so that
$\Index~\phi_{*}[\Dirac_{E}]=0$ over $X-U$. This is the physical
meaning of the definition of the reduced K-homology group
(\ref{redKhom}), which measures charges of D-branes relative to that
of the vacuum. The K-homology of $X$ in this case should then be
defined by the {\it relative} K-homology group of Section~\ref{RelK}
as $\K_\sharp^{\rm t}(X^\infty,\pt)\cong\K_\sharp^{\rm t}(X)$, where
we have used excision. Since the six-term exact sequence for this
relative K-homology has connecting homomorphism $\partial:\K_i^{\rm
  t}(X^\infty,\pt)\to\K_{e(i+1)}^{\rm t}(\pt)$ acting by
$\partial[W,F,\psi]=[\partial W,F|_{\partial W},\psi|_{\partial W}]$,
boundaries $\partial W$ of spin$^c$ manifolds $W$ map into $\pt$ and
so any D-brane $(M,E,\phi)$ which is bordant to the
trivial K-cycle $(\emptyset,\emptyset,\emptyset)$ on $X$ should carry the
same charge as the vacuum. This is guaranteed by the bordism
relation. The argument extends to a ``compactification'' of spacetime
in which $X=Q\times Y$, with $Q$ locally compact and $Y$ a compact
topological space without boundary. D-branes will now either be
located at particular points in $Y$ or will wrap submanifolds of
$Y$. Requiring that such configurations be equivalent to the vacuum
asymptotically in $Q^\infty$ thus requires adding a copy of $Y$ at
infinity in $Q^\infty$, and so we look for K-homology classes with a
trivialization on $Y$ at infinity. Using relative bordism and
excision, the classes of D-brane K-cycles are now seen to live in the
group $\K_\sharp^{\rm t}(X,Y)\cong\K_\sharp^{\rm
  t}(X/Y,\pt)\cong\K^{\rm t}_\sharp(Q^\infty,\pt)$.

Let us now turn to direct sum. It represents ``gauge
symmetry enhancement for coincident branes'' which occurs when several
D-branes wrap the same submanifold $M\subset X$~\cite{Witten2}. In
this case, the Chan-Paton bundles $E_i\to M$ of the constituent
D-branes are augmented to the higher-rank Chan-Paton bundle
$E=\bigoplus_i\,E_i$ of the combined brane configuration. Such a
combination is called a {\it bound state} of D-branes. The branes are
bound together by open string excitations corresponding to classes of
bundle morphisms $[f_{ij}]\in\Hom(E_i,E_j)$. Other open string degrees
of freedom are described by the higher cohomology groups
$\Ext^p(E_i,E_j)$, $p\geq1$.

Finally, let us look at vector bundle modification. Consider the
K-cycle $(\pt,\id_\pt^\complex,\iota)$  on $X$ where
$\id_\pt^\complex=\pt\times \mathbb{C}$ and $\iota: \pt\hookrightarrow X$ is
the inclusion of a point. Let $F= \pt\times\mathbb{R}^{2n}$ with $n \geq
1$. Then $\widehat{M}=\pt\times\S^{2n} \cong\S^{2n}$ is a
$2n$-dimensional sphere, and $\pi=\varepsilon:\S^{2n}
\rightarrow \textrm{pt}$ is the collapsing map. From the clutching
construction of Section~\ref{ss2},
$H(F) = H(F)|_{\S^{2n}}= H(F)|_{\pi^{-1}(\textrm{pt})}$ is the
Bott generator for the K-theory of $\S^{2n}$. Using vector bundle
modification one then has
$[\pt,\id_\pt^\complex,\iota]=[\S^{2n},H(F)\otimes\id_\pt^\complex,
\iota\circ\pi]=[\S^{2n},H(F),\varepsilon]$, where
$\iota\circ \pi=\varepsilon$ is a collapsing map. This equality
represents the ``blowing up'' of a D$(-1)$-brane (also known as a {\it
  D-instanton}) into a collection of
spherical D$(2n-1)$-branes and is a simple example of what is known
in string theory as the ``dielectric effect''~\cite{Myers1}. This is
described in more generality in Section~\ref{Polar} below. The
equality also illustrates the crucial point that topological
K-homology naturally encodes the fact that D-branes are typically not
static objects but will ``decay'' into stable configurations of
branes. Suppose that the spacetime has a split $X=\real\times X'$,
where $\real$ is thought of as parametrizing ``time'' and $X'$ is
``space''. Consider a D-brane in $X$ whose worldvolume is initially of
the form $M=\real\times M'$ with $M'\subset X'$. In the distant
future (at large time), this brane will ``decay'' into stable D-branes
of lower dimension which wrap non-trivial homology cycles of $X'$. In
the example above, if we regard the $2n$-sphere $\S^{2n}$ as the
one-point compactification of the locally compact space $\real^{2n}$,
then the spherical D$(2n-1)$-brane can be thought of as decaying into
a bound state of D$(-1)$-branes. There are no conserved charges for
higher branes because $\real^{2n}$ is contractible. The precise
formulation of this notion of ``stability'' in K-homology is one of
the goals of the present paper.

Type~II superstring theory on the spacetime manifold $X$ itself splits
into two string theories, called Type~IIA and Type~IIB. The former
contains stable supersymmetric D$p$-branes for all even $0\leq
p\leq8$ while the latter contains stable supersymmetric D$p$-branes
for all odd $-1\leq p\leq9$. Thus Type~IIA D-branes are naturally
classified by the group $\K_1^{\rm t}(X)$ while $\K_0^{\rm t}(X)$
classifies Type~IIB branes. Since $\dim_\real X$ is even in the
present case, by Poincar\'e duality this is in complete
agreement with the corresponding K-theory
classification~\cite{16,Witten1}. We shall now proceed to examine
properties of these D-branes by giving more rigorous mathematical
calculations in K-homology, beginning with an accurate statement of
what is meant by ``stable'' in the above discussion.

\subsection{Stability\label{Stability}}

In the absence of background form fields, stable supersymmetric
D-branes are known to wrap the non-trivial spin$^c$ homology cycles of
the spacetime manifold $X$ in which they live~\cite{FW1}. In
K-homology, this is asserted by the following fundamental result that
will play an important role throughout the rest of this paper.

\begin{definition}
Let $X$ be a space and $Y \subset X$. We say that a D-brane
$[M,E,\phi] \in K^t_{\sharp}(X)$ \textit{wraps} $Y$ if $\phi(M)
\subset Y$.
\end{definition}

The idea behind this definition is that the condition above
determines a continuous map $\psi : M \rightarrow Y$ such that $i
\circ \psi = \phi$, where $i : Y \rightarrow X$  is the canonical
inclusion and $i_*([M,E,\psi])=[M,E,\phi]$, and so we can interpret
the initial D-brane as one over $Y$.

On the other hand, the notion of stability is, mathematically, a
heuristic one, since we interpret that a D-brane $[M,E,\phi]\in
K^t_{\sharp}(X)$ is \textit{unstable} if it is decomposable
(non-trivially) in the form
$$[M,E,\phi]=\sum_{p=0}^n~\sum_{i=1}^{m_p}\,d^{~}_{pi}(M,E)~
\big[M_i^p,\id_{M_i^p},\phi_i^p\big],$$ where the branes
$\big[M_i^p,\id_{M_i^p},\phi_i^p\big]$ are linearly independent
generators of $ K^t_{\sharp}(X)$. Namely, the stable objects are the
possible linearly independent generators.

\begin{theorem}\label{bigt}
Let $X$ be a compact connected finite CW-complex of dimension $n$
whose rational homology can be presented as
$$\H_\sharp(X;\mathbb{Q})= \bigoplus_{p=0}^{n}~\bigoplus_{i=1}^{m_p}
\,\big[M^{p}_{i}\big] {}~\mathbb{Q} \ , $$ where $M_i^{p}$ is a
$p$-dimensional compact connected spin$^c$ submanifold of $X$
without boundary and with orientation cycle $[M_i^{p}]$ given by the
spin$^c$ structure. Suppose that the canonical inclusion map
$\iota_i^{p}:M_i^p\hookrightarrow X$ induces, for each $i,p$, a
homomorphism
$(\iota_i^p)_*:\H_p(M_i^p;\rat)\to\H_p(X;\rat)\cong\rat^{m_p}$ with
the property \beq
\big(\iota_i^p\big)_*\big[M_i^p\big]=\kappa_{ip}\,\big[M_i^p\big]
\label{bigtcond}\eeq for some $\kappa_{ip}\in\rat\setminus\{0\}$.
Then the lattice $\Lambda_{\K_\sharp^{\rm t}(X)}:=\K_\sharp^{\rm
t}(X)\,/\,\tor_{\K_\sharp^{\rm
    t}(X)}$ is generated by the classes of K-cycles
$$\big[M_{i}^p,\id^\complex_{M_i^p},\iota_i^p\big] \ , \quad 0 \leq p\leq n
\ , ~~ 1 \leq i\leq m_p \ . $$
\end{theorem}
\begin{proof}
Fixing $0 \leq p\leq n$, $1 \leq i\leq m_p$, let
$\{x_{ab}^{ip}\}_{b\in[n_a^{ip}]}$ be cohomology classes in degree~$a$
generating the rational cohomology group
$\H^a(M_i^p;\rat)\cong\rat^{n_a^{ip}}$ for each $a=0,1,\dots,p$. Since
$M_i^p$ is oriented and connected, one has
$\H^p(M_i^p;\rat)\cong\rat\cong\H^0(M_i^p;\rat)$ and hence
$n_p^{ip}=1=n_0^{ip}$. Without loss of generality we may assume that
$x_{01}^{ip}=1$, and we set $x_p^{ip}:=x_{p1}^{ip}$. The rational
cohomology ring of the submanifold $M_i^p\subset X$ can thus be
presented as
$$\H^\sharp\big(M_i^p;\rat\big)=1~\rat~\oplus~\Big(\,\bigoplus_{a=0}^{p-1}~
\bigoplus_{b=1}^{n_a^{ip}}\,x_{ab}^{ip}~\rat\Big)~\oplus~x_p^{ip}~\rat
\ . $$
In particular, the Todd class $\td(TM_i^p)\in\H^{\rm even}(M_i^p;\rat)$ may be
expressed in the form $$\td\big(TM_i^p\big)=1+\sum_{a=1}^{p-1}~
\sum_{b=1}^{n_a^{ip}}\,d_{ab}^{ip}~x_{ab}^{ip}+\delta_{e(p),0}~x_p^{ip}$$ for
some
$d_{ab}^{ip}\in\rat$ with $d_{ab}^{ip}=0$ whenever $a$ is odd. Let us
now use the Chern character (\ref{Chernchar}) to compute
\bea
\ch_\bullet\big(M_i^p,\id_{M_i^p}^\complex,\iota_i^p\big)&=&\big(
\iota_i^p\big)_*\big(\td(TM_i^p)\cap[M_i^p]\big) \nn\\ &=&
\big(\iota_i^p\big)_*\Big([M_i^p]+\sum_{a=1}^{p-1}~
\sum_{b=1}^{n_a^{ip}}\,d_{ab}^{ip}~x_{ab}^{ip}\cap[M_i^p]+
r_{ip}~[\pt]\Big) \nn\\ &=& \kappa_{ip}~[M_i^p]+\sum_{a=1}^{p-1}~
\sum_{b=1}^{n_a^{ip}}\,d_{ab}^{ip}~\big(\iota_i^p\big)_*
\big(x_{ab}^{ip}\cap[M_i^p]\big)+
r_{ip}~\big(\iota_i^p\big)_*[\pt]
\label{Cherncompute}\eea
for some $r_{ip}\in\rat$ with $r_{ip}=0$ for $p$ odd. We have used
$\ch^\bullet(\id_{M_i^p}^\complex)=1$ and
(\ref{bigtcond}). For each $1\leq a<p$, $1\leq b\leq n_a^{ip}$ one has
$(\iota_i^p)_*(x_{ab}^{ip}\cap[M_i^p])\in\H_{p-a}(X;\rat)$.

The ordered collection $$\vec c=\begin{matrix}
\big(\,\scriptstyle[\pt]&
\underbrace{\,\scriptstyle[M_1^1]\cdots[M^1_{m_1}]\,}&
\dots&\underbrace{\,\scriptstyle[M_1^n]\cdots[M^n_{m_n}]\,}
\,\big)\end{matrix}$$
of homology cycles is a basis of $\H_\sharp(X;\rat)$ as a rational vector
space. On the other hand, if we set
$$\vec h=\begin{matrix}\big(\,\scriptstyle
\ch_\bullet(\pt,\id_\pt^\complex,\iota_1^0)&
\underbrace{\,\scriptstyle\ch_\bullet(M_1^1,\id_{M_1^1}^\complex,\iota_1^1)
\cdots\ch_\bullet(M^1_{m_1},\id_{M^1_{m_1}}^\complex,
\iota^1_{m_1})\,}&\dots&\underbrace{\,\scriptstyle
\ch_\bullet(M_1^n,\id_{M_1^n}^\complex,\iota_1^n)\cdots
\ch_\bullet(M^n_{m_n},\id_{M^n_{m_n}}^\complex,
\iota^n_{m_n})\,}\,\big)\end{matrix}$$
then from (\ref{Cherncompute}) it follows that
$\vec h=\vec c~{\mbf\Omega}$, where $\mbf\Omega$ is an upper
triangular matrix whose diagonal elements are the non-zero rational
numbers $\kappa_{ip}$. Thus $\det\mbf\Omega\neq0$ and so the collection
$\vec h$ is also a basis of $\H_\sharp(X;\rat)$ as a rational vector
space. Since $X$ is a finite CW-complex, $\ch_{\bullet}\otimes
\Id_{\mathbb{Q}} :\K_\sharp^{\rm t}(X) \otimes_\zed\mathbb{Q} \rightarrow
\H_\sharp(X;\mathbb{Q})$ is an isomorphism and hence
$(\ch_\bullet\otimes\Id_\rat)^{-1}\vec h$ is a set of generators
for $\K_\sharp^{\rm t}(X) \otimes_\zed \mathbb{Q}$.
\end{proof}

\begin{remark}
Some comments are in order concerning the statement of the previous
Theorem. The inclusions considered are topological embeddings
(which, between manifolds, may turn out to be smooth maps), since
$X$ is not assumed to be a manifold. Since embedding manifolds into
other manifolds can be very troublesome, our assumptions are of a
purely combinatorial nature, stated by the shape of the rational
homology of $X$ and the behavior of the maps induced in homology by
the inclusion maps. One instance in which this is a straightforward
result is when $X$ has a CW-structure given by manifolds satisfying
the stated properties whose skeletons have big enough gaps, for
instance if $X^{2k}=X^{2k+1}$, for all $k$, as is the case for
projective spaces other than real projective
spaces.\hfill{$\lozenge$}

\end{remark}

\begin{remark}\label{bigtrem}
We do not know if this theorem can be proven by replacing the
assumption (\ref{bigtcond}) with a weaker condition. The crucial issue
is whether or not there is a non-trivial linear relation
$\sum_{p,i}\,n_{pi}~[M_{i}^p,\id^\complex_{M_i^p},\iota_i^p]=0$ over
$\zed$ among the ``lifts''
$[M_{i}^p]\mapsto[M_{i}^p,\id^\complex_{M_i^p},\iota_i^p]$ of the
non-trivial homology cycles of $X$ to K-homology. If such a relation
exists, then the lifts of some non-trivial singular homology classes
in $\K_\sharp^{\rm t}(X)$ are~$0$. This means
that some D-brane state is unstable, even though it wraps a
non-trivial spin$^c$ homology cycle. It either decays into the closed string
vacuum state, or is not completely unstable but decays into other
D-branes according to the solutions of the linear equation
$\sum_{p,i}\,n_{pi}~[M_{i}^p,\id^\complex_{M_i^p},\iota_i^p]=0$ over
$\zed$. The same argument as that used in Section~\ref{BwBs} below
shows that such a decay is always into branes wrapped on manifolds of
lower dimension than that of the original D-brane. The condition
(\ref{bigtcond}) guarantees that this does not occur. We shall analyse
this feature from a different perspective in
Section~\ref{Lifts}. This analysis illustrates the fact that D-branes
need not generally simply correspond to subspaces of
spacetime. \hfill{$\lozenge$}
\end{remark}

Another perspective on the problem of finding the K-homology
generators is given by Spin$^c$ bordism. Let MSpin$_{\sharp}^c(X)$
be the Spin$^c$ bordism group of a finite  CW-complex $X$. This
group is the set of equivalence classes, through bordism, of pairs
$(M,f)$, where $M$ is a compact Spin$^c$ manifold without boundary
and $f : M \rightarrow X$ is a continuous map. Disjoint union
induces an abelian group structure on MSpin$_{\sharp}^c(X)$. This
group is $\mathbb{Z}$-graded by the dimension of $M$. This
construction gives a generalized homology theory
MSpin$_{\sharp}^c(\,\cdot \,)$ on the category of finite
CW-complexes, which can also be defined in terms of a spectrum
\textbf{MSpin$^c$}. For more details see \cite{CoFl2,CoFl1,8}.

Conner and Floyd (\cite{CoFl1}) constructed an orientation map
$\textbf{MU} \rightarrow \rm{K}^{\rm{U}}$ between the complex
bordism and the K-theory spectra which was latter generalized to a
map \textbf{MSpin$^c$} $\rightarrow \rm{K}^{\rm{U}}$ by
Atiyah-Bott-Shapiro (\cite{4}). Hence, if $X$ is a finite
CW-complex, this translates into a homomorphism MSpin$^c_{\sharp}(X)
\rightarrow \rm{K}^t_{\sharp}(X)$, given by $[M,f] \mapsto
[M,\id_{M}^\complex,f]$ which is a natural transformation of
homology theories inducing an MSpin$^c_{\sharp}(\rm{pt})$-module
structure on $\rm{K}^t_{\sharp}(\rm{pt})$ . The relevant result for
our purposes is the following theorem by Hopkins and Hovey
(\cite{HopHo}).

\begin{theorem}
The map $\rm{MSpin}^c_{\sharp}(X)
\otimes_{\rm{MSpin}^c_{\sharp}(\rm{pt})}\rm{K}^t_{\sharp}(\rm{pt})
\rightarrow \rm{K}^t_{\sharp}(X)$ induced by the Atiyah-Bott-Shapiro
orientation is an isomorphism of
$\rm{K}^t_{\sharp}(\rm{pt})$-modules.
\end{theorem}

This immediately implies the following result, reducing the problem
of calculating the K-homology generators to the analogous problem in
Spin$^c$ bordism.

\begin{theorem}
Let $X$ be a finite CW-complex. Suppose $[M_i,f_i]$, $1 \leq i \leq
m$ are generators of $\rm{MSpin}^c_{\sharp}(X)$ as an
$\rm{MSpin}^c_{\sharp}(\rm{pt})$-module. Then
$[M_i,\id_{M_i}^\complex,f_i]$, $1 \leq i \leq m$, generate
$\rm{K}^t_{\sharp}(X)$ as a $\rm{K}^t_{\sharp}(\rm{pt})$-module.
\end{theorem}

Namely, for all $n=0,1$, $\rm{K}^t_{n}(X)$ is generated by the
elements $[M_i,\id_{M_i}^\complex,f_i]$, $1 \leq i \leq m$, such
that $\rm{dim}\, M_i=n$. In the case where $X$ is also a manifold,
we can consider differentiable bordism, namely we can consider
bordism of pairs $(M,f)$, where $M$ is a compact Spin$^c$ manifold
without boundary and $f : M \rightarrow X$ is a differentiable map.
We then obtain another abelian group
$\rm{MSpin^c}^{\rm{Diff}}_{\sharp}(X)$, which is canonically
isomorphic to $\rm{MSpin}^c_{\sharp}(X)$ (\cite{CoFl2}).

\subsection{Branes within Branes\label{BwBs}}

A D$p$-brane also generally has, in addition to its $p$-brane charge,
lower-dimensional $q$-brane charges with $q=p-2,p-4,\dots$ which
depend on the Chan-Paton bundle over its
worldvolume~\cite{Douglas2}. Let $M\subset X$ be a compact connected
spin$^c$ manifold without boundary and let $E$ be a complex vector
bundle over $M$. Then $[M,E,\iota]\in\K^{\rm t}_\sharp(X)$ where
$\iota:M\hookrightarrow X$ is the natural inclusion. Under the assumptions of
Theorem~\ref{bigt}, if the brane is torsion-free then it
has an expansion in terms of the lifted homology basis for the lattice
$\Lambda_{\K^{\rm t}_\sharp(X)}$ of the form
\beq
[M,E,\iota]=\sum_{p=0}^n~\sum_{i=1}^{m_p}\,d^{~}_{pi}(M,E)~
\big[M_i^p,\id_{M_i^p},\iota_i^p\big]
\label{BwBexp}\eeq
with $d_{pi}(M,E)\in\zed$. The crucial point here is that the branes on the
right-hand side of (\ref{BwBexp}) are of {\it lower} dimension than
the original brane on the left-hand side, and have even codimension
with respect to the worldvolume $M$.

\begin{lemma}\label{BwBlemma}
If $[M_i^p,\id_{M_i^p}^\complex,\iota_i^p]\in\im(\iota_i^p)_*$ for each
$0\leq p\leq n$, $1\leq i\leq m_p$, then
$d_{pi}(M,E)=0$ for all $p\neq\dim_\real(M)-2j$ with
$j=0,1,\dots,\big\lfloor\frac{\dim_\real(M)}2\big\rfloor$.
\end{lemma}
\begin{proof}
We apply the Chern character (\ref{Chernchar}) to both sides of
(\ref{BwBexp}). Then
$\ch_\bullet(M_i^p,\id_{M_i^p}^\complex,\iota_i^p)$ is a sum of
homology cycles in $\H_{\rm even}(X;\rat)$ (resp. $\H_{\rm
  odd}(X;\rat)$) for $p$ even (resp. odd) of degree at most $p$.
Since $[M,E,\iota]=\iota_*[M,E,\Id_M]$, the conclusion then follows
from the commutative diagram
$$\xymatrix {\K_\sharp^{\rm t}(M) \ar[r]^{\iota_*}
  \ar[d]_{\textrm{\ch}_{\bullet}}
&  \K_\sharp^{\rm t}(X) \ar[d]^{\ch_{\bullet}}\\
            \H_\sharp(M;\mathbb{Q}) \ar[r]_{\iota_*} &\H_\sharp(X;\mathbb{Q})
            \ . }$$
\end{proof}

\begin{remark}\label{BwBrem}
This relationship represents the possibility of being able to
construct stable states of D-branes as bound states in
higher-dimensional branes by placing non-trivial Chan-Paton bundles on
the higher-dimensional worldvolume. Note that this works even when $E$
is a non-trivial {\it line} bundle. In general, it is difficult to
determine the lower $(p-1)$-brane charges $d_{pi}(M,E)\in\zed$ in
(\ref{BwBexp})
explicitly. We will return to this issue in
Section~\ref{Lifts}. \hfill{$\lozenge$}
\end{remark}

\subsection{Polarization\label{Polar}}

There is an ``opposite'' effect to the one just described wherein a
D$p$-brane can expand or ``polarize'' into a {\it higher} dimensional
brane~\cite{Myers1}. This is the dielectric effect which was mentioned in
Section~\ref{Qual} and it is intrinsically due to the nonabelian
structure groups that higher rank Chan-Paton bundles possess. We will
now describe this process in more detail. Let $M\subset X$ be a
compact spin$^c$ manifold without boundary. Then $M$ is $\K_{\rm
  t}^\sharp$-oriented. Let $F$ be a real $\C^\infty$ spin$^{c}$
  vector bundle over $M$ of even rank $2r$ with structure group ${\rm
    O}(2r)$ and Thom class $\tau^{~}_F$. Then $F\oplus\id_M^\real$ is a
  smooth ${\rm O}(2r+1)$ vector bundle which admits a nowhere zero section
  $\sec^F : M \rightarrow F\oplus\id_M^\real$ given by
  $\sec^F(x):=0_x\oplus 1$ for $x\in M$. The map $\sec^F$ may be regarded
  as a section of the sphere bundle defined by
  (\ref{hatMdef},\ref{pidef}). Then the Poincar\'{e} duality
  isomorphism yields a functorial homomorphism
  $$ \sec_{!}^{F} \,:\, \K_{\rm t}^i\big(M\big)~
  \longrightarrow~\K_{\rm t}^i\big(\,\widehat{M}\,\big)$$
given for $i=0,1$ and $\xi\in\K_{\rm t}^i(M)$ by
$$\sec_{!}^{F}(\xi):=\Phi_{\widehat{M}}^{-1}
  \circ \sec^{F}_{*} \circ \Phi^{~}_{M}(\xi) \ . $$
This is called the \textit{Gysin homomorphism}, and by
using (\ref{hatMBF}) it may be represented as the
composition~\cite{14}
  $$\sec_{!}^{F}\,:\,\K^i_{\rm t}\big(M\big) ~
\stackrel{\mathfrak{T}^{~}_{M,F}}{\longrightarrow}~
\K_{\rm t}^i\big(\B^{+}(F)\,,\,\S(F)\big)~
\stackrel{{\rm exc}}{\longrightarrow}~\K_{\rm t}^i\big(\,
\widehat{M}\,,\,\B^{-}(F)\big)~\stackrel{\varsigma^*}
{\longrightarrow}~\K_{\rm t}^i\big(\,\widehat{M}\,\big)$$
where the first map is the Thom isomorphism of $F$, the second map is
excision and the third map is
restriction induced by the inclusion $\varsigma:(\,\widehat{M}\,,\,\emptyset)
\hookrightarrow(\,\widehat{M}\,,\,\B^{-}(F))$.

Consider now the complex vector bundle $H(F)\to\widehat{M}$ defined in
(\ref{HFdef}). With $\pi$ the bundle projection (\ref{pidef}),
$[H(F)]|_{\pi^{-1}(x)}=[H(F)|_{\pi^{-1}(x)}]$ is the Bott generator of
$\widetilde{\K}_{\rm t}^0(\B(F)_{x}\,/\,\S(F)_{x})$ for all $x\in M$. It
follows that the K-theory class $[H(F)]\in\K_{\rm
  t}^0(\,\widehat{M}\,)$ is related to the Thom
class of $F$ by $\varsigma^*\circ{\rm exc}(\tau^{~}_{F})=
[H(F)]$. Since $\varsigma^*$ and ${\rm exc}$ are ring
isomorphisms, from the explicit expression for the Thom isomorphism
(\ref{ThomisodefV}) one has
$\sec_{!}^{F}[E]=\pi^{*}[E]\cup[H(F)]=[\pi^{*}(E) \otimes H(F)]$ for
any complex vector bundle $E\to M$. Thus given a K-cycle $(M,E,\phi)$
on $X$, we can rewrite vector bundle modification as the equivalence
relation
\beq
\big(M,E,\phi\big) \sim
\big(\,\widehat{M},\sec_{!}^{F}[E], \phi \circ \pi\big) \ .
\label{vecmodrewrite}\eeq

The {\it charge} of the D-brane $[M,E,\phi]\in\K_\sharp^{\rm t}(X)$ is, by
definition, the index (\ref{IndexphiE}). From (\ref{vecmodrewrite}) we
see that we can rewrite it using $$\mu^{\rm a}\big[M,E,\phi\big]=\mu^{\rm
  a}\big[\,\widehat{M},\sec_!^F[E],\phi\circ\pi\big]$$ to get
$$\Index~\phi_{*}\big[\Dirac_{E}
\big]=\ch^\bullet\,\sec_!^F\big[E\big]\cup\td\big(T
\widehat{M}\,\big)\big[\,\widehat{M}\,\big] \ . $$
Thus the charge of a polarized D$p$-brane can be expressed entirely in
terms of characteristic classes associated with the higher-dimensional
spherical D$(p+2r)$-brane into which it has dissolved to form a bound
state. A similar formula was noted in a specific context
in~\cite{11}. This is a general feature of bound states of D-branes, and
they can always be expressed entirely in terms of quantities intrinsic
to the ambient space $X$~\cite{MM1}, as we will now proceed to show.

\subsection{Tachyon Condensation\label{BSC}}

The constructions we have thus far presented involve {\it stable}
states of D-branes. We can also consider configurations of branes
which are {\it a priori} unstable and decay into stable D-branes. This
requires us to start considering virtual elements in K-theory, with
the stable states being associated to the positive cone of the
K-theory group. Using these elements we can construct an explicit set
of generators for $\K_\sharp^{\rm t}(X)$, including its torsion
subgroup. The decay mechanism is then represented as a change of basis
for the topological K-homology group after we compare these generators
with those obtained in Theorem~\ref{bigt}.

\begin{proposition}\label{cpn}
Let $X$ be an $n$-dimensional compact connected spin$^c$ manifold
without boundary whose degree~$0$ reduced topological K-theory group
can be presented as the split
  $$\widetilde{\K}_{\rm t}^{0}(X)=
  \Big(\,\bigoplus_{j=1}^{m}\,\xi^{~}_{j}~\mathbb{Z} \Big) ~\oplus~
  \Big(\,\bigoplus_{i=1}^{k}~\bigoplus_{l=1}^{p_{i}}\,\gamma_{l}^{i}~
   \mathbb{Z}_{n_{i}}\Big) \ , $$
  where $n_{i} \geq 2$,  $p_{i} \geq 1$ and
  $\xi^{~}_j,\gamma_l^i\in\widetilde{\K}_{\rm t}^0(X)$ for each $1 \leq j
  \leq m$, $1 \leq i \leq k$, $1 \leq l \leq p_{i}$. Choose
  representatives for the generators $\xi^{~}_{j} = [E^{~}_{j}]-[F^{~}_{j}]$
and
  $\gamma_{l}^{i} = [G_{l}^{i}]-[H_{l}^{i}]$ in terms of complex
  vector bundles over $X$. Then $\widetilde{\K}^{\rm t}_{e(n)}(X)$ is
  generated as an abelian group by the elements
\bea
\big[X,E^{~}_{j}, \Id^{~}_X\big] - \big[X,F^{~}_{j}, \Id^{~}_X\big] \quad &,&
\quad 1 \leq j \leq m \ , \nn\\ \big[X,G_{l}^{i},\Id_X\big]-
\big[X,H_{l}^{i},\Id_X\big] \quad &,& \quad 1
  \leq i \leq k \ ,  \quad 1 \leq l \leq p_{i}\ . \nn
\eea
\end{proposition}
\begin{proof}
We explicitly construct the Poincar\'{e} duality isomorphism
$\Phi^{\rm t}_{X} :\K_{\rm t}^{0}(X) \rightarrow\K_{e(n)}^{\rm t}(X)$
induced from (\ref{PhiXdef}) in this case. Recall from Section~\ref{RelK}
that the K-cycle class $[X, \id_X^\complex, \textrm{id}^{~}_{X}]$ is the
fundamental class of $X$ in $\K_{e(n)}^{\rm t}(X)$ and that the
isomorphism $\mu^{\rm s}$ is compatible with cap products. It follows
that the map $\Phi_X^{\rm t}:=(\mu^{\rm s})^{-1}\circ\Phi^{~}_X$ is
given explicitly by
\bea
\Phi^{\rm t}_X\big(\xi^{~}_j\big)&=&\big([E^{~}_j]-[F^{~}_j]\big)\cap
\big[X, \id_X^\complex, \Id^{~}_{X}\big]\nn\\
&=&\big[X, \id_X^\complex\otimes
E^{~}_j, \Id^{~}_{X}\big]-\big[X, \id_X^\complex\otimes
F^{~}_j, \Id^{~}_{X}\big]~=~\big[X,E^{~}_j, \Id^{~}_{X}\big]-
\big[X,F^{~}_j, \Id^{~}_{X}\big] \ . \nn
\eea
The conclusion now follows by Poincar\'{e} duality.
\end{proof}

\begin{remark}\label{IIBrem}
Suppose that $X$ satisfies the conditions of
both Theorem~\ref{bigt} and Proposition~\ref{cpn}. Let
$[M_i^p,\id_{M_i^p}^\complex,\iota_i^p]$ be generators of the lattice
$\Lambda_{\widetilde{\K}_{e(n)}^{\rm t}(X)}=\widetilde{\K}_{e(n)}^{\rm
  t}(X)\,/\,\tor_{\widetilde{\K}_{e(n)}^{\rm t}(X)}$ given by
Theorem~\ref{bigt},
and $ [X,E_j, \textrm{id}_X] - [X,F_j, \textrm{id}_X]$ the generators
given by Proposition~\ref{cpn}. Since these are bases of the same free
$\mathbb{Z}$-module $\Lambda_{\widetilde{\K}_{e(n)}^{\rm t}(X)}$, there
are uniquely defined integers $a_{pi}^j$ such that
\beq
\big[X,E^{~}_j, \textrm{id}^{~}_X\big] - \big[X,F^{~}_j,
\textrm{id}^{~}_X\big]=
\sum_{\stackrel{\scriptstyle p=1}{e(p)=e(n)}}^n~\sum _{i=1}^{m_p}
\,a_{pi}^j~\big[M_i^p,\id_{M_i^p}^\complex,\iota_i^p\big]
\label{IIBchangebasis}\eeq
for $1\leq j\leq m$. In the string theory setting, $X$ is a
ten-dimensional spin manifold and (\ref{IIBchangebasis}) represents
a change of basis on $\Lambda_{\widetilde{\K}_0^{\rm t}(X)}$. The
right-hand side is an expansion in terms of the stable torsion-free
Type~IIB D-branes which wrap the non-trivial spin$^c$ homology cycles of $X$ in
even degree. The left-hand side is the difference between a pair of
{\it spacetime-filling} D9-branes wrapping the entire ambient space
$X$. The relative sign difference indicates that one of these branes
should be regarded as oppositely charged relative to the other, i.e. it
is an {\it antibrane}. The left-hand side thus represents a
brane-antibrane system. It is unstable and (\ref{IIBchangebasis})
describes its decay into lower-dimensional stable D-branes in $X$. Note that
any stable D-brane of even degree can be constructed from such 9-brane
pairs. We have thereby reproduced the construction of Type~IIB stable
supersymmetric D-branes from spacetime filling brane-antibrane
pairs~\cite{16,Sen1,Witten1}. As before, however, it is in general
quite difficult to explicitly determine the $(p-1)$-brane charges
$a_{pi}^j\in\zed$ in (\ref{IIBchangebasis}). We remark on the
analogous construction in Type~IIA string theory in
Section~\ref{IIArem}.~\hfill{$\lozenge$}
\end{remark}

\begin{example}[{\it ABS Construction}]\label{ABSEx}
Let $X$ be a compact connected ten-dimensional $\C^\infty$ spin
manifold without boundary. Let $(M,E,\phi)$ be a K-cycle on $X$ such that
$\phi:M\hookrightarrow X$ is a proper imbedding with $M$ connected and
of even codimension $2k$ in $X$ (so that the corresponding K-homology class
describes a D$p$-brane in $X$ with $p=9-2k$). Choosing a riemannian
metric on $X$, the normal bundle $NM\to M$ of $M$ in $X$ fits into an
exact sequence of real vector bundles as $$0~\longrightarrow~TM~
\stackrel{\phi_*}{\longrightarrow}~TX~\longrightarrow~NM~
\longrightarrow~0 \ . $$ Let $w_i(F)\in\H^i(M;\zed_2)$ denote the
$i$-th Stiefel-Whitney class of a real vector bundle $F\to M$ with
$w_0(F)=1$. If the metric on $X$ restricts non-degenerately to the
worldvolume $M$, then one has the Whitney sum formula
$$\phi^*w_i(TX)=\sum_{j=0}^i\,w_j(TM)\cup w_{i-j}(NM) \ . $$
Since $X$ and $M$ are orientable, we have $w_1(TX)=w_1(TM)=0$ and
hence $w_1(NM)=0$. Since $X$ is spin, we also have $w_2(TX)=0$ and
hence $w_2(NM)=w_2(TM)$. Thus endowing $M$ with a spin$^c$ structure
is equivalent to endowing its normal bundle with a spin$^c$
structure. It follows that $NM\to M$ is a real spin$^c$ $\C^\infty$
vector bundle with structure group ${\rm SO}(2k)$.
Applying the clutching construction of Section~\ref{ss2} to $F=NM$,
vector bundle modification then identifies the D-branes
\beq
\big[\,\widehat{M},H(NM)\otimes\pi^*(E),\phi\circ\pi\big]
=\big[M,E,\phi\big] \ .
\label{NMclutch}\eeq

To proceed further we need the following elementary result~\cite{4}.

\begin{lemma}\label{l4}
Let $W$ be a compact manifold and $Z$ a connected manifold which are
both non-empty and have the same dimension. Then any embedding $f :
W\rightarrow Z$ is a diffeomorphism.
\end{lemma}
\begin{proof}
Let $V=Z\,\setminus\,f(W)$. Since $W$ and $Z$ have the same
dimension, $f(W)$ is open in $Z$. On the other hand, since $W$ is compact,
$f(W)$ is compact in $Z$ and hence closed. Thus one concludes that $V$
is both open and closed in $Z$. Since $Z$ is connected, it follows
that either $V=\emptyset$ or $V=Z$. Since $W$ and $Z$ are non-empty,
one has $V=\emptyset$.
\end{proof}
\noindent
Let us apply this lemma to the sphere bundle $\widehat{M}$, which in
the present case is a compact ten-dimensional submanifold of $X$. It follows
that $\widehat{M}\cong X$ and so the left-hand side of (\ref{NMclutch})
represents a configuration of spacetime-filling D9-branes. To see that it
is an element of the basis set provided by Proposition~\ref{cpn},
we appeal to the Atiyah-Bott-Shapiro (ABS) construction in topological
K-theory~\cite{ABS1}. For this, we use the metric on $X$ to construct a
tubular neighbourhood $M'$ of $M$ in $X$. Let $\overline{M}$ denote
the closure of $M'$ in $X$ and $M^\flat$ its boundary. The neighbourhood
$M'$ may be identified with the total space of the normal bundle
$NM\to M$. Without loss of generality, we can identify $M'$ with the
interior of the unit ball bundle $\B(NM)\setminus\S(NM)$ (whose fibres
consist of normal vectors with norm $<1$). Then we have the identifications
$\overline{M}=\B(NM)$ and $M^\flat=\S(NM)$. Let $\rho:M'\to M$ be the
retraction of the regular neighbourhood $M'$ onto $M$, and denote the
twisted spinor bundles over $\overline{M}$ by
$\Delta^\pm_E:=\Delta^\pm(NM)\otimes\rho^*(E)$. Set $X'=X\setminus M'$.

Suppose first that the bundle $\Delta_E^-$ admits an extension over
$X'$, also denoted $\Delta_E^-$. Via the (extended) Clifford multiplication map
$\sigma_E:=\sigma\otimes\Id_{\rho^*(E)}$, the bundle $\Delta_E^+$ is isomorphic
to $\Delta_E^-$ on $M^\flat$, and so it can also be extended over $X'$
by declaring that it be isomorphic to $\Delta_E^-$ over $X'$. This
gives a pair of bundles $\Delta_E^\pm\to X$ that determine an element
$[\Delta_E^+]-[\Delta_E^-]$ of the reduced K-theory group $\widetilde{\K}_{\rm
  t}^0(X)$ which vanishes on $X'$. On the other hand, from
(\ref{HFdef}) we see that this K-theory
element is just the Gysin homomorphism
$$\big[\Delta_E^+\big]-\big[\Delta_E^-\big] ~=~
\Omega_!^{NM}\big[E\big] ~=~ \pi^*\big[E\big]\cup\big[H(NM)\big]
\ . $$ Using this fact along with the diffeomorphism $\widehat{M}\cong
X$, the identification (\ref{NMclutch}) becomes
\beq
\big[X,\Delta_E^+,\Id^{~}_X\big]-\big[X,\Delta_E^-,\Id^{~}_X\big]
=\pm\,\big[M,E,\phi\big]
\label{9braneid}\eeq
where the sign depends on whether or not the spin$^c$ structures on
$\widehat{M}$ and $X$ coincide. This is the standard construction of a
Type~IIB D-brane $[M,E,\phi]$ in terms of spacetime-filling
brane-antibrane pairs~\cite{16,Witten1}. In this context the Clifford
multiplication map $\sigma_E$ is called the {\it tachyon
field} and the decay mechanism (\ref{9braneid}) is known as {\it
tachyon condensation}. If the bundle $\Delta_E^-$ does not admit
an extension over $X'$, we use Swan's theorem to construct a complex
vector bundle $G\to M$ such that $G\oplus(S_1(NM)\otimes E)$ is
trivial over $M$, and hence whose pullback $\rho^*(G)\oplus\Delta^-_E$
is trivial over $\overline{M}$. Then $\rho^*(G)\oplus\Delta^-_E$ can
be extended over the whole of $X$ as a trivial bundle. The bundle
$\rho^*(G)\oplus\Delta^+_E$ is isomorphic to
$\rho^*(G)\oplus\Delta^-_E$ on $M^\flat$ under the vector bundle map
$\Id_{\rho^*(G)}\oplus\sigma_E$, and so it can
also be extended over $X$ by setting it equal to $\rho^*(G)\oplus\Delta^-_E$
over $X'$. The resulting K-theory class is again trivial over $X'$
(but not over $X$), and by the direct sum relation the Poincar\'e dual
K-homology class coincides with
(\ref{9braneid}). \hfill{$\lozenge$}\end{example}

\subsection{Unstable 9-Branes\label{IIArem}}

The crux of the constructions of Section~\ref{BSC} is that one can use virtual
elements which signal instability of the given configurations of
branes. Using Corollary~\ref{corKclass} one can replace $\Gamma(X)$
with the collection of isomorphism classes of triples $(M,\xi,\phi)$,
where $M$ and $\phi$ are as in Definition~\ref{kcycle} and
$\xi\in\K_{\rm t}^0(M)$ is a class in the degree~$0$ topological
K-theory of $M$. Clearly both definitions lead to the same group
$\K_\sharp^{\rm t}(X)$. One can further extend the definition to
triples $(M,\xi,\phi)$ with $\xi\in\K_{\rm t}^\sharp(M)=\K_{\rm
  t}^0(M)\oplus\K_{\rm t}^1(M)$~\cite{3}. The $\zed_2$-grading on
$\K_\sharp^{\rm t}(X)$ is then defined by taking $\K_0^{\rm t}(X)$
(resp. $\K_1^{\rm t}(X)$) to be the subgroup given by classes of
K-cycles $(M,\xi,\phi)$ such that $\xi|_{M_l}\in\K_{\rm t}^{i_l}(M)$ for
some $i_l=0,1$ and $\dim_\real(M_l) + i_l$ is an even (resp. odd)
integer for all connected components $M_l$ of $M$. Vector bundle
modification is now generically described as the equivalence relation
(\ref{vecmodrewrite}) using the Gysin homomorphism associated to a
real spin$^c$ vector bundle $F\to M$ whose rank has the same parity as
that of the dimension of (the connected components of) $M$.

To see that this definition is in fact equivalent to our previous one,
let $[M,\xi,\phi]\in\K_i^{\rm t}(X)$ with $i=0,1$, $M$ connected, and
$\xi$ a non-zero element of $\K_{\rm t}^j(M)$ for some $j=0,1$. If
$m:=\dim_\real(M)$, then one has $m+j\equiv i~{\rm mod}~2$ by
definition. Consider the trivial spin$^c$ bundle
$F=\id_M^{\real^{i+m+1}}$ and the associated Gysin map
$\Omega_!^F:\K_{\rm t}^j(M)\to\K_{\rm
  t}^{e(j+i+m)}(\,\widehat{M}\,)$. Since $j+i+m\equiv j+m+j+m~{\rm
  mod}~2\equiv0~{\rm mod}~2$, one has $\K_{\rm
  t}^{e(j+i+m)}(\,\widehat{M}\,)\cong\K_{\rm
  t}^0(\,\widehat{M}\,)$. It follows that there are complex vector
bundles $E,H\to M$ with $\Omega_!^F(\xi)=[E]-[H]$, and by vector
bundle modification one has
$[M,\xi,\phi]=[\,\widehat{M},E,\phi\circ\pi]-
[\,\widehat{M},H,\phi\circ\pi]$ in $\K_i^{\rm t}(X)$. Notice that by
using the usual cup product on the K-theory ring $\K^\sharp_{\rm t}(X)$, the
cap
product (\ref{capprodext}) may now be alternatively defined by
$\xi'\cap[M,\xi,\phi]=[M,\phi^*\xi'\cup\xi,\phi]$ for $\xi'\in\K_{\rm
  t}^i(X)$ and K-cycle classes $[M,\xi,\phi]\in\K^{\rm t}_j(X)$.

Suppose that $X$ is a compact spin$^c$ manifold without boundary
obeying the conditions of Theorem~\ref{bigt}. Let $E$ be a complex
vector bundle over $X$ and $\alpha$ an automorphism of $E$. This
defines a degree~$1$ K-theory class $[E,\alpha]\in\K_{\rm t}^1(X)$
which we assume to be torsion-free. Applying the Poincar\'e duality
isomorphism as before then gives the
analog of the expansion (\ref{IIBchangebasis}) as
\beq
\big[X,(E,\alpha),\Id^{~}_X\big]=
\sum_{\stackrel{\scriptstyle p=1}{e(p)=e(n+1)}}^n~\sum _{i=1}^{m_p}
\,b_{pi}^{~}(E,\alpha)~\big[M_i^p,\id_{M_i^p}^\complex,\iota_i^p\big]
\label{IIAchangebasis}\eeq
in $\Lambda_{\widetilde{\K}^{\rm t}_{e(n+1)}(X)}$. In the string theory
setting, this is a relation in $\Lambda_{\K^{\rm t}_1(X)}$
expressing the decay of an {\it unstable D9-brane} into stable
Type~IIA D-branes~\cite{Horava1,16}. $E\to X$ is the Chan-Paton bundle
on the 9-brane and now the automorphism $\alpha:E\to E$ plays the role
of the tachyon field. As an explicit example of the decay mechanism
(\ref{IIAchangebasis}),
we may construct the Type~IIA version of the ABS construction of
Example~\ref{ABSEx}. Now we consider a K-cycle $(M,E,\phi)$ on $X$
with $M$ of odd codimension $2k+1$ in $X$, so that the corresponding
normal bundle $NM\to M$ is an ${\rm SO}(2k+1)$ vector bundle. By
Lemma~\ref{l4} one again has a diffeomorphism $\widehat{M}\cong
X$. Define $\Delta^{~}_E:=\Delta(NM)\otimes\rho^*(E)$, where
$\Delta(NM)=S(NM)|_{\overline{M}}$ is the pull-back of the unique
irreducible spinor bundle $S(NM)$ over $NM$, and assume that it admits
an extension over $X'$. Then the Gysin homomorphism gives
$\Omega_![E]=[\Delta^{~}_E,\exp\sigma^{~}_E]$ in $\K_{\rm t}^1(X)$, and so by
vector bundle modification one has the identification
$$\big[X,(\Delta^{~}_E,\exp\sigma^{~}_E),\Id^{~}_X\big]=\pm\,\big[M,E,\phi\big]
\
. $$ This is the standard construction of a Type~IIA D-brane
$[M,E,\phi]$ in terms of unstable spacetime-filling
9-branes~\cite{Horava1,16}.

\begin{remark}\label{IIAremimp}
It is important to realize that one can stick to our original
definition and thus avoid $\K_{\rm t}^1$-classes entirely. One of the
great advantages of the geometric formulation of K-homology, in
contrast to other homology theories, is that it is naturally defined
in terms of {\it stable} objects and one need never consider virtual
elements. While brane-antibrane systems are straightforward to
construct, the unstable D-branes defined by virtual K-theory classes
in degree~$1$ are not so natural in this framework. This reflects
the difficulties encountered in the description of these D-brane
states directly in string theory.

To illustrate this point further, let $X$ be as in
Proposition~\ref{cpn} and consider the Gysin homomorphism in K-theory
$ \sec_{!}^{F_n}:\K_{\rm t}^i(X)
  \rightarrow\K_{\rm t}^{e(i+n)}(\,\widehat{X}\,)$,
  where $F_n=X \times \mathbb{R}^n$. Then $\widehat{X}=X \times \S^n$
  and the Gysin homomorphism becomes a map
 $$ \sec_{!}^{F_n} \,:\, \K_{\rm t}^i(X)~
  \longrightarrow~\K_{\rm t}^{e(i+n)}(X \times \S^n) \ . $$
The sphere bundle projection $\pi : X \times \S^n \rightarrow X$ in
this case is the projection onto the first factor. Since $\pi_!
\circ \sec_{!}^{F_n} = \Id_{\K_{\rm t}^\sharp(X)}$, it follows that
$\sec_{!}^{F_n}$ is a monomorphism and $\pi_!$ is an epimorphism. By
definition one has $\sec_{!}^{F_n} = ( \Phi_{X \times \S^n}^{\rm t}
) ^{-1} \circ \sec^{F_n}_{*} \circ \Phi_{X}^{\rm t}$ and $\pi_! =
( \Phi_{X}^{\rm t}) ^{-1} \circ \pi_* \circ \Phi_{X \times
\S^n}^{\rm t}$. Because the Poincar\'{e} duality maps are isomorphisms,
one concludes that the induced maps $\sec^{F_n}_{*}$ and $\pi_*$ in
K-homology are also a monomorphism and an epimorphism, respectively.

Assume that the degree~$1$ topological K-theory group of $X$ admits a
split
 $${\K}_{\rm t}^{1}(X)=
  \Big(\,\bigoplus_{j=1}^{m}\,\xi^{~}_{j}~\mathbb{Z} \Big) ~\oplus~
  \Big(\,\bigoplus_{i=1}^{k}~\bigoplus_{l=1}^{p_{i}}\,\gamma_{l}^{i}~
   \mathbb{Z}_{n_{i}}\Big) \ . $$
There is a commutative diagram
$$\xymatrix {\K^1_{\rm t}(X) \ar[r]^{\sec_{!}^{F_1}}
  \ar[d]_{\Phi_{X}^{\rm t}}
&  \K^0_{\rm t}(X \times \S^1) \ar[d]^{\Phi_{X \times \S^1}^{\rm t}}\\
            \K^{\rm t}_{e(n-1)}(X) \ar[r]_{\!\!\!\!\!\!\!\sec^{F_1}_*}
&\K^{\rm t}_{e(n-1)}(X \times \S^1)
            \ . }$$
Let  $\sec_{!}^{F_1}(\xi^{~}_{j}) = [E^{~}_{j}]-[F^{~}_{j}]$ and
  $\sec_{!}^{F_1}(\gamma_{l}^{i}) = [G_{l}^{i}]-[H_{l}^{i}]$ in terms of
complex
  vector bundles over $X \times \S^1$. Then for all $1 \leq j \leq
  m$ one has
\bea
[ X \times \S^1,E^{~}_{j}, \pi] - [ X \times \S^1,
F^{~}_{j}, \pi]&=&
\pi_* \big([ X \times \S^1,E^{~}_{j}, \Id^{~}_{X \times \S^1}] -
[ X \times \S^1,F^{~}_{j}, \Id^{~}_{X \times \S^1}] \big)\nonumber\\
&=&\pi_* \circ \sec^{F_1}_*[ X,\tau^{~}_{j}, \Id^{~}_{X}] \nonumber
\eea
for some $\tau^{~}_{j} \in \K^0_{\rm t}(X)$. Since $\pi_* \circ \sec^{F_1}_* =
\Id^{~}_{\K_\sharp ^{\rm t}(X)}$, by Poincar\'{e} duality it
follows that
$$[ X \times \S^1,E^{~}_{j}, \pi] - [ X \times \S^1,F^{~}_{j}, \pi]
\quad , \quad 1 \leq j \leq m$$
is a set of generators for the torsion-free part of the K-homology group
$\K^{\rm t}_{e(n-1)}(X)$. One can use the same procedure for the other
set of generators of $\K^1_{\rm t}(X)$ to conclude that
$$[ X \times \S^1,G_{l}^{i}, \pi] -[ X \times \S^1,H_{l}^{i}, \pi]
\quad , \quad 1\leq i\leq k \ , ~~ 1\leq l\leq p_i$$
is a set of generators for the torsion subgroup of $\K^{\rm
  t}_{e(n-1)}(X)$. In the string theory setting, this gives an
``M-theory'' realization of unstable Type~IIA 9-branes in terms of
brane-antibrane systems on an 11-dimensional extension $X\times\S^1$
of the spacetime manifold $X$~\cite{16,Witten1}. More generally, one
can start from any real spin$^c$ line bundle $F\to X$ and describe
these unstable D-brane states in terms of a spin$^c$ circle bundle
over $X$.~\hfill{$\lozenge$}
\end{remark}

\newsection{Torsion-Free D-Branes}

In this section we will describe some elementary applications of the
formalism of the previous section. While for the most part we will
arrive at the anticipated results, this simple analysis will illustrate how the
known properties of D-branes arise within a mathematically precise
formalism. We will only look at examples of torsion-free K-homology
groups, defering the analysis of torsion D-branes to the next
section.

\subsection{Spherical D-Branes\label{Sphere}}

An important role will be played by the D-branes which wrap images
of $n$-dimensional spheres $\S^n$ in $X$. We will first consider the
case where $X$ is an arbitrary finite CW-complex. Let $n \geq 0$ and
let $E$ be a complex vector bundle over $\S^n$. Using
Lemma~\ref{hom} we can construct a homomorphism $$\gamma_{n,E}\, :\,
[\S^n,X]~ \longrightarrow~\K_\sharp^{\rm t}(X)$$ given by $$
\gamma_{n,E}[\psi]:=[\S^n,E,\psi] \ . $$ We can also construct a
  homomorphism $$h_{n,E} \,:\, \pi_n(X) ~\longrightarrow~\K_\sharp^{\rm t}(X)$$
  given by $$h_{n,E}[f]:=[\S^n,E,f] \ . $$  The subgroup of
$\K_\sharp^{\rm t}(X)$ generated by the K-cycle classes of the form
$[\S^n,E,\psi]$ for $n\geq1$ and $[\pt,F,\phi]$ is denoted $\S_\sharp^{\rm
  t}(X)$. It has a natural $\zed_2$-grading $\S_\sharp^{\rm t}(X)=\S_0^{\rm
  t}(X)\oplus\S_1^{\rm t}(X)$ by the parity $e(n)$ of the sphere
dimensions $n$.

\begin{proposition}\label{hur}
Let $n \geq 0$ and let $E$ be a complex vector bundle over $\S^n$.
\begin{alphlist}
\item If $X$ is path connected and simply connected, then ${\rm
    im}\,\gamma_{n,E}={\rm im}\,h_{n,E}$ in $\K_\sharp^{\rm t}(X)$.
\item If $E= \id_{\S^n}^\complex$, then $h_n:=h_{n,\id_{\S^n}^\complex}$ is the
  Hurewicz homomorphism in K-homology.
\end{alphlist}
\end{proposition}
\begin{proof}
(a) follows immediately from the fact that $[\S^n,X]=\pi_n(X)$ in this
case~\cite{6}. (b) follows from the fact that
$[\S^n,E,f]=f_*[\S^n,\id_{\S^n}^\complex,\Id_{\S^n}^{~}]$ with
$[\S^n,\id_{\S^n}^\complex,\Id_{\S^n}^{~}]$ the fundamental class of
$\S^n$ in $\K_\sharp^{\rm t}(\S^n)$.
\end{proof}

\begin{remark} \label{r1}
Let $f : \S^1 \rightarrow \S^1$ be a continuous map with
$f(s_0)=x_0\neq s_0$. Regarding $\S^1 \subset \mathbb{C}$, let $\theta
\in (0,1)$ be defined by $x_0=\e^{2\pi\ii\theta}\,s_0$. Define $f_0 :
\S^1 \rightarrow \S^1$ by $f_0(z):=\e^{-2\pi\ii\theta}\,f(z)$. Then
$f_0$ is continuous with $f_0(s_0)=s_0$, and so it is a based map
at $s_0$. The map $H : [0,1]\times \S^1 \rightarrow \S^1$ defined by
$H(t,z):=\e^{-2\pi\ii t\,\theta}\,f(z)$ is a homotopy between
$f$ and $f_0$. Since homotopy of maps is an equivalence
relation, we conclude that the map $G : [\S^1,\S^1] \rightarrow \pi_{1}(\S^1)$
given by the assignment $[f] \mapsto  [f_0]$ is well-defined and a
bijection. Thus we may consider $\gamma_{1,E} : [\S^1,X]
\rightarrow \K_\sharp^{\rm t}(X)$ as a homomorphism $\gamma_{1,E}:
\pi_{1}(X) \rightarrow\K_\sharp^{\rm t}(X)$ for any topological space
$X$. \hfill{$\lozenge$}
\end{remark}

We shall now specialize to the case $X=\S^n$, whose cellular structure
consists of a single $k$-cell in dimensions $k=0,n$. We will find generators
for the subgroup of $\K_\sharp^{\rm t}(\S^n)$ generated by the lower
dimensional spheres $\S^k$ with $1\leq k\leq n$ and $\pt$.

\begin{proposition}\label{prop2}
$\K_\sharp^{\rm t}(\S^0)=\S_\sharp^{\rm t}(\S^0) \cong \mathbb{Z}\oplus\zed$
with $\S_{0}^{\rm t}(\S^0) \cong\mathbb{Z}\oplus\zed$ and $\S_{1}^{\rm
  t}(\S^0) \cong 0$.
\end{proposition}
\begin{proof}
By definition of the group $\S_\sharp^{\rm t}(\pt)$ it follows immediately that
$\S_{1}^{\rm t}(\textrm{pt})$ is the free abelian group generated by
the trivial K-cycle class $[\emptyset,\emptyset,\emptyset]$, i.e.
$\S_{1}^{\rm t}(\textrm{pt})=0$. A complex vector bundle $E$
over pt is just a finite dimensional vector space, i.e. there exists
an integer $m>0$ such that $E=\textrm{pt} \times \mathbb{C}^m=
\bigoplus_{i=1}^{m}\,\id_\pt^\complex$. Since the unique map
$\phi : \textrm{pt} \rightarrow \textrm{pt}$ is the identity, it
follows that $[\textrm{pt},E,\phi]=m~[\textrm{pt},\id_\pt^\complex
,\Id^{~}_{\textrm{pt}}]$. Thus $\S_{0}^{\rm t}(\textrm{pt}) \cong
[\textrm{pt}, \id_\pt^\complex ,\Id^{~}_{\textrm{pt}}]~\mathbb{Z}
\cong \mathbb{Z}$ and the conclusion now follows from Lemma~\ref{l2}.
\end{proof}
\noindent
As a consequence of Lemma~\ref{l3}, Theorem~\ref{bigt} and
Proposition~\ref{prop2} we have the following result.

\begin{lemma} \label{l6}
If $n \geq 1$ then $\S_\sharp^{\rm t}(\S^n)$ is generated by classes of the
form
$[\S^k,E,\phi]$ and $[\pt,\id_\pt^\complex ,\iota]$ where $1 \leq k
\leq n$, $E$ is a generating vector bundle for the K-theory of $\S^k$,
and $\iota :\pt\hookrightarrow\S^n$ is the inclusion of a point.
\end{lemma}

Recall that the (complex) K-theory of the spheres is given by
\begin{displaymath}
\K_{\rm t}^0(\S^n)~\cong~\left \{ \begin{array}{ll} \mathbb{Z} \oplus
    \mathbb{Z}
&, \quad n~\textrm{even}\\
      \quad \mathbb{Z} & , \quad n~\textrm{odd}\end{array} \right. \ ,
\qquad \K_{\rm t}^1(\S^n) ~\cong~ \left \{ \begin{array}{ll} 0 & , \quad
n~\textrm{even}\\
      \mathbb{Z} & , \quad n~\textrm{odd} \end{array} \right. \ .
\end{displaymath}
The trivial line bundle $\id_{\S^n}^\complex=\S^n\times \mathbb{C}$ is
always a degree~$0$ generator, given by the monomorphism $\K_{\rm
  t}^0(\pt) \hookrightarrow \K_{\rm t}^0(\S^n)$ induced by the
inclusion of a point. The non-trivial generator of $\K^0_{\rm
  t}(\S^2)$ is obtained from the homeomorphism $\mathbb{C}\P^1 \cong
\S^2$ by taking the class of the canonical line bundle $\mathcal{L}_1$ over the
complex projective line $\mathbb{C}\P^1$. The non-trivial generator
$[\mathcal{L}_1]^p$ of $\K_{\rm t}^0(\S^{2p})$, $p \in \mathbb{N}$ is
obtained from $[\mathcal{L}_1]$ by using the K-theory cup
product~\cite{7}. The $\K_{\rm
  t}^1$-groups are obtained through suspension
$\S^{n+1}\cong\Sigma\S^n=\S^n\wedge\S^1$. By Poincar\'e duality one
concludes that
\bea
  \K_{0}^{\rm t}(\S^n)&\cong& \K_{\rm t}^{e(n)}(\S^n) ~\cong~
 \left \{ \begin{array}{ll}
      \mathbb{Z} \oplus \mathbb{Z} &, \quad
      n~\textrm{even}\\
      \quad \mathbb{Z} & , \quad n~\textrm{odd} \end{array} \right. \
  , \nonumber\\
\K_{1}^{\rm t}(\S^n) &\cong& \K_{\rm t}^{e(n-1)}(\S^n) ~\cong~
 \left \{ \begin{array}{ll} 0
  &, \quad n~\textrm{even}\\
      \mathbb{Z} &, \quad n~\textrm{odd}\end{array} \right. \nonumber
\eea
for $n\geq 1$, and hence $$\K_{\sharp}^{\rm t}(\S^n) \cong  \mathbb{Z}
\oplus \mathbb{Z} \ . $$

\begin{proposition}\label{KS}
  Let $n\geq 1$.
\begin{alphlist}
\item $\K_\sharp^{\rm t}(\S^n)$ is generated by the classes
$[\pt,\id_\pt^\complex
  ,\iota]$ and $[\S^n,\id_{\S^n}^\complex ,\Id^{~}_{\S^n}]$.
\item $\S_\sharp^{\rm t}(\S^n)=\K_\sharp^{\rm t}(\S^n)$ as
  $\zed_2$-graded abelian groups.
\end{alphlist}
  \end{proposition}
  \begin{proof}
(a) follows from calculating the Chern characters of the classes.
    Since  $\ch^\bullet(\id_X^\complex)=1$ for any space $X$ and
    $\td(T\textrm{pt})=1$, it follows that
    $\ch_\bullet(\textrm{pt},\id_\pt^\complex,\iota)=\iota_{*}(
\ch^\bullet(\id_\pt^\complex)\cup\td(T\textrm{pt})
    \cap[\textrm{pt}])=\iota_{*}[\textrm{pt}]=1$.
    We also have $T\S^n \oplus
    \id_{\S^n}^\complex=\id_{\S^n}^{\complex^{n+1}}$, so that
    $1=\td(\id_{\S^n}^{\complex^{n+1}})=\td(T\S^n)
    \cup\td(\id_{\S^n}^{\complex})= \td(T\S^n)$ and
    $\ch_\bullet(\S^n,\id_{\S^n}^\complex,\Id_{\S^n}^{~})=
\ch^\bullet(\id_{\S^n}^\complex)\cup \td(T\S^n) \cap [\S^n]=[\S^n]$.
Thus $\ch_\bullet$ maps the pertinent classes to distinct non-torsion
elements of $\H_\sharp(\S^n;\zed)$, and the conclusion follows by
Poincar\'{e} duality. (b)~follows from the fact that the generators of
$\K_\sharp^{\rm t}(\S^n)$ are in $\S_\sharp^{\rm t}(\S^n)$.
  \end{proof}
\begin{remark}
Proposition~\ref{KS}(a) is just a special case of Theorem~\ref{bigt}
and it allows us to conclude, without the assumption of
  Poincar\'{e} duality, that the torsion-free part of $\K_\sharp^{\rm
    t}(\S^n)$ is generated by the said classes. This is also true of
  the equality $\K_\sharp^{\rm t}(\textrm{pt})=\S_\sharp^{\rm
    t}(\textrm{pt})$ considered in
  Proposition~\ref{prop2}.~{$\lozenge$}
  \end{remark}
\begin{cor}\label{prop4}
  Let $n\geq 0$.
\begin{alphlist}
\item $\S_0^{\rm t}(\S^{n}) \cong  \mathbb{Z} \oplus \mathbb{Z}$ is
  generated by the classes $[\pt,\id_\pt^\complex ,\iota]$ and
  $[\S^n,\id_{\S^n}^\complex,\Id_{\S^n}^{~}]$, while $\S_1^{\rm t}(\S^n) \cong
0$.
\item The Hurewicz homomorphism in reduced K-homology $h_n:
  \pi_n(\S^n) \rightarrow\widetilde{\K}_\sharp^{\rm t}(\S^n)$ is a
  bijection.
\end{alphlist}
\end{cor}
\begin{proof}
(a) follows immediately along the lines of the proof of
Proposition~\ref{KS}, since $\ch_\bullet$ is an isomorphism in this
case. (b) follows from the fact that
$h_n=\ch_\bullet^{-1}\circ\rho_n$, where $\rho_n : \pi
  _n(\S^n) \stackrel{\approx}{\longrightarrow} \H_n(\S^n;\zed)$ is the
  Hurewicz isomorphism.
  \end{proof}

\begin{remark}
{}From the discussion of Section~\ref{Qual} we see that the classes
$[\S^{2k},\id_{\S^{2k}}^\complex,\iota^{~}_{\S^{2k}}]$, with $2k < n$ and
$\iota^{~}_{\S^{2k}}:\S^{2k}\hookrightarrow\S^n$ the inclusion, are
all identified in $\K_\sharp^{\rm t}(\S^n)$ through
vector bundle modification. Looking at the proof of
Proposition~\ref{KS}, one has
$[\S^n,\id_{\S^n}^{\complex^n},\Id_{\S^n}^{~}] =
[\S^n,\id_{\S^n}^\complex,\psi^{~}_{n}]$ where $\psi_{n}:\S^n\to\S^n$
is a map of winding number $n$. In particular, by
Corollary~\ref{prop4}(b) a K-cycle class
$[\S^n,\id_{\S^n}^\complex,\phi] \in\K_\sharp^{\rm t}(\S^n)$ depends
only on the degree of the map $\phi$. Finally, from
(\ref{IndexphiE}) the charge of the D-brane
$[\S^n,\id_{\S^n}^\complex,\Id^{~}_{\S^n}]$ is
$\ch^\bullet(\id_{\S^n}^\complex) \cup \td(T\S^n)[\S^n]=1$, and
similarly for the ``vacuum'' D-brane $[\textrm{pt},\id_\pt^\complex
,\iota]$. Thus the mathematical analysis above reproduces the
well-known physical property that D-branes in flat space carry no
lower-dimensional D-brane charges and thus have a simple additive
charge. \hfill{$\lozenge$}
\end{remark}

\subsection{T-Duality\label{suspension}}

Using the reduced version of the exterior product of
Section~\ref{ExtProd} and the consequent
K\"{u}nneth theorem, we can investigate the relationship between
the groups $\widetilde{\K}^{\rm t}_{0}(\Sigma^nX)$ and
$\widetilde{\K}^{\rm t}_{e(n)}(X)$, where $\Sigma^nX=\S^n \wedge X$ is
the $n$-th reduced suspension of the topological space $X$. By Bott
periodicity and induction one immediately concludes that
$\widetilde{\K}^{\rm t}_{0}(\Sigma^{2n}X) \cong \widetilde{\K}^{\rm t}_{0}(X)=
\widetilde{\K}^{\rm t}_{e(2n)}(X)$. On the other hand, a simple application of
the K\"{u}nneth theorem in its reduced version yields
$\widetilde{\K}^{\rm t}_{0}(\Sigma^{1}X)\cong \widetilde{\K}^{\rm t}_{1}(X)
\otimes \widetilde{\K}^{\rm t}_{1}(\S^1) \cong  \widetilde{\K}^{\rm
  t}_{1}(X)$, and by induction we conclude that $\widetilde{\K}^{\rm
  t}_{0}(\Sigma^{2n+1}X)\cong \widetilde{\K}^{\rm
  t}_{1}(X)=\widetilde{\K}^{\rm t}_{e(2n+1)}(X)$.

Let us now consider the group $\K^{\rm t}_0(X \times \S^1)$. The
K\"{u}nneth theorem gives $\K^{\rm t}_0(X \times \S^1) \cong \K^{\rm t}_0(X)
\oplus \K^{\rm t}_1(X)$, and therefore
\beq
\widetilde{\K}^{\rm t}_0(X \times \S^1) \cong
\widetilde{\K}^{\rm t}_0(X) \oplus \K^{\rm t}_1(X) \ .
\label{9braneKunneth}\eeq
The inclusion $\iota : X\cong X\times\pt\hookrightarrow X \times \S^1$
induces a homomorphism $\iota_* : \widetilde{\K}^{\rm t}_0(X) \rightarrow
\widetilde{\K}^{\rm t}_0(X \times \S^1)$.
{}From the decomposition (\ref{9braneKunneth}) it follows that
$\iota_*(\alpha)=\alpha\oplus0$ for all $\alpha\in\widetilde{\K}^{\rm t}_0(X)$,
and hence ${\rm im} ~ \iota_*
=\widetilde{\K}^{\rm t}_0(X)$ and ${\rm coker} ~\iota_* =
\widetilde{\K}^{\rm t}_0(X \times \S^1)/\widetilde{\K}^{\rm t}_0(X) \cong
\K^{\rm t}_1(X)$ where we have identified $\K^{\rm t}_1(X)$ (resp.
$\widetilde{\K}^{\rm t}_0(X)$) with the subgroup of
$\widetilde{\K}^{\rm t}_0(X \times \S^1)$
consisting of K-cycle classes $[M,E,\phi]$ such that up to homotopy
$\phi(M)\nsubseteq X \times \rm{pt}$ (resp. $\phi(M)\subseteq X \times
\rm{pt}$). This construction can be used to provide an alternative
``M-theory'' definition of the unstable 9-branes in Type~IIA
superstring theory introduced in Section~\ref{IIArem} which does not
require virtual K-theory elements. For $X$ a ten-dimensional compact
spin manifold without boundary, they are identified with the classes
$[X,E,\phi]$ on the 11-dimensional space $X\times\S^1$ for which
$E\neq\id_X^\complex$ and $\phi(X)\nsubseteq X\times\pt$. This is
consistent with the construction presented in Remark~\ref{IIAremimp}.

Another application of these simple observations is to the description
of T-duality in topological K-homology. Let $Q$ be a finite CW-complex
and let $\torus^n\cong(\S^1)^n$ be an $n$-dimensional torus. By the
K\"unneth theorem one has $\K^{\rm t}_{i}(\torus^n) \cong \K^{\rm
  t}_{i}(\S^1)^{\oplus2^{n-1}}\cong\zed^{\oplus2^{n-1}}$ for
$i=0,1$. Generalizing the computation of (\ref{9braneKunneth}) thus
gives the isomorphisms
\bea
\K^{\rm t}_0(Q \times \torus^n) &\cong& \left(
\widetilde{\K}^{\rm t}_0 (Q) \oplus
\K^{\rm t}_{1}(Q) \oplus  \zed\right)^{\oplus2^{n-1}} \ , \nonumber\\
\K^{\rm t}_1 (Q \times \torus^n) &\cong& \left(
\K^{\rm t}_{1}(Q) \oplus
\widetilde{\K}^{\rm t}_0 (Q) \oplus  \zed\right)^{\oplus2^{n-1}} \nonumber
\eea
and therefore
\beq
\K^{\rm t}_0 (Q \times \torus^n) \cong \K^{\rm t}_1 (Q \times
\torus^n) \ .
\label{Tdualiso}\eeq
This isomorphism describes a relationship between
Type~IIB and Type~IIA D-branes on the spacetime
$X=Q\times\torus^n$ called {\it T-duality}. From the identifications
above we see that the isomorphism exchanges wrapped D-branes
$[M,E,\phi]$ (having $\phi(M)\nsubseteq Q\times\pt$) with unwrapped
D-branes (having $\phi(M)\subseteq Q\times\pt$). The powers of
$2^{n-1}$ give the expected multiplicity of D$p$-brane charges arising
from wrapping all higher stable D-branes on various cycles of the
torus~$\torus^n$.

A more geometrical derivation of the T-duality isomorphism
(\ref{Tdualiso}) may be given as follows. Let $\Lambda\cong\zed^n$ be
a lattice of rank $n$ in a real vector space $V$ of dimension $n$, and
let $\Lambda^\vee\subset V^\vee$ be the dual lattice. Consider the
real torus $\torus^n=V/\Lambda$ and the corresponding dual torus
$\widetilde{\torus}{}^n=V^\vee/\Lambda^\vee$. The lattices $\Lambda$
and $\Lambda^\vee$ may then be identified with the first homology
lattices $\H_1(\torus^n;\zed)$ and
$\H_1(\,\widetilde{\torus}{}^n;\zed)$, while the first homology lattice
  of $\torus^n\times\widetilde{\torus}{}^n$ coincides with
  $\Lambda\otimes\Lambda^\vee$. There is a unique line
bundle $\mathcal{P}$ over the product space
$\torus^n\times\widetilde{\torus}{}^n$, called the {\it Poincar\'e
line bundle}, such that for any point $t\in\widetilde{\torus}{}^n$ the
restriction $\mathcal{P}_t=\mathcal{P}|_{\torus^n\times\{t\}}$ represents an
element of the Picard group of $\torus^n$ corresponding to $t$, and
such that the restriction
$\mathcal{P}|_{\{0\}\times\widetilde{\torus}{}^n}$ is the trivial
complex line bundle over $\widetilde{\torus}{}^n$. This bundle defines
a class in $\K_{\rm t}^0(\torus^n\times\widetilde{\torus}{}^n)$ which is a
  K-theory cup product of odd degree generators for the K-theory of
  the tori $\torus^n$ and $\widetilde{\torus}{}^n$.

Consider now the projections $$\xymatrix{&\ar[ld]_p Q\times\torus^n
\times\widetilde{\torus}{}^n\ar[rd]^{\widetilde{p}}\\ Q\times\torus^n&
&Q\times\widetilde{\torus}{}^n \ . }$$ The T-duality isomorphism in
topological K-theory~\cite{BEM1,Hori1} $$T_!\,:\,\K_{\rm
  t}^i\big(Q\times\torus^n\big)~
\stackrel{\approx}{\longrightarrow}~\K_{\rm t}^{e(i+n)}\big(Q\times
\widetilde{\torus}{}^n\big)$$ is given for $\xi\in\K_{\rm
  t}^i(Q\times\torus^n)$ and $i=0,1$ by $$T_!(\xi)=\widetilde{p}_!
(p^*\xi\otimes\mathcal{P}) \ , $$ where $\widetilde{p}_!:\K_{\rm t}^i(Q
\times\torus^n\times\widetilde{\torus}{}^n)\to\K_{\rm t}^{e(i+n)}(Q
\times\widetilde{\torus}{}^n)$ is the push-forward map in K-theory
which is given by the topological index. Since we assume that
the spacetimes $X=Q\times\torus^n$ and
$\widetilde{X}=Q\times\widetilde{\torus}{}^n$ are spin (equivalently
$Q$ is spin), they are
$\K_{\rm t}^\sharp$-oriented and thus obey Poincar\'e duality. The
K-homology of $Q\times\torus^n$ thereby has a set of generators given by
$[Q\times\torus^n,\xi,\Id^{~}_{Q\times\torus^n}]$ where $\xi\in\K_{\rm
  t}^\sharp(Q\times\torus^n)$ is a generator, and similarly for
$Q\times\widetilde{\torus}{}^n$. It follows that the map $$T^!\,:\,
\K_i^{\rm t}\big(Q\times\torus^n\big)~\longrightarrow~\K_{e(i+n)}^{\rm
  t}\big(Q\times\widetilde{\torus}{}^n\big)$$ given by $$T^!
\bigl[Q\times\torus^n,\xi,\Id^{~}_{Q\times\torus^n}\big]=\big
[Q\times\widetilde{\torus}{}^n,\widetilde{p}_!(p^*\xi\otimes\mathcal{P}),
\Id^{~}_{Q\times\widetilde{\torus}{}^n}\big]$$ is a well-defined group
homomorphism. Since $T^!=\Phi^{\rm t}_{Q\times\widetilde{\torus}{}^n}\circ
T_!\circ(\Phi^{\rm t}_{Q\times\torus^n})^{-1}$, it is an isomorphism. This
isomorphism is the T-duality isomorphism in topological
K-homology. While this map is defined in terms of virtual K-theory
elements, one can straightforwardly obtain a picture with only stable
isomorphism classes appearing by applying vector bundle modification
along the lines explained in Section~\ref{IIArem}. If $n$ is even, the
T-duality isomorphism maps a spacetime-filling brane-antibrane pair on
$X$ to a spacetime-filling brane-antibrane pair on the 11-dimensional
``M-theory'' extension $\widetilde{X}\times\S^1$, as spelled out by
the construction of Remark~\ref{IIAremimp}. In particular, this
description can be used to provide a more general construction of
T-duality in the case of a spin$^c$ torus bundle over
$Q$~\cite{BEM1}. The construction also thereby provdes a topological
K-homology realization of brane descent relations among
D-branes~\cite{OS1}.

\subsection{Projective D-Branes\label{CPn}}

The simplest example of flat D-branes in a curved background is provided by
the complex projective spaces $\complex\P^n$ of real dimension
$2n$. Being complex manifolds they are automatically spin$^c$. The
cellular structure in this case may be described by the
stratification of $\complex\P^n$ into linearly embedded subspaces as
\beq
\complex\P^0~\subset~\complex\P^1~\subset~\cdots~
\subset~\complex\P^{n-1}~\subset~\complex\P^n
\label{CPnfilt}\eeq
where $\complex\P^0=\pt$. Since $\mathbb{C}\P^n$ satisfies the hypotheses of
Theorem~\ref{bigt}, a set of generators for its reduced topological K-homology
group $\widetilde{\K}_\sharp^{\rm t}(\mathbb{C}\P^n)$ is given by
\beq
\big[\mathbb{C}\P^k,\id_{\complex\P^k}^\complex,\iota^{~}_{k}\big] \quad
, \quad 1 \leq k \leq n
\label{CPnfiltbasis}\eeq
where $\iota_{k}:\mathbb{C}\P^k\hookrightarrow\mathbb{C}\P^n$ is the canonical
inclusion. On the other hand, let $\mathcal{L}_n$ denote the canonical
line bundle over $\complex\P^n$ and $\mathcal{L}_n^\vee$ its dual line
bundle. The reduced K-theory of $\complex\P^n$ is then given by
$\widetilde{\K}^{\sharp}_{\rm
  t}(\mathbb{C}\P^n)=\widetilde{\K}_{\rm
t}^{0}(\mathbb{C}\P^n)=\bigoplus_{i=1}^{n}
([\id_{\complex\P^n}^\complex]-[\mathcal{L}_{n}^\vee])^i~\mathbb{Z}$.
{}From Proposition~\ref{cpn} it follows that the K-cycle classes
$([\id_{\complex\P^n}^\complex]-[\mathcal{L}_{n}^\vee])^{i} \cap
[\mathbb{C}\P^n,\id_{\complex\P^n}^\complex ,\Id^{~}_{\complex\P^n}]$
describe spacetime-filling D-branes on complex projective space and we
arrive at the following result.

\begin{proposition}
For $n \geq 1$, $\widetilde{\K}^{\rm t}_\sharp(\mathbb{C}\P^n)=
\widetilde{\K}^{\rm t}_{0}(\mathbb{C}\P^n)\cong\zed^{\oplus n}$ is the
free abelian group with generators
$$
   \sum_{\stackrel{\scriptstyle k=0}{\scriptstyle
       k~{\rm even}}}^{i}
\Big[\mathbb{C}\P^n\,,\,\mbox{$\bigoplus\limits_{l=0}^{\binom{i}{k}}$}\, (
\mathcal{L}_{n}^\vee)^{\otimes(i-k)}\,,\,\Id^{~}_{\complex\P^n}\Bigr]-
\sum_{\stackrel{\scriptstyle k=0}{\scriptstyle k~{\rm
odd}}}^{i}\Big[\mathbb{C}\P^n\,,\,
\mbox{$\bigoplus\limits_{l=0}^{\binom{i}{k}}$}\,
(\mathcal{L}_{n}^\vee)^{\otimes(i-k)}
\,,\,\Id^{~}_{\complex\P^n}\Big] \quad , \quad 1 \leq i \leq n \ . $$
\label{CPncanbasis}\end{proposition}

\begin{remark}
The decay of the brane-antibrane system provided by
Proposition~\ref{CPncanbasis} into the stable D-branes
described by the K-cycle classes (\ref{CPnfiltbasis}) is rather
intricate to describe. Since $T\mathbb{C}\P^n \oplus
\id_{\complex\P^n}^\complex \cong
(\mathcal{L}_{n}^\vee)^{\oplus(n+1)}$, one has
$\td(T\mathbb{C}\P^n)=f(-c_1(\mathcal{L}_{n}))^{n+1}$ where
$c_1(\mathcal{L}_n)$ is the first Chern class of the canonical line
bundle $\mathcal{L}_n\to\complex\P^n$ and
$$f(x) = 1 + \frac{x}{2} + \sum _{k \geq 1}\,(-1)^{k-1}~
\frac{B_k}{2k!}~x^{2k}$$
with $B_k\in\rat$ the $k$-th Bernoulli number. This fact may be used
to attempt to find the change of basis map (\ref{IIBchangebasis})
between these two sets of generators of
$\widetilde{\K}^{\rm t}_\sharp(\mathbb{C}\P^n)$ via the Chern character and
the homology of $\mathbb{C}\P^n$. However, both the Chern character and
the Todd class lead directly to a strictly rational-valued change of
basis matrix. The obstruction to this explicit procedure is encoded in
whether or not the Chern character admits an integral lift, i.e. an
extension of the usual map into rational homology to a map into
integer homology. With the exception of the projective line
$\complex\P^1\cong\S^2$, one easily checks that such a lift is not
possible on complex projective spaces. The problem of integral lifts
is discussed more thoroughly in the next Section. \hfill{$\lozenge$}
\end{remark}

\begin{remark}\label{otherprojs}
These results generalize straightforwardly to all
three spin$^c$ projective spaces $\dalg\P^n$, where $\dalg$ is one of
the three division algebras generated by the complex numbers
$\complex$, the quaternions $\quat$ or the Cayley numbers $\oct$. Let
$r:=\dim_\real(\dalg)$, so that $\dalg\P^1\cong\S^r$. Then the only
non-trivial reduced homology groups of $\dalg\P^n$ are
$\widetilde{\H}_{r\,k}(\dalg\P^n;\zed)\cong\zed$, $1\leq k\leq n$. The
cellular structure is determined by a stratification of $\dalg\P^n$ into
$(r\,k)$-cells for $k=0,1,\dots,n$ described by linearly embedded
subspaces, analogously to (\ref{CPnfilt}). The K-theory ring is
generated by $\frac
r2\,[\id_{\dalg\P^n}^\complex]-[\mathcal{L}_{\dalg\P^n}^\vee]$, with
$\mathcal{L}_{\dalg\P^n}\to\dalg\P^n$ the canonical complex line
bundle. In all three instances one arrives at the isomorphism
$\widetilde{\K}_\sharp^{\rm
  t}(\dalg\P^n)\cong\widetilde{\H}^{~}_\sharp(\dalg\P^n;\zed)$, with generators
constructed as above. The equivalence between K-homology and singular
homology in these cases can be understood through the appropriate
spectral sequence and the sparseness of the cellular structure of
$\dalg\P^n$. Spectral sequences will play an important role in the
investigations of subsequent sections. The real projective spaces
$\real\P^n$, based on the algebraically open field $\dalg=\real$, have a
more intricate K-homology and will be treated in the next
section. \hfill{$\lozenge$}
\end{remark}

\newsection{Torsion D-Branes}

Let us now turn to the somewhat more interesting situation in which
D-branes are described by K-cycles which generally produce torsion elements in
K-homology. In these cases one can encounter K-homology groups which
do not coincide with the corresponding integral homology groups, and
here the K-homology classification makes some genuine predictions that
cannot be detected by ordinary homological methods. We will first
consider the general problem of finding explicit homology cycles in the
spacetime which are wrapped by D-branes. This analysis extends that of
Section~\ref{Stability} to examine general circumstances under which a
spin$^c$ homology cycle has a non-trivial lift to K-homology and
hence is wrapped by a stable D-brane. Then we turn to a number of explicit
examples illustrating how the K-cycle representatives of torsion
charges are constructed in practice.

\subsection{Stability\label{Lifts}}

Since $\K^{\rm t}_\sharp$ is a homology theory defined
by means of a ring spectrum, it satisfies the wedge
axiom~\cite{8}. One consequence of this fact is that we can
immediately obtain the groups $\widetilde{\K}^{\rm
  t}_\sharp(\bigvee_{\alpha}\S^n_{\alpha})$ using the K-homology of the
spheres, since then
$\widetilde{\K}^{\rm t}_\sharp(\bigvee_{\alpha}\S^n_{\alpha})= \bigoplus
_{\alpha}\widetilde{\K}^{\rm t}_\sharp(\S_\alpha^n)$. Another consequence
is that we can use the Atiyah-Hirzebruch-Whitehead (AHW) spectral
sequence~\cite{8} to compute the $\K^{\rm t}_\sharp$-groups of
CW-complexes.

Let $X$ be a connected finite CW-complex, and let $X^{[n]}$ denote
its $n$-skeleton with $X^{[0]}=\pt$. By the Whitehead cellular
approximation theorem, the inclusion $\iota_n:X^{[n]}\hookrightarrow
X$ induces an isomorphism in integral homology up to degree $n-1$.
Consider the AHW spectral sequence
$\{\E_{p,q}^r,\dd^r\}_{r\in\nat;p,q\in\zed}$ for reduced K-homology
satisfying \beq \E^2_{p,q} = \widetilde{\H}_p\big(X\,;\,\K^{\rm
t}_{e(q)}(\pt)\big) ~ \Longrightarrow~  \widetilde{\K}^{\rm
t}_{e(p+q)}\big(X\big) \label{E2pqgen}\eeq with
$\widetilde{\H}_p(X;\zed):=\H_p(X,\pt;\zed)$. Convergence of the
spectral sequence means that there is a filtration
$\{\F_{p,n-p}\}_{p\in\nat_0}$ of $\widetilde{\K}^{\rm t}_{e(n)}(X)$
given by
$$0~=~\F_{0,n}~\subseteq~\F_{1,n-1}~\subseteq~\cdots~\subseteq~
\F_{p,n-p}~\subseteq~\cdots~\subseteq~\F_{n,0}~\subseteq~\cdots~\subseteq~
\bigcup_{p+q=n}\F_{p,q}~=~\widetilde{\K}^{\rm t}_{e(n)}(X) \ , $$
where $\F_{p,q} = {\rm im}~(\iota_p)_*:\widetilde{\K}^{\rm t}_{e(p+q)}(X^{[p]})
\to\widetilde{\K}^{\rm t}_{e(p+q)}(X)$ and $\F_{p+1,n-p-1}/\F_{p,n-p} \cong
\E^{\infty}_{p+1,n-p-1}$.

For each $j=1,2,3$ there is a natural epimorphism
$$\beta_j^X \,:\, \H_j(X;\zed)=\E^2_{j,0} ~\longrightarrow~
\E^{\infty}_{j,0}= \F_{j,0}/\F_{j-1,1} \ . $$
In particular, since $\F_{0,1}=0=\F_{1,1}$ the cases $j=1,2$ yield an
epimorphism
$$\beta_j^X \,:\, \H_j(X;\zed)=\E^2_{j,0} ~\longrightarrow~
\E^{\infty}_{j,0}= \F_{j,0} ~\hookrightarrow~\widetilde{\K}^{\rm t}_{e(j)}(X) \
. $$ By analysing the spectral sequence, one concludes that
$\beta_j^X$ is injective if and only if no
non-zero differential $\dd^r:\E^r_{p,q}\to\E^r_{p-r,q+r-1}$ reaches
$\E^{k}_{j,0}$ for $k\geq2$. Thus if the reduced singular homology
$\widetilde{\H}_\sharp(X;\zed)$ is concentrated in odd (resp. even)
degree, except for possibly $\H_2(X;\zed)$ (resp. $\H_1(X;\zed)$ and
$\H_3(X;\zed)$), then $\beta_1^X$ (resp. $\beta_2^X$) is injective.

{}From these considerations one concludes that if $X$ is a connected CW-complex
of dimension $\leq4$, then there are natural short exact sequences
$$0 ~\longrightarrow~ \H_1(X;\zed) ~\longrightarrow~ \K_1^{\rm t}(X)
{}~\longrightarrow~ \H_3(X;\zed)~ \longrightarrow~ 0 \ , $$
$$0 ~\longrightarrow~ \H_2(X;\zed) ~\longrightarrow~
\widetilde{\K}_0^{\rm t}(X) ~\longrightarrow~ \H_4(X;\zed)
{}~\longrightarrow~ 0 \ . $$
The latter sequence splits, yielding an isomorphism
$$\textrm{ch}_{\rm even}^{\zed} \,:\, \widetilde{\K}_0^{\rm t}(X)
{}~\stackrel{\approx}{\longrightarrow}
{}~ \widetilde{\H}_{\textrm{even}}(X;\zed) \ . $$
If $X$ is simply connected, then the map $\textrm{ch}_3^{\zed} :
\K_1^{\rm t}(X) \rightarrow \H_3(X;\zed)$ is a bijection and we
thereby obtain an isomorphism
\beq
\textrm{ch}_{\bullet}^{\zed} \,:\, \widetilde{\K}_\sharp^{\rm t}(X)
{}~\stackrel{\approx}{\longrightarrow}~ \widetilde{\H}_\sharp(X;\zed)
\label{chlift}\eeq
of $\zed_2$-graded abelian groups such that the diagram
\beq
\xymatrix{ \widetilde{\K}^{\rm t}_\sharp(X)
\ar[r]^{\rm{ch}_{\bullet}^ \mathbb{Z}}  \ar[rd]_{\rm{ch}_{\bullet}} &
\widetilde{\H}_\sharp(X;\mathbb{Z}) \ar[d]^{\Lambda}\\
      & \widetilde{\H}_\sharp(X;\mathbb{Q})}
\label{chliftdiag}\eeq commutes, where $\Lambda :
\widetilde{\H}_\sharp(X;\mathbb{Z}) \rightarrow
\widetilde{\H}_\sharp(X;\mathbb{Q})$ is the homomorphism induced by
the inclusion of abelian groups $\mathbb{Z}\hookrightarrow
\mathbb{Q}$. (These calculations were first carried out in
\cite{KamSch}.)

The isomorphism (\ref{chlift}) is called an {\it integral lift} of the
Chern character in K-homology. It extends the usual Chern character
map between stable D-branes and non-trivial homology cycles in $X$ to
include torsion classes. For
$X$ of dimension $\leq 3$, this isomorphism even exists without the
assumption of simple connectivity~\cite{17}. For CW-complexes $X$ of higher
dimension, the problem of determining which homology cycles lift to
stable D-branes is much more difficult, because then the analysis of
the spectral sequence is not so clear cut. For instance, in general
$\F_{2,1}\cong\H_1(X;\zed)$ is non-trivial, thereby making this kind
of analysis generically impossible.

\begin{remark}
The filtration groups $\F_{p,q}$ approximating the full K-homology
group consist of D-branes $[M,E,\phi]$ in $X$ whose worldvolumes are
supported in the $p$-skeleton, i.e. $\phi(M)\subseteq X^{[p]}$. The
extension groups $\E^\infty_{p,q}$ between successive approximants
consist of those D-branes in the $p$-skeleton which are not supported
on the $(p-1)$-skeleton, i.e. $\phi(M)\nsubseteq X^{[p-1]}$, or in other words
$\E_{p,q}^\infty$ consists of D$(p-1)$-branes which carry no
lower-dimensional brane charges. By definition, the approximations
$\E_{p,q}^r$ for $r\geq2$ compute the homology of the differential
$\dd^{r-1}$ ($\E_{p,q}^1$ is the group of singular $p$-chains on $X$
with values in $\K_{e(q)}^{\rm t}(\pt)$ with $\dd^1$ the usual
simplicial boundary homomorphism).

Let $M\subset X$ be a $p$-dimensional compact spin$^c$ manifold
without boundary which defines a non-trivial homology class
$[M]\in\E_{p,q}^2$ in (\ref{E2pqgen}). If $[M]$ extends through the
spectral sequence as a non-trivial element of all homology groups
$\E_{p,q}^r$, then it can represent a non-trivial element of
$\E_{p,q}^\infty$ and hence have a non-trivial lift to K-homology. In
this case there exists a D-brane $[M,E,\phi]$ wrapping $M$ on the
$p$-skeleton of $X$ which is stable and carries no lower brane
charges. Conversely, suppose that $$[M]=\dd^r\omega$$ for some
$r\in\nat$ and $\omega\in\widetilde{\H}_\sharp(X;\zed)$. Then the homology
class $[M]$ can be lifted to K-homology, but the lift is trivial as it
vanishes in $\E_{p,q}^\infty$. This means that there exists a D-brane
wrapping $M$ in $X^{[p]}$ with no lower brane charges, but this D-brane
is unstable. Thus the AHW spectral sequence in this context
keeps track of the possible obstructions for a homology cycle
of $\widetilde{\H}_p(X;\zed)$, starting from (\ref{E2pqgen}), to
survive to~$\E^\infty_{p+1,n-p-1}$. Then, the solution of the
K-homology extension problem required to get the filtration groups
$\F_{p,q}$ from $\E_{p,q}^\infty$ identifies the lower brane charges
carried by D-branes and changes the additive structure in K-homology
from that of the singular homology classes. The spectral sequence in
this regard measures the possible obstructions to extending $[M]$
non-trivially over higher-dimensional simplices of
$X$. \hfill{$\lozenge$}
\end{remark}

\subsection{A-Branes\label{LensSpaces}}

Let $p,q_1, \ldots,q_n$ be integers with $p \geq 1$ and gcd$(p,q_i)=1$
for all $i=1, \ldots,n$. There is a free ${\rm C}^\infty$ action
$$ G \,:\, \mathbb{Z}_p \times \S^{2n+1} ~\longrightarrow~ \S^{2n+1}$$
given by
$$G\big(\e^{2\pi\ii k/p},(z_0,z_1,\ldots,z_n)\big)=
\big(\e^{2\pi\ii k/p}\,z_0,\e^{2\pi\ii q_1\,k/p}\,z_1,\ldots,\e^{2\pi\ii
  q_n\,k/p}\,z_n \big) $$ where we regard $\zed_p\subset\S^1$ and
$\S^{2n+1}\subset\complex^{n+1}$.
The corresponding quotient space ${\mathbb{L}}(p;q_1,\ldots,q_n)$ is a compact
connected ${\rm C}^\infty$ manifold of dimension $2n+1$ called a {\it
  Lens space}. For definiteness we will consider only the case
$n=1$. The corresponding D-branes are then a particular instance of
topological A-model D-branes (or {\it A-branes} for
short)~\cite{Douglas1,MMS2,Witten3} which are
mirror duals to the B-branes described in Example~\ref{BBranes} and
belong to the derived Fukaya category of the
spacetime~\cite{Aspinwall1}. The mirror manifold to the algebraic
variety $X$ is taken to be the non-compact Calabi-Yau threefold which
is the total space of the rank~$2$ complex vector bundle
$(\mathcal{L}_1^\vee)^{\otimes
  p}\oplus(\mathcal{L}_1^{~})^{\otimes(p-2)}\to\complex\P^1$. For $q=1$ the
Lens
space ${\mathbb{L}}(p;1)$ may be identified with the boundary
of~$(\mathcal{L}_1^\vee)^{\otimes p}$. Higher-dimensional Lens spaces
are similarly identified with the boundaries of the total spaces of
the line bundles $(\mathcal{L}_n^\vee)^{\otimes p}\to\complex\P^n$.

The Lens space ${\mathbb{L}}(p;q)$ is a compact connected spin three-manifold which
admits a CW-complex structure with one $n$-cell for each dimension
$n=0,1,2,3$~\cite{6}. Its singular homology is given by
\bea
\H_0\big({\mathbb{L}}(p;q)\,;\,\zed\big)&\cong&\zed~\cong~
\H_3\big( {\mathbb{L}}(p;q)\,;\,\zed\big) \ , \nonumber\\
\H_1\big({\mathbb{L}}(p;q)\,;\,\zed\big)&\cong&\zed_p \ , \nonumber\\
\H_2\big({\mathbb{L}}(p;q)\,;\,\zed\big)&\cong&0 \ .
\label{lenshom}\eea
Since we know the singular homology of ${\mathbb{L}}(p;q)$, we can work out the
spectral sequence in this case and thus calculate the topological
K-homology $\K^{\rm t}_\sharp({\mathbb{L}}(p;q))$.

\begin{proposition}
$\K^{\rm t}_0\big({\mathbb{L}}(p;q)\big)~\cong~\zed \ , \quad
\K^{\rm t}_1\big({\mathbb{L}}(p;q)\big)~\cong~\zed\oplus\zed_p \ . $
\label{lensKhom}\end{proposition}
\begin{proof}
There exists an AHW spectral sequence
$\{\E^r_{n,m},\dd^r\}_{r\in\nat;n,m\in\zed}$ converging to $\K^{\rm
  t}_\sharp({\mathbb{L}}(p;q))$ with
\begin{displaymath}
\E^2_{n,m} ~=~ \H_n\big({\mathbb{L}}(p;q)\,;\,\K^{\rm t}_{e(m)}({\rm pt})\big)~
\cong~ \left \{ \begin{array}{ll}
           \H_n\big({\mathbb{L}}(p;q)\,;\,\mathbb{Z}\big)  & ,\quad m \textrm{ even}\\
           \qquad 0 & ,\quad m \textrm{ odd} \end{array} \right. \ .
\end{displaymath}
{}From (\ref{lenshom}) it follows that the only non-zero groups are
$\E^2_{0,2k} \cong \mathbb{Z} \cong \E^2_{3,2k}$ and
$\E^2_{1,2k} \cong \mathbb{Z}_p$ with $k \in \mathbb{Z}$. The next
sequence of homology groups of the differential module is defined by
$$\E^3_{n,m}:= \frac{\ker\,\dd^2\,:\,\E^2_{n,m} ~\longrightarrow~
\E^2_{n-2,m+1}}{\textrm{im}~\dd^2\,:\,
\E^2_{n+2,m-1} ~\longrightarrow~ \E^2_{n,m}} \ . $$
If $m$ is odd, $n \geq 4$ or $n<0$, then $\E^2_{n,m}=0$ so that
$\ker\,\dd^2=0={\rm im}~\dd^2$ and hence $\E^3_{n,m}=0$. For the
remaining cases with $m=2k$ and $n=0,1,2,3$, the pertinent part of the
differential bicomplex is of the form $$
\begin{matrix}\cdots&\stackrel{\dd^2}{\longrightarrow} & \E^2_{n+2,2k-1}
& \stackrel{\dd^2}{\longrightarrow}& \E^2_{n,2k} & \stackrel{\dd^2}
{\longrightarrow}& \E^2_{n-2,2k+1}
&\stackrel{\dd^2}{\longrightarrow}&\cdots \ , \\[2pt] & &\parallel&
&\parallel& &\parallel& & \\[2pt] & &0& &\H_n\big({\mathbb{L}}(p;q)\,;\,
\zed\big)& &0& & \end{matrix} $$ implying that ${\rm im}~\dd^2=0$ and
hence $\E^3_{n,2k}=\ker\,\dd^2\cong\H_n({\mathbb{L}}(p;q);\zed)$.

By induction we conclude from this data that
$\E^r_{0,2k}\cong\mathbb{Z}\cong\E^r_{3,2k}$ and
$\E^r_{1,2k}\cong\mathbb{Z}_p$ for every $r\geq2$, with all other
homology groups vanishing. We therefore have
  \begin{displaymath}
\E^{\infty}_{n,m} ~=~\lim_{\stackrel{\scriptstyle\longrightarrow}
{\scriptstyle r}}\,\E^r_{n,m} ~\cong~ \left \{ \begin{array}{ll}
           \H_n\big({\mathbb{L}}(p;q)\,;\,\mathbb{Z}\big)=\mathbb{Z}  & ,\quad m \textrm{
             even} \ , ~~ n=0,3\\
           \H_1\big({\mathbb{L}}(p;q)\,;\,\mathbb{Z}\big)=\mathbb{Z}_p & ,\quad m
           \textrm{ even} \ , ~~ n=1\\
            \H_2\big({\mathbb{L}}(p;q)\,;\,\mathbb{Z}\big)=0 & ,
\quad \textrm{otherwise}\end{array}\right. \ .
     \end{displaymath}
For each $l\in\zed$ let $\F_{0,l}=\E^{\infty}_{0,l}$. Then solving the
extension problems
\bea
0 &\longrightarrow& \F_{n-1,1-n} ~\longrightarrow~ \F_{n,-n}
{}~\longrightarrow~\E^{\infty}_{n,-n}  ~\longrightarrow~ 0 \ , \nonumber\\
0 &\longrightarrow& \F_{n-1,2-n} ~\longrightarrow~ \F_{n,1-n}
{}~\longrightarrow~\E^{\infty}_{n,1-n}  ~\longrightarrow~ 0 \nonumber
\eea
for every $n\in\nat$ will produce groups $\F_{n,-n}$ and
$\F_{n,1-n}$ such that $\{\F_{n,-n} \}_{n\in\nat_0}$
(resp. $\{\F_{n,1-n} \}_{n\in\nat_0}$) is a filtration of $\K^{\rm
  t}_0({\mathbb{L}}(p;q))$ (resp. $\K^{\rm t}_1({\mathbb{L}}(p;q))$). Starting from the data
above, it is straightforward to compute $
     \F_{n,-n}=\mathbb{Z}$ for all $n\in\nat_0$, and hence
     $\K^{\rm t}_0({\mathbb{L}}(p;q))=\mathbb{Z}$. Furthermore, one finds
     $\F_{0,1}\cong0$, $\F_{1,0}\cong\zed_p\cong\F_{2,-1}$ and
       $\F_{n,1-n}\cong\zed\oplus\zed_p$ for all $n\geq3$, so that
       $\K^{\rm t}_1({\mathbb{L}}(p;q))=\mathbb{Z} \oplus \mathbb{Z}_p$.
     \end{proof}

Let us now work out D-brane representatives for the K-homology groups
of Proposition~\ref{lensKhom}. Since $\widetilde{\K}^{\rm
  t}_0({\mathbb{L}}(p;q))=0$, it is immediate that
$[\textrm{pt},\id_{\pt}^\complex, \iota]$ is the generator of $\K^{\rm
  t}_0({\mathbb{L}}(p;q))= \mathbb{Z}$. Furthermore, since ${\mathbb{L}}(p;q)$ is an
odd-dimensional spin manifold, it is $\K_{\rm t}^\sharp$-orientable
and so it has a fundamental class
$[{\mathbb{L}}(p;q),\id_{{\mathbb{L}}(p;q)}^\complex,\Id^{~}_{{\mathbb{L}}(p;q)}]$ which is the free
generator of $\K^{\rm t}_1({\mathbb{L}}(p;q))=\mathbb{Z} \oplus \mathbb{Z}_p$.
If we take $q=1$ and let $\mathcal{L}_1$ denote as before the
canonical line bundle over $\mathbb{C}\P^1$, then we can identify the
sphere bundle of $\mathcal{L}_1^{\otimes p}$ with the Lens space
${\mathbb{L}}(p;1)$. In this case, from the K-theory of ${\mathbb{L}}(p;1)$~\cite{7} we can
identify the torsion generator of $\K^{\rm t}_1({\mathbb{L}}(p;1))$ with the
K-cycle class
\beq
\big[{\mathbb{L}}(p;1),\id_{{\mathbb{L}}(p;1)}^\complex,\Id^{~}_{{\mathbb{L}}(p;1)}\big]-
\big[{\mathbb{L}}(p;1),\pi^{*}(\mathcal{L}_1^\vee),\Id^{~}_{{\mathbb{L}}(p;1)}\big]
\label{lensanti}\eeq
where $\pi:\S(\mathcal{L}_1^{\otimes p})\to\complex\P^1$ is the bundle
projection.

To describe the decay of the spacetime-filling brane-antibrane pair
(\ref{lensanti}) into stable D-branes, we note that
$\H_{1}({\mathbb{L}}(p;q);\mathbb{Z}) \cong \mathbb{Z}_p \cong \pi_1({\mathbb{L}}(p;q))$ and
that the Hurewicz homomorphism $\rh_1 :
\pi_1({\mathbb{L}}(p;q)) \rightarrow \H_{1}({\mathbb{L}}(p;q);\mathbb{Z})$ given by
$\rho_1[f]=f_*[\S^1]$ is a bijection. In addition, the Hurewicz
homomorphism in K-homology $h_1 : \pi_1({\mathbb{L}}(p;q)) \rightarrow
\K^{\rm t}_1({\mathbb{L}}(p;q))$ is given by
$h_1[f]=f_*[\S^1,\id_{\S^1}^\complex,\Id^{~}_{\S^1}]=
[\S^1,\id_{\S^1}^\complex,f]$. Since ${\mathbb{L}}(p;q)$ is a compact
three-dimensional manifold, the homological Chern
character admits an integral lift (\ref{chlift}) fitting into the
commutative diagram (\ref{chliftdiag}) for $X={\mathbb{L}}(p;q)$. Furthermore,
$\ch_{\rm odd}^ {\mathbb{Z}}:=\ch_{1}^
{\mathbb{Z}} \oplus \ch_{3}^ {\mathbb{Z}} : \K^{\rm t}_1({\mathbb{L}}(p;q))
\rightarrow \H_{1}({\mathbb{L}}(p;q);\mathbb{Z}) \oplus \H_{3}({\mathbb{L}}(p;q);\mathbb{Z})$ is an
isomorphism. In particular, $\ch_{1}^ {\mathbb{Z}} :
\Tor_{\K^{\rm t}_1({\mathbb{L}}(p;q))} \rightarrow \H_{1}({\mathbb{L}}(p;q);\mathbb{Z}) $ is
an isomorphism. Its inverse is given by the isomorphism $\beta_1 :
\H_{1}({\mathbb{L}}(p;q);\mathbb{Z}) \rightarrow \Tor_{\K^{\rm t}_1({\mathbb{L}}(p;q))}$
which fits into the commutative diagram~\cite{17}
$$\xymatrix{ \pi_1\big({\mathbb{L}}(p;q)\big) \ar[r]^{h_1} \ar[d]_{\rh_1} & \Tor_{\K^{\rm
      t}_1({\mathbb{L}}(p;q))} \ . \\
       \H_{1}\big({\mathbb{L}}(p;q)\,;\,\mathbb{Z}\big) \ar[ur]_{\beta_1}  } $$
It follows that $h_1 : \pi_1({\mathbb{L}}(p;q)) \rightarrow
\Tor_{\K^{\rm t}_1({\mathbb{L}}(p;q))}$ is an isomorphism. The generator of the
fundamental group $[f] \in \pi_1({\mathbb{L}}(p;q))=\zed_p$ may be taken to be
any loop obtained by projecting a path on the universal cover $\S^3\to
{\mathbb{L}}(p;q)$ connecting two points on $\S^3$ that are related by the
$\zed_p$-action defining the Lens space. Then
$[\S^1,\id_{\S^1}^\complex,f]$ is the torsion generator of $\K^{\rm
  t}_1({\mathbb{L}}(p;q))$. For $q=1$ it coincides with the generator
(\ref{lensanti}).

\begin{remark}
Examining the proof of Proposition~\ref{lensKhom},
we see that this construction of the stable D-brane states in ${\mathbb{L}}(p;q)$
follows from the form of the homology groups $\E^\infty_{n,m}$ of the
differential module in the AHW spectral sequence. In the present
example, all homology cycles have non-trivial lifts to K-homology and
are thus wrapped by stable states of D-branes. \hfill{$\lozenge$}
\end{remark}

\subsection{Projective D-Branes}

We will now complete the calculation initiated in Section~\ref{CPn} by
exhibiting the D-branes in the fourth and final {\it real} projective
spaces $\real\P^m$, which arise in certain orbifold spacetimes of
string theory~\cite{Gukov1}. They can be realized as the quotient of the
$m$-sphere $\S^m$ by the antipodal map. Let $q_m:\S^m\to\real\P^m$ be
the quotient map. With the exception of the
projective line $\real\P^1\cong\S^1$, the corresponding K-homology
groups contain torsion subgroups. Analogously to (\ref{CPnfilt}), the
CW-complex structure of $\mathbb{R}\P^m$ may be given by the
stratification provided by linearly embedded
subspaces, so that its set of $k$-cells consists of the single element
$\mathbb{R}\P^{k}$ for $k=0,1,\dots,m$ with
$\mathbb{R}\P^{k}/\mathbb{R}\P^{k-1}\cong\S^{k}$. The singular
homology of $\real\P^m$ is given by
\bea
\H_0(\real\P^{2n+1};\zed)&\cong&\zed~\cong~\H_m(\real\P^m;\zed)
\ , \nonumber\\ \H_{m-2i}(\real\P^m;\zed)&\cong&\zed_2 \ , \quad
i=1,\dots,\left\lfloor
\mbox{$\frac m2$}\right\rfloor \ . \nonumber
\eea
Let $\mathcal{L}_{\real\P^m}=\S^m\times\real/\zed_2$ be the canonical
flat line bundle over $\real\P^m$, and
$\iota_k:\real\P^k\hookrightarrow\real\P^m$ the inclusion of the
$k$-cell.

\subsubsection{$\real\P^{2n+1}$}

We begin with the odd-dimensional cases
$m=2n+1$. In this instance $\mathbb{R}\P^{2n+1}$ is a spin$^c$
manifold. Thus it is $\K_{\rm t}^\sharp$-oriented and we can apply
Poincar\'e duality to compute its topological K-homology
$\K_\sharp^{\rm t}(\real\P^{2n+1})$ from its known K-theory $\K_{\rm
  t}^\sharp(\real\P^{2n+1})$~\cite{7}.

\begin{proposition}
$\K_0^{\rm t}(\real\P^{2n+1})~\cong~\zed \ , \quad \K_1^{\rm
t}(\real\P^{2n+1})~\cong~\zed \oplus\zed_{2^n} \ . $
\label{KhomRP2n1}\end{proposition} \noindent Applying
Proposition~\ref{cpn} to the example at hand, one finds that the
generating D-brane of $\K^{\rm t}_0(\real\P^{2n+1})$ is
$[\pt,\id_\pt^\complex,\iota]$, while the spacetime-filling D-brane
$[\real\P^{2n+1},\id_{\real\P^{2n+1}}^\complex,\Id^{~}_{\real\P^{2n+1}}]$
is the free generator of $\K_1^{\rm t}(\real\P^{2n+1})$. The torsion
generator of $\K_1^{\rm t}(\real\P^{2n+1})$ is the spacetime-filling
brane-antibrane pair \beq
\big[\real\P^{2n+1},\id_{\real\P^{2n+1}}^\complex,
\Id^{~}_{\real\P^{2n+1}}\big]-\big[\real\P^{2n+1},
\mathcal{L}^{~}_{\real\P^{2n+1}}\otimes\complex,
\Id^{~}_{\real\P^{2n+1}}\big] \ . \label{RP2n1antigen}\eeq

As in the examples of Lens spaces, the decay products of the
brane-antibrane system (\ref{RP2n1antigen}) cannot be determined
through Theorem~\ref{bigt} due to the torsion. The difference between
K-homology and singular homology here can be
understood by appealing to the AHW spectral sequence. After some
calculation one finds the filtration groups
$\F_{2n-3,4-2n}\cong\zed_{2^{n-1}}$ and
$\F_{2n-1,2-2n}\cong\zed_{2^n}$~\cite{BEJMS1,8}, which thereby alter the
additive structure in $\H_{\rm
  odd}(\real\P^{2n+1};\zed)$. The lift of the generator
$[\real\P^{2n-1}]\in\H_{2n-1}(\real\P^{2n+1};\zed)\cong\zed_2$ to
K-homology is the stable D-brane
$\omega_0=[\real\P^{2n-1},\id_{\real\P^{2n-1}}^\complex,\iota_{2n-1}^{~}]
\in\K_1^{\rm t}(\real\P^{2n+1})$. While $[\real\P^{2n-1}]$ is of
order~$2$ in $\H_{\rm odd}(\real\P^{2n+1};\zed)$, $\omega_0$ is of
order~$2^n$ in $\K_1^{\rm t}(\real\P^{2n+1})$ and is thus
equal to (\ref{RP2n1antigen}). For every $k=0,1,\dots,n$,
$\omega_k:=2^k\,\omega_0
=[\real\P^{2n-2k-1},\id_{\real\P^{2n-2k-1}}^\complex,\iota_{2n-2k-1}^{~}]$
(with $\omega_n:=0$) corresponds to the order~$2$ generator
$[\real\P^{2n-2k-1}]\in\H_{2n-2k-1}(\real\P^{2n+1};\zed)\cong\zed_2$.
These associations illustrate that an integral lift $\ch_{\rm
  odd}^\zed$ of the Chern character, along the lines described in
Section~\ref{Lifts}, does not generally exist in the present class of
examples.

\begin{remark}\label{RP2n1decay}
This example furnishes a nice illustration of D-brane
decay~\cite{BEJMS1}. Placing together
$2^k$ D$(2n-2)$-branes wrapping $\real\P^{2n-1}\subset\real\P^{2n+1}$
creates an unstable state that decays into a D$(2n-2k-2)$-brane
wrapping $\real\P^{2n-2k-1}\subset\real\P^{2n-1}$, due to the
triviality of the singular homology classes $2^k\,[\real\P^{2n-2k-1}]$
in $\H_{\rm odd}(\real\P^{2n+1};\zed)$. Similarly, stacks of $2^j$ of
these D$(2n-2k-2)$-branes for $1\leq j<n-k$ decay into a
D$(2n-2k-2j-2)$-brane, and so on. \hfill{$\lozenge$}
\end{remark}

\begin{remark}
For $n=1$ this construction of the torsion generator of $\K_1^{\rm
  t}(\real\P^3)$ coincides with the construction which is completely
analogous to that used in Section~\ref{LensSpaces} to construct the
torsion D0-brane wrapping $\S^1\cong\real\P^1$ on the Lens spaces
${\mathbb{L}}(p;q)$. \hfill{$\lozenge$}
\end{remark}

\subsubsection{$\mathbb{R}\P^{2n}$}

The even-dimensional real projective spaces $\real\P^{2n}$ are more
difficult to deal with because they are not orientable. In
particular, they are not spin$^c$ and so most of the techniques used
thus far cannot be applied to this case. In fact, this space
provides an exotic example whereby not only does the K-homology
differ from singular homology, but also where Poincar\'e duality
breaks down and the K-homology differs from the dual K-theory which
in the present case is given by $\K^0_{\rm
t}(\real\P^{2n})\cong\zed\oplus\zed_{2^{n-1}}$, $\K^1_{\rm
t}(\real\P^{2n})\cong 0$.

\begin{proposition}
$\K_0^{\rm t}(\real\P^{2n})~\cong~\zed \ , \quad
\K_1^{\rm t}(\real\P^{2n})~\cong~\zed_{2^{n-1}} \ . $
\label{KhomRP2n}\end{proposition}
\begin{proof}
A simple application of the AHW spectral sequence shows that
$\widetilde{\K}_0^{\rm t}(\real\P^{2n})=0$. Since $\H_{\rm
  odd}(\real\P^{2n};\zed)=0$, via the Chern character we conclude that
$\K_1^{\rm t}(\real\P^{2n})$ has no free part. Finally, by applying
the universal coefficient theorem of Section~\ref{UCT} to $X=
\real\P^{2n}$ one concludes that $\tor_{\K_1^{\rm
    t}(\real\P^{2n})}\cong\tor_{\K_{\rm
    t}^0(\real\P^{2n})}\cong\zed_{2^{n-1}}$.
\end{proof}

As always, the generator of $\K_0^{\rm t}(\real\P^{2n})\cong\zed$ is
the D-instanton $[\pt,\id_\pt^\complex,\iota]$. The remaining torsion
generators of $\K_1^{\rm t}(\real\P^{2n})\cong\zed_{2^{n-1}}$ are more
difficult to find. They may be constructed as follows. By excision,
the quotient map $p_{2n}: (\mathbb{R}\P^{2n},\mathbb{R}\P^{2n-1}) \rightarrow (
\mathbb{R}\P^{2n}/\mathbb{R}\P^{2n-1}, \rm{pt})$ induces
an isomorphism $(p_{2n})_* : \K^{\rm
t}_\sharp(\mathbb{R}\P^{2n},\mathbb{R}\P^{2n-1})
\rightarrow \K^{\rm t}_\sharp(\mathbb{R}\P^{2n}/\mathbb{R}\P^{2n-1},
\rm{pt})$ giving
$$\K^{\rm t}_\sharp(\mathbb{R}\P^{2n},\mathbb{R}\P^{2n-1}) ~\cong~
\K^{\rm t}_\sharp(\mathbb{R}\P^{2n}/\mathbb{R}\P^{2n-1}, \pt
) ~\cong~ \widetilde{\K}^{\rm t}_\sharp(
\mathbb{R}\P^{2n}/\mathbb{R}\P^{2n-1}) ~\cong~
\widetilde{\K}^{\rm t}_\sharp(\S^{2n})$$ and one concludes that
\beq
\K^{\rm t}_0(\mathbb{R}\P^{2n},\mathbb{R}\P^{2n-1}) ~\cong~
\mathbb{Z} \ , \qquad
\K^{\rm t}_1(\mathbb{R}\P^{2n},\mathbb{R}\P^{2n-1}) ~\cong~ 0 \ .
\label{KRP2nsrel}\eeq
The six-term exact sequence associated to the pair
$(\mathbb{R}\P^{2n},\mathbb{R}\P^{2n-1})$ is given by
$$\xymatrix{\K_{0}^{\rm t}(\mathbb{R}\P^{2n-1})
  \ar[r]^{(\iota_{2n-1})_*} & \K_{0}^{\rm t}(\mathbb{R}\P^{2n})
  \ar[r]^{\!\!\!\!\!\!\!\!\!\!\!\!\varsigma_*} & \K_{0}^{\rm
    t}(\mathbb{R}\P^{2n},\mathbb{R}\P^{2n-1}) \ar[d]^\partial \\
\K_{1}^{\rm t}(\mathbb{R}\P^{2n},\mathbb{R}\P^{2n-1})\ar[u]^\partial &
\K_{1}^{\rm t}(\mathbb{R}\P^{2n}) \ar[l]^{\qquad\varsigma_*} & \K_{1}^{\rm
  t}(\mathbb{R}\P^{2n-1}) \ar[l]^{(\iota_{2n-1})_*} \ . }$$
The homomorphism $(\iota_{2n-1})_* : \K_{0}^{\rm
  t}(\mathbb{R}\P^{2n-1}) \rightarrow \K_{0}^{\rm
  t}(\mathbb{R}\P^{2n})$ is induced by the inclusion of the
$(2n-1)$-skeleton in $\real\P^{2n}$. Since both groups $\K_{0}^{\rm
  t}(\mathbb{R}\P^{2n-1})=\K_{0}^{\rm t}(\mathbb{R}\P^{2n})=\zed$ are
generated by $[\pt,\id_\pt^\complex,\iota]$, it follows that
$(\iota_{2n-1})_*[\pt,\id_\pt^\complex,\iota]=[\pt,\id_\pt^\complex,\iota]$
and hence $(\iota_{2n-1})_*$ is an isomorphism. Combining this fact
with (\ref{KRP2nsrel}), we conclude that the six-term exact sequence
truncates to the short exact sequence given by
$$0 ~\longrightarrow ~\K_{0}^{\rm
    t}(\mathbb{R}\P^{2n},\mathbb{R}\P^{2n-1}) ~\stackrel{\partial}
{\longrightarrow} ~\K_{1}^{\rm t}(\mathbb{R}\P^{2n-1}) ~
\stackrel{(\iota_{2n-1})_*}{\longrightarrow}~
\K_{1}^{\rm t}(\mathbb{R}\P^{2n}) ~\longrightarrow~0 \ . $$

It finally follows that
\beq
\K_{1}^{\rm t}(\mathbb{R}\P^{2n}) ~\cong~
\K_{1}^{\rm t}(\mathbb{R}\P^{2n-1})\,/~{\rm im}~\partial~ \cong~
\mathbb{Z} \oplus \mathbb{Z}_{2^{n-1}}\,/~{\rm im}~\partial \ .
\label{torsion2n1skel}\eeq
Comparing with Proposition~\ref{KhomRP2n} we conclude that the
connecting homomorphism has range ${\rm im}~\partial\cong\zed$. The
torsion D-branes in this case are thus supported in the
$(2n-1)$-skeleton which is a spin$^c$ submanifold of
$\real\P^{2n}$. Their explicit K-cycle representatives, along with
the pertinent decay products, can be constructed exactly as in our
previous example above. Note that there are no spacetime filling
branes in $\real\P^{2n}$.

\begin{remark}
To understand the geometrical meaning of the quotient in
(\ref{torsion2n1skel}), consider the commutative diagram
$$\xymatrix{& (\S^{2n},\textrm{pt}) \\
(\B^{2n},\S^{2n-1}) \ar[ru]^{p_{2n}'} \ar[rd]_{f_{2n}} & \\
 & (\mathbb{R}\P^{2n},\mathbb{R}\P^{2n-1}) \ar[uu]_{p_{2n}} }$$ where
$f_{2n}$ is the characteristic map of the $2n$-cell. This induces the
commutative diagram in K-homology given by
$$\xymatrix{& \widetilde{\K}^{\rm t}_0(\S^{2n}) \\
\K^{\rm t}_0(\B^{2n},\S^{2n-1}) \ar[ru]^{(p'_{2n})_*} \ar[rd]_{(f_{2n})_*} & \\
 & \K^{\rm t}_0(\mathbb{R}\P^{2n},\mathbb{R}\P^{2n-1})
 \ar[uu]_{(p_{2n})_*}}$$ where the induced maps $(f_{2n})_*$ and
$(p'_{2n})_*$ are isomorphisms. It follows that
$[\B^{2n},\id_{\B^{2n}}^\complex,f_{2n}^{~}]$ is the generator of
$\K^{\rm t}_0(\mathbb{R}\P^{2n},\mathbb{R}\P^{2n-1})\cong\zed$ with $\partial
[\B^{2n},\id_{\B^{2n}}^\complex,f^{~}_{2n}]
=[\S^{2n-1},\id_{\S^{2n-1}}^\complex,q^{~}_{2n-1}]
=h_{2n-1}[q_{2n-1}]$, where
$h_{2n-1}:\pi_{2n-1}(\real\P^{2n-1})\to\K_1^{\rm t}(\real\P^{2n-1})$
is the Hurewicz homomorphism in K-homology and
the quotient map $q_{2n-1} : \S^{2n-1} \rightarrow
\mathbb{R}\P^{2n-1}$ is the generator $[q_{2n-1}] \in
\pi_{2n-1}(\mathbb{R}\P^{2n-1}) \cong \mathbb{Z}$. We thus conclude
that ${\rm im}~ \partial ={\rm im}~ h_{2n-1}$, and hence the quotient
by the image of the boundary homomorphism in (\ref{torsion2n1skel})
projects out the integrally charged D-brane
$[\S^{2n-1},\id_{\S^{2n-1}}^\complex,q^{~}_{2n-1}]$ which fills the entire
$(2n-1)$-cell of $\real\P^{2n}$. \hfill{$\lozenge$}
\end{remark}

\subsubsection{$\mathbb{R}\P^{2n+1} \times \mathbb{R}\P^{2k+1}$}

When dealing with torsion K-homology groups, the structure of D-branes
on product manifolds becomes an interesting problem. Let us first
consider the representative example $\mathbb{R}\P^{2n+1} \times
\mathbb{R}\P^{2k+1}$ wherein the factors each support torsion
D-branes.

\begin{proposition}
$\K_0^{\rm t}(\mathbb{R}\P^{2n+1} \times \mathbb{R}\P^{2k+1})~\cong~
\zed\oplus\zed\oplus\zed_{2^k}\oplus\zed_{2^n}\oplus\zed_{2^p}~\cong~
\K_1^{\rm t}(\mathbb{R}\P^{2n+1} \times
\mathbb{R}\P^{2k+1})$ where $p={\rm gcd}(n,k)$.
\label{KhomRPprod}\end{proposition}
\begin{proof}
We apply the K\"{u}nneth theorem of Section~\ref{ExtProd}
(Theorem~\ref{Kunneththm}). The torsion
extension for the $\K^{\rm t}_0$-group is given by
$$\bigoplus_{i+j=1}\,\Tor\big(\K^{\rm t}_{i}(\mathbb{R}\P^{2n+1})\,,\,
\K^{\rm t}_{j}(\mathbb{R}\P^{2k+1})\big)~=~
\Tor\big(\mathbb{Z}\,,\,\mathbb{Z} \oplus \mathbb{Z}_{2^k}\big) ~\oplus~
\Tor\big(\mathbb{Z} \oplus \mathbb{Z}_{2^n}\,,\,\mathbb{Z}\big)~=~0 \ , $$
and so there is an isomorphism
\bea
\K^{\rm t}_0(\mathbb{R}\P^{2n+1} \times \mathbb{R}\P^{2k+1})&=&
  \big(\K^{\rm t}_0(\mathbb{R}\P^{2n+1}) \otimes
  \K^{\rm t}_0(\mathbb{R}\P^{2k+1}) \big) ~\oplus~ \big(
\K^{\rm t}_1(\mathbb{R}\P^{2n+1}) \otimes
  \K^{\rm t}_1(\mathbb{R}\P^{2k+1}) \big) \nonumber\\ &=&
\mathbb{Z}\oplus\zed \oplus  \mathbb{Z}_{2^k} \oplus
  \mathbb{Z}_{2^n} \oplus \mathbb{Z}_{2^p} \ .
\label{KRPprodiso}\eea
On the other hand, the torsion extension for the $\K^{\rm t}_1$-group
is $\mathbb{Z}_{2^p}$. Again the short exact sequence of
Theorem~\ref{Kunneththm} for the present space
splits and we find the same isomorphism as in (\ref{KRPprodiso}) for
the K-homology group $\K^{\rm t}_1(\mathbb{R}\P^{2n+1} \times
\mathbb{R}\P^{2k+1})$.
\end{proof}

The generating D-branes are straightforward to work out as before. For
the various subgroups of $\K_0^{\rm t}(\mathbb{R}\P^{2n+1} \times
\mathbb{R}\P^{2k+1})$ given by Proposition~\ref{KhomRPprod} one finds
the generators
\bea
\underline{\zed\oplus\zed}&:&
\scriptstyle{\big[\textrm{pt}\,,\,\id_\pt^\complex\,,\,\iota\big]
\quad , \quad \big[\mathbb{R}\P^{2n+1} \times
  \mathbb{R}\P^{2k+1}\,,\,\id_{\mathbb{R}\P^{2n+1} \times
  \mathbb{R}\P^{2k+1}}^\complex\,,\,\textrm{id}^{~}_{\mathbb{R}\P^{2n+1} \times
  \mathbb{R}\P^{2k+1}}\big]}  \ , \nonumber\\
\underline{\zed_{2^n}}&:& \scriptstyle{\big[\mathbb{R}\P^{2n+1} \times
\mathbb{R}\P^{2k+1}\,,\,\mathcal{L}^{~}_{\real\P^{2n+1}}\otimes\complex
\,\boxtimes \id_{\real\P^{2k+1}}^\complex\,,\,
\textrm{id}^{~}_{\mathbb{R}\P^{2n+1} \times
  \mathbb{R}\P^{2k+1}}\big]} \ , \nonumber\\
\underline{\zed_{2^k}}&:& \scriptstyle{\big[\mathbb{R}\P^{2n+1} \times
\mathbb{R}\P^{2k+1}\,,\,\id_{\mathbb{R}\P^{2n+1}}^\complex
\boxtimes\mathcal{L}^{~}_{\real\P^{2k+1}}\otimes\complex \,,\,
  \textrm{id}^{~}_{\mathbb{R}\P^{2n+1} \times
  \mathbb{R}\P^{2k+1}}\big]} \ , \nonumber\\
\underline{\zed_{2^p}}&:& \scriptstyle{\big[\mathbb{R}\P^{2n+1} \times
\mathbb{R}\P^{2k+1}\,,\,(\mathcal{L}^{~}_{\real\P^{2n+1}}\otimes\complex
\,\boxtimes \mathcal{L}^{~}_{\real\P^{2k+1}}\otimes\complex)
  \oplus\id_{\mathbb{R}\P^{2n+1} \times
  \mathbb{R}\P^{2k+1}}^\complex \,,\,
\textrm{id}^{~}_{\mathbb{R}\P^{2n+1} \times
  \mathbb{R}\P^{2k+1}}\big]} \nonumber\\ && \qquad\scriptstyle{-\,
\big[\mathbb{R}\P^{2n+1} \times
  \mathbb{R}\P^{2k+1}\,,\,(\mathcal{L}^{~}_{\real\P^{2n+1}}\otimes\complex
\,\boxtimes\id_{\mathbb{R}\P^{2n+1}}^\complex) \oplus (
\id_{\mathbb{R}\P^{2k+1}}^\complex  \boxtimes
\mathcal{L}^{~}_{\real\P^{2k+1}}\otimes\complex)
  \,,\,\textrm{id}^{~}_{\mathbb{R}\P^{2n+1} \times
  \mathbb{R}\P^{2k+1}}\big]} \ . \nonumber
\eea
The $2^n$-torsion and $2^k$-torsion charges come from stable
spacetime-filling D-branes on $\mathbb{R}\P^{2n+1}
\times\mathbb{R}\P^{2k+1}$. Since they carry non-trivial line bundles
on their worldvolumes, they can be decomposed into lower-dimensional
D-branes carrying trivial line bundles along the lines of
Section~\ref{BwBs}. The precise nature of the constituent D-branes can
again be deduced upon careful examination of the AHW spectral
sequence. The decomposition of the $2^p$-torsion
spacetime-filling brane-antibrane pairs is analogous to that of the
$\real\P^{2n+1}$ example studied earlier.

For the first four subgroups of $\K_1^{\rm t}(\mathbb{R}\P^{2n+1} \times
\mathbb{R}\P^{2k+1})$ one finds the generators
\bea
\underline{\zed\oplus\zed}&:&[\mathbb{R}\P^{2n+1},
\id_{\mathbb{R}\P^{2n+1}}^\complex,\iota^{~}_{2n+1}] \quad , \quad
[\mathbb{R}\P^{2k+1},\id_{\mathbb{R}\P^{2k+1}}^\complex,
\iota^{~}_{2k+1}]  \ , \nonumber\\
\underline{\zed_{2^n}}&:&[\mathbb{R}\P^{2n+1},
\mathcal{L}^{~}_{\real\P^{2n+1}}\otimes\complex,\iota^{~}_{2n+1}]-
[\mathbb{R}\P^{2n+1},\id_{\mathbb{R}\P^{2n+1}}^\complex,
\iota^{~}_{2n+1}] \ , \nonumber\\
\underline{\zed_{2^k}}&:&[\mathbb{R}\P^{2k+1},
\mathcal{L}^{~}_{\real\P^{2k+1}}\otimes\complex,\iota^{~}_{2k+1}]-
[\mathbb{R}\P^{2k+1},\id_{\mathbb{R}\P^{2k+1}}^\complex,
\iota^{~}_{2k+1}] \ . \nonumber
\eea
The $2^p$-torsion class is more difficult to determine in this case
because it arises from the torsion extension in the K\"unneth formula. It
can be found by again comparing to singular homology and identifying
it as the lift of the remaining cycles in $\H_{\rm
  odd}(\mathbb{R}\P^{2n+1} \times \mathbb{R}\P^{2k+1};\zed)$ after the
other decay products have been determined from the AHW spectral
sequence along the lines of the $\real\P^{2n+1}$ example
above~\cite{BEJMS1}.

\subsubsection{$\mathbb{R}\P^{2n+1} \times \S^k$}

For our final example of projective D-branes we consider a product
spacetime in which one factor carries only torsion-free D-branes. For
the representative example $\mathbb{R}\P^{2n+1} \times \S^k$ we
proceed exactly as in the previous case.

\begin{proposition}
$\K^{\rm t}_0(\mathbb{R}\P^{2n+1} \times\S^k) ~\cong~ \left \{
\begin{array}{ll}
           {\quad\mathbb{Z} \oplus \mathbb{Z}} &
{, \quad k~{\rm even}}\\
           {\mathbb{Z} \oplus \mathbb{Z} \oplus
             \mathbb{Z}_{2^n}} & {, \quad
           k~{\rm odd}}\end{array} \right. \ , \qquad
\K^{\rm t}_1(\mathbb{R}\P^{2n+1} \times \S^k) \cong \left \{ \begin{array}{ll}
{\mathbb{Z} \oplus \mathbb{Z} \oplus \mathbb{Z}_{2^n}
\oplus \mathbb{Z}_{2^n}}& {,\quad k~{\rm even}}\\
{\quad \mathbb{Z} \oplus \mathbb{Z} \oplus
  \mathbb{Z}_{2^n}}&{
           , \quad k~{\rm odd}}\end{array} \right. \ . $
\label{KhomRPSprod}\end{proposition}
\noindent
The generators of the $\K_0^{\rm t}$-groups are given by
     $$\left\{\begin{array}{rl}\scriptstyle{\big[\pt\,,\,\id_\pt^\complex
\,,\,\iota\big] \ ,
\quad \big[\S^k\,,\,\id_{\S^k}^\complex\,,\,\iota_{S^k}^{~}\big]} & ;
\scriptstyle{\quad
k~{\rm even}}\\ {}\scriptstyle{\big[\pt\,,\,\id_\pt^\complex\,,\,\iota\big] \ ,
\quad \big[\mathbb{R}\P^{2n+1} \times \S^k\,,\,
\id_{\mathbb{R}\P^{2n+1} \times \S^k}^\complex\,,\,
\Id^{~}_{\mathbb{R}\P^{2n+1} \times \S^k}\big] \ ,} & \\
{}\scriptstyle{\big[\mathbb{R}\P^{2n+1} \times \S^k\,,\,
\mathcal{L}_{\real\P^{2n+1}}\otimes
\complex\,\boxtimes\id_{\S^k}^\complex\,,\,\Id^{~}_{\mathbb{R}\P^{2n+1}
  \times \S^k}\big]-\big[\mathbb{R}\P^{2n+1} \times \S^k\,,\,
\id_{\mathbb{R}\P^{2n+1} \times
  \S^k}^\complex\,,\,\Id^{~}_{\mathbb{R}\P^{2n+1} \times \S^k}\big]} &
; \scriptstyle{\quad
k~{\rm odd}}\end{array}\right. \ , $$
while the $\K^{\rm t}_1$-groups are generated by the D-branes
     $$\left\{\begin{array}{rl}\scriptstyle{\big[\mathbb{R}\P^{2n+1}\,,\,
\id_{\real\P^{2n+1}}^\complex\,,\,
\iota^{~}_{2n+1}\big] \ , \quad \big[\mathbb{R}\P^{2n+1} \times \S^k\,,\,
\id_{\mathbb{R}\P^{2n+1} \times \S^k}^\complex\,,\,
\Id^{~}_{\mathbb{R}\P^{2n+1} \times \S^k}\big] \ ,} & \\
{}\scriptstyle{\big[\mathbb{R}\P^{2n+1}\,,\,\mathcal{L}_{\real\P^{2n+1}}
\otimes\complex\,,\,
\iota_{2n+1}\big]-\big[\mathbb{R}\P^{2n+1}\,,\,\id_{\real\P^{2n+1}}^\complex\,,\,
\iota^{~}_{2n+1}\big] \ ,} & \\
{}\scriptstyle{\big[\mathbb{R}\P^{2n+1} \times \S^k\,,\,
\mathcal{L}_{\real\P^{2n+1}}^{~}
\otimes\complex\,\boxtimes
\id_{\mathbb{R}\P^{2n+1} \times \S^k}^\complex\,,\,
\Id^{~}_{\mathbb{R}\P^{2n+1} \times \S^k}\big]-\big[\mathbb{R}\P^{2n+1}
\times \S^k\,,\,\id_{\mathbb{R}\P^{2n+1} \times \S^k}^\complex\,,\,
\Id^{~}_{\mathbb{R}\P^{2n+1} \times \S^k}\big]} & ;
 \scriptstyle{    \quad k~{\rm even}} \\
{}\scriptstyle{\big[\mathbb{R}\P^{2n+1}\,,\,\id_{\real\P^{2n+1}}^\complex\,,\,
\iota^{~}_{2n+1}\big] \ , \quad \big[\S^k\,,\,\id_{\S^k}^\complex\,,\,
\iota_{\S^k}^{~}\big]\ , } & \\ \scriptstyle{
     \big[\mathbb{R}\P^{2n+1}\,,\,\mathcal{L}_{\real\P^{2n+1}}
\otimes\complex\,,\,\iota_{2n+1}\big]-\big[\mathbb{R}\P^{2n+1}\,,\,
\id_{\real\P^{2n+1}}^\complex\,,\,\iota^{~}_{2n+1}\big]} & ;
 \scriptstyle{    \quad k~{\rm odd}}\end{array}\right. \ . $$

\subsection{D-Branes on Calabi-Yau Spaces}

We conclude this section by further indicating how our analysis of
torsion D-branes can be applied to the topological A-model and B-model
examples studied earlier, and ultimately to a K-cycle description of
mirror symmetry~\cite{Aspinwall1}. For definiteness, let us work with
the Fermat quintic threefold $Y$ defined by
$$Y=\Big\{(z_1,\ldots ,z_5) \in \mathbb{C}\P^4~  \Big| ~
\mbox{$\sum\limits_{i=1}^5$}\, z_i^5=0 \Big\} \  . $$
This is a simply-connected complex projective algebraic variety of
real dimension~6. It is therefore a spin$^c$ manifold and satisfies
Poincar\'e duality. Let $\mathcal{H}_Y$ be the hyperplane line bundle on $Y$,
i.e. the restriction to $Y$ of the line bundle which is associated
with any hyperplane $\mathcal{H}$ in $\complex\P^4$. Let ${D}$
the corresponding hyperplane divisor whose zero set in $\complex\P^4$
is precisely the original hyperplane $\mathcal{H}$. Let
$C\cong\complex\P^1$ be a degree~$1$ rational curve on $Y$. Then $\K_{\rm
  t}^0(Y)\cong\mathbb{Z}^{\oplus4}$~\cite{18,19} and by Poincar\'e
duality we have $\K_0^{\rm t}(Y)\cong\mathbb{Z}^{\oplus4}$. From the
known K-theory generators we can thus identify the generating A-branes
\bea
\big[\pt,\id_\pt^\complex,\iota\big] &,& \quad
\big[Y,\id_Y^\complex,\Id_Y^{~}\big] \ , \nonumber\\
\big[Y,\mathcal{H}_Y,\Id_Y\big]-\big[Y,\id_Y^\complex,\Id_Y^{~}
\big]~=~\big[D,\id_D^\complex,\iota_D^{~}\big] &,&
\quad \big[Y,(\iota^{~}_C)_![\id_C^\complex],\Id^{~}_Y\big]~=~
\big[C,\id_C^\complex,\iota^{~}_C\big] \ . \nonumber
\eea
In addition, one has $\K_{\rm
  t}^1(Y)\cong\mathbb{Z}^{\oplus204}$ so that $\K^{\rm
  t}_1(Y)\cong\mathbb{Z}^{\oplus204}$. The corresponding D-branes wrap
the $204$ independent three-cycles of $\H_3(Y;\zed)$ and are
constructed using Theorem~\ref{bigt}. As expected, the A-branes all
wrap lagrangian submanifolds with flat line bundles.

The corresponding B-branes live in the multiply-connected non-singular
Calabi-Yau threefold $X$ obtained by quotienting $Y$ by the
$\zed_5$-action generated by $z_i\to\zeta^{i-1}\,z_i$, $i=1,\dots,5$
where $\zeta^5=1$~\cite{Aspinwall1,GP1,Witten2a}. This is also a
complex projective algebraic variety
of real dimension~$6$, and hence a spin$^c$ manifold. Let
$\mathcal{H}_X$ be the hyperplane line bundle restricted to $X$, and
let $\mathcal{L}_X\to X$ be the flat line bundle $\mathcal{L}_X=Y
\times \mathbb{C}/\mathbb{Z}_5$ with respect to the $\zed_5$-action
above. Then by Poincar\'e duality one has $\K_0^{\rm t}(X)\cong\K_{\rm
  t}^0(X)\cong\mathbb{Z}^{\oplus4} \oplus\mathbb{Z}_5$. Using the
known K-theory generators~\cite{19} we may write down the D-branes
corresponding to the free part as
$$\left\{\begin{array}{r}[\pt,\id_\pt^\complex,\iota] \ , \\
{}[X,\mathcal{H}_X,\Id_X]-[X,\id_X^\complex,\Id^{~}_X] \ , \\
{}[X,\mathcal{H}_X\otimes\mathcal{H}_X\oplus\id_X^\complex,\Id^{~}_X]-
[X,\mathcal{H}_X\oplus\mathcal{H}_X,\Id_X] \ , \\
{}\big[X,(\mathcal{H}_X)^{\otimes3}\oplus(\mathcal{H}_X)^{\oplus3},\Id_X\big]
-\big[X,(\mathcal{H}_X\otimes\mathcal{H}_X)^{\oplus3}\oplus\id_X^\complex,
\Id^{~}_X\bigr] \ , \end{array} \right.$$ while the torsion generator is the
spacetime-filling brane-antibrane pair
$$[X,\mathcal{L}_X,\Id_X]-[X,\id_X^\complex,\Id^{~}_X] \ . $$
This brane-antibrane system decays into a stable torsion D4-brane at
the Gepner point of the given Calabi-Yau moduli
space~\cite{19}. Furthermore, one has $\K_1^{\rm t}(X)\cong\K_{\rm
  t}^1(X)\cong\mathbb{Z}^{\oplus44} \oplus \mathbb{Z}_5$, with the
free part generated by lifting the $44$ three-cycles in $\H_3(X;\zed)$ and the
torsion generated by applying the Hurewicz homomorphism to the
generator of the fundamental group $\pi_1(X)\cong\zed_5$.

\newsection{Flux Stabilization of D-Branes\label{FibreD}}

In this final section we shall consider D-branes which live on the
total space $X$ of a fibration $$\xymatrix{F\ar[r]&X\ar[d]\\ &B}$$
where we assume that the base space $B$ is a simply connected finite
CW-complex (we can consider $B$ to be only path-connected, which
then requires the use of local coefficients). There are many such
situations in which one is interested in the classification of
D-branes in $X$. In fact, many of the examples we have considered
previously fall into this category. For instance, both the Lens
spaces ${\mathbb{L}}(p;q_1,\dots,q_n)$ and the real projective spaces
$\real\P^{2n+1}$ are circle bundles over $\complex\P^n$.
Furthermore, the very application of vector bundle modification
identifies those D-branes such that one is a spherical fibration
over the other. Both our descriptions of ``M-theory'' definitions of
unstable 9-branes and T-duality also have natural extensions to
torus bundles.

Since K-homology satisfies both the wedge and weak homotopy
equivalence axioms for finite CW-complexes, we may apply the
Leray-Serre spectral sequence to calculate the topological
K-homology groups in these instances. The Leray-Serre
theorem~\cite{8} states that there is a spectral sequence
$\{\E^{r}_{p,q},\dd^{r} \}_{r\in\nat;p,q\in\zed}$ converging to
$\K^{\rm t}_\sharp(X)$ and satisfying $$\E^2_{p,q}=
\H_p\big(B\,;\,\K^{\rm t}_{e(q)}(F)\big)$$ for all $p,q\in\zed$.
This spectral sequence relates D-branes on $X$ to homology cycles of
the base and fibres. Stability criteria can be formulated along the
lines of Section~\ref{Lifts} to determine which homology classes in
$\H_p(B;\K^{\rm t}_{e(q)}(F))$ lift to non-trivial classes in
$\K^{\rm
  t}_\sharp(X)$. Part of the homology basis of $X$ wrapped by stable
D-branes may then contain cycles of the base $B$ embedded in $X$ as
the zero section of the fibre bundle along with cycles from the
inclusion of some of the fibres~$F$.

The worldvolumes will typically be labelled by the characteristic
class of the fibre bundle. Some of these
singular homology cycles may be trivial in the homology of $X$ (but
not in the homology of $B$ or $F$), and on its own a D-brane wrapping
the given cycle in $X$ would be unstable. However, regarding $X$ as the
total space of a non-trivial fibration can effectively render the D-brane
state stable in a process somewhat reverse to the decay of branes wrapping
non-trivial homology cycles that we studied earlier
(Sections~\ref{Stability} and~\ref{Lifts}). We refer to this process
as ``flux stabilization''~\cite{BDS1,BRS1,MMS1}, with the
characteristic class of the fibration playing the role of a ``flux''
on the D-brane worldvolume. These fluxes act as conserved topological
charges on the D-branes which give an obstruction for them to decay to
the vacuum state. For example, a circle bundle $X$ over $B$ is classified
entirely by its first Chern class $c_1(X)\in\H^2(B;\zed)$.

\subsection{Spherically-Fibred D-Branes\label{SFD-B}}

We begin with the class of fibrations wherein the fibre spaces
$F=\S^{2n}$ are even-dimensional spheres.

\begin{proposition}
Let $\S^{2n}\hookrightarrow X\rightarrow B$ be a spherical fibration
such that the base space $B$ is a simply connected finite CW-complex
with freely generated singular homology concentrated in even degree.
Then
\begin{displaymath}
\K^{\rm t}_0(X) ~\cong~\H_\sharp(B;\mathbb{Z})\oplus\H_\sharp(B;\mathbb{Z})
\ , \qquad\K^{\rm t}_1(X) ~\cong~ 0 \ .
\end{displaymath}
\end{proposition}
\begin{proof}
The second term of the Leray-Serre spectral sequence is given by
\begin{displaymath}
\E^2_{p,q} \cong  \left \{ \begin{array}{ll}
     \H_p(B; \mathbb{Z} \oplus \mathbb{Z})  &
      , \quad\textrm{$q$ even}\\
      \qquad 0 & , \quad\textrm{$q$ odd} \end{array} \right. \ .
\end{displaymath}
By the universal coefficient theorem for singular homology
one has $\H_p(B;\mathbb{Z}^{\oplus n}) \cong\H_p(B;
\mathbb{Z})^{\oplus n}$, and so
\begin{displaymath}
\E^2_{p,q} \cong  \left \{ \begin{array}{ll}
      \H_p(B; \mathbb{Z})\oplus\H_p(B; \mathbb{Z})  &, \quad
      \textrm{$q$ even}\\
      \qquad\qquad 0 & , \quad\textrm{$q$ odd} \end{array} \right. \ .
\end{displaymath}
Under the assumptions on the homology of $B$, one has $\E^2_{p,q}=0$ if
either $p$ or $q$ are odd and hence $\E^r_{p,q}
\cong \E^2_{p,q}$ for all $r \geq 2$. It follows that
$\E^{\infty}_{p,q} \cong \E^2_{p,q}$. Then it is easy to see that
$\F_{p,1-p}=\F_{0,1}=\E^{\infty}_{0,1}=0$ for all $p \geq 0$, and so
$\K^{\rm t}_1(X)=0$. Furthermore, one has the filtration groups
\begin{displaymath}
\F_{p,-p} \cong  \left \{ \begin{array}{ll}
      \mbox{$\bigoplus\limits_{q=0}^p$}\,\H_q(B; \mathbb{Z})
\oplus\H_q(B; \mathbb{Z}) &
      , \quad\textrm{$p$ even}\\
       \mbox{$\bigoplus\limits_{q=0}^{p-1}$}\,\H_q(B; \mathbb{Z})
\oplus\H_q(B; \mathbb{Z}) &
      , \quad\textrm{$p$ odd} \end{array} \right.
\end{displaymath}
and hence $\K^{\rm t}_0(X)=
\H_\sharp(B;\mathbb{Z})\oplus\H_\sharp(B;\mathbb{Z})$.
\end{proof}

\subsection{Fractional Branes}

Next we look at the cases where the fibre $F$ is a finite set of
points. In this case, stable D-branes again wrap cycles in the base $B$. Any
such D-brane is also accompanied by a set of $|F|$ mirror images
called {\it fractional branes}~\cite{DM1} which live on the various
leaves of the covering space $X$.

\begin{proposition}
Let $F \hookrightarrow X\rightarrow B$ be a covering such that the
base space $B$ is a simply connected finite CW-complex with freely
generated singular homology concentrated in even degree. Then
\begin{displaymath}
\K^{\rm t}_0(X) ~\cong~\H_\sharp(B;\mathbb{Z})^{\oplus|F|} \ , \qquad
\K^{\rm t}_1(X) ~\cong~0 \ .
\end{displaymath}
\end{proposition}
\begin{proof}
Since the functor $\K^{\rm t}_\sharp$ satisfies the infinite wedge
axiom~\cite{3}, the second term of the Leray-Serre spectral sequence
is given by
\begin{displaymath}
\E^2_{p,q} \cong  \left \{ \begin{array}{ll}
      \H_p\Big(B\,;\,\mbox{$\bigoplus\limits_{\alpha \in F}$}\,
\mathbb{Z}\Big)  &, \quad
      \textrm{$q$ even}\\
      \qquad 0 & , \quad\textrm{$q$ odd} \end{array} \right. \ .
\end{displaymath}
Applying the universal coefficient theorem allows us to rewrite this
term as
\begin{displaymath}
\E^2_{p,q} \cong  \left \{ \begin{array}{ll}
      \H_p(B; \mathbb{Z})^{\oplus|F|}  &, \quad
      \textrm{$q$ even}\\
      \qquad 0 & , \quad\textrm{$q$ odd} \end{array} \right. \ .
\end{displaymath}
Under the stated assumptions on the homology of $B$ we thereby
conclude that
\begin{displaymath}
\E^{\infty}_{p,q} ~\cong~ \E^2_{p,q} ~\cong~  \left \{ \begin{array}{ll}
      \H_p(B; \mathbb{Z})^{\oplus|F|}  &, \quad
      \textrm{$q$ even, $p\geq0$ even}\\
      \qquad 0 & , \quad\textrm{otherwise} \end{array} \right. \ .
\end{displaymath}
One easily shows that $\F_{p,1-p}=\F_{0,1}=\E^{\infty}_{0,1}=0$ for all
$p \geq 0$, and so $\K^{\rm t}_1(X)=0$. One also concludes that
\begin{displaymath}
\F_{p,-p} \cong  \left \{ \begin{array}{ll}
      \mbox{$\bigoplus\limits_{\alpha\in F}~\bigoplus\limits_{q=0}^p$}
\,\H_q(B; \mathbb{Z}) &, \quad
      \textrm{$p$ even}\\
      \mbox{$\bigoplus\limits_{\alpha\in
          F}~\bigoplus\limits_{q=0}^{p-1}$}\,\H_q(B; \mathbb{Z}) &, \quad
      \textrm{$p$ odd} \end{array} \right.
\end{displaymath}
and hence $\K^{\rm t}_0(X)= \bigoplus_{\alpha\in
  F}\,\H_\sharp(B;\mathbb{Z})$.
\end{proof}

There is a relative version of the Leray-Serre spectral sequence
of the form
$$\E^2_{p,q}=\widetilde{\H}_p\big(B\,;\,\K^{\rm t}_{e(q)}(F)
\big) ~\Longrightarrow~  \K^{\rm t}_{e(p+q)}\big(X,F\big) \ . $$
Using an analysis similar to the one just made allows one to
determine D-brane states analogous to those of Section~\ref{SFD-B} in
the case where the homology of $B$ is supported complimentarily to
that above and the fractional branes are all identified with one
another.

\begin{proposition}
Let $F \hookrightarrow X\rightarrow B$ be a covering such that the
base space $B$ is a simply connected finite CW-complex with freely
generated reduced singular homology concentrated in odd degree. Then
\begin{displaymath}
\K^{\rm t}_0(X,F) ~\cong~0 \ , \qquad
\K^{\rm
  t}_1(X,F)~\cong~\H_\sharp(B;\mathbb{Z})\oplus\H_\sharp(B;\mathbb{Z})
\ .
\end{displaymath}
\end{proposition}

\begin{cor}
Let $F \hookrightarrow X\rightarrow B$ be a covering such that the
base space $B$ is a simply connected finite CW-complex with freely
generated reduced singular homology concentrated in degree $ n~{\rm
mod} ~ 2$. Then
\begin{displaymath}
\widetilde{\K}^{\rm t}_{e(n)}(X)
{}~\cong~\widetilde{\H}_\sharp(B;\mathbb{Z}) \ , \qquad
\widetilde{\K}^{\rm t}_{e(n+1)}(X)~\cong~0 \ .
\end{displaymath}
\end{cor}

\begin{remark}
Since $B$ is path-connected, an application of these results to the
trivial fibration $\Id_B:B\to B$ gives as a corollary that $\K_0^{\rm t}(B)$
(resp. $\K_1^{\rm t}(B)$) is isomorphic to $\H_\sharp(B;\zed)$ while
$\K_1^{\rm t}(B)$ (resp. $\K_0^{\rm t}(B)$) is trivial when $B$ has
non-trivial homology only in even (resp. odd)
degree.~\hfill{$\lozenge$}
\end{remark}

\subsection{Spherically-Based D-Branes}

For our final general class of fibre bundles, we consider the cases
where the base space $B$ is a sphere. In such instances the stable
D-branes on $X$ are determined by images of K-cycles on the fibre
$F$. Our first result completely determines the case of coverings of
even-dimensional spheres.

\begin{proposition}
Let $F \hookrightarrow X\rightarrow \S^{2n}$, $n \geq 1$ be a
fibration over the $2n$-sphere such that the topological K-homology
group of the fibre
$\K_\sharp^{\rm t}(F)=\K^{\rm t}_0(F)$ is freely generated. Then
\begin{displaymath}
\K^{\rm t}_0(X) ~\cong~\K^{\rm t}_0(F)\oplus\K^{\rm t}_0(F) \ , \qquad
\K^{\rm t}_1(X) ~\cong~0 \ .
\end{displaymath}
\end{proposition}
\begin{proof}
The second term of the Leray-Serre spectral sequence is given by
\begin{displaymath}
\E^2_{p,q} \cong  \left \{ \begin{array}{ll}
     \K^{\rm t}_{e(q)}(F)  &, \quad
      \textrm{$p=0,2n$}\\
      \quad 0 & , \quad\textrm{otherwise} \end{array} \right. \ .
\end{displaymath}
Since $\K^{\rm t}_1(F)=0$, this becomes
\begin{displaymath}
\E^2_{p,q} \cong  \left \{ \begin{array}{ll}
     \K^{\rm t}_0(F)  &, \quad
      \textrm{$q$ even, $p=0,2n$}\\
      \quad 0 & , \quad\textrm{otherwise} \end{array} \right. \ .
\end{displaymath}
Since $\E^2_{p,q}=0$ if either $p$ or $q$ is odd, it follows that
$\dd^r=0$ for all $r,p,q$ and so $\E^{\infty}_{p,q}=\E^2_{p,q}$. One easily
concludes that $\F_{p,1-p}=\F_{0,1}=\E^{\infty}_{0,1}=\K^{\rm t}_1(F)=0$ for
all $p$, and so $\K^{\rm t}_1(X)=0$. On the other hand, for $p \geq
2n-1$ one has $\F_{p,-p}=\F_{0,0}=\E^{\infty}_{0,0}=\K^{\rm t}_0(F)$
and the only extension problem arises in the exact sequence
$$0 ~\longrightarrow~\F_{2n-1,1-2n}~ \longrightarrow ~\F_{2n,-2n}~
\longrightarrow ~\E^{\infty}_{2n,-2n}~ \longrightarrow~ 0 \ . $$
Since $\Ext(\K^{\rm t}_0(F),\K^{\rm t}_0(F))=0$ it follows that $\F_{2n,-2n}=
\K^{\rm t}_0(F)\oplus\K^{\rm t}_0(F)$. For $p >2n$ one has
$\F_{p,-p}=\F_{2n,-2n}$, and so we
conclude that $\K^{\rm t}_0(X)=\K^{\rm t}_0(F)\oplus\K^{\rm t}_0(F)$.
\end{proof}

\begin{proposition}\label{S1fibre}
Let $F \hookrightarrow X\rightarrow \S^{1}$ be a fibration over the
circle such that the topological K-homology group of the fibre obeys
$\Ext(\K^{\rm t}_i(F),\K^{\rm t}_{e(i+1)}(F))=0$ for $i=0,1$. Then
$$\K^{\rm t}_0(X)~=~\K^{\rm t}_1(X)~=~\K^{\rm t}_0(F) \oplus \K^{\rm
  t}_1(F) \ . $$
\end{proposition}
\begin{proof}
The second term in the Leray-Serre spectral sequence is given by
\begin{displaymath}
\E^2_{p,q} \cong  \left \{ \begin{array}{ll}
     \K^{\rm t}_{e(q)}(F)  &, \quad
      \textrm{$p=0,1$}\\
      \quad 0 & , \quad\textrm{otherwise} \end{array} \right. \ .
\end{displaymath}
The spectral sequence collapses at the second level, and so
\begin{displaymath}
\E^{\infty}_{p,q} ~\cong~ \E^2_{p,q} ~\cong~  \left \{ \begin{array}{ll}
     \K^{\rm t}_{e(q)}(F)  &, \quad
      \textrm{$p=0,1$}\\
      \quad 0 & , \quad\textrm{otherwise} \end{array} \right. \ .
\end{displaymath}
We therefore have $\F_{0,q}\cong\E^2_{0,q}\cong\K^{\rm t}_{e(q)}(F)$.
Since $\Ext(\K^{\rm t}_i(F),\K^{\rm t}_{e(i+1)}(F))=0$, $i=0,1$ one
finds $\F_{p,-p}=\K^{\rm t}_0(F) \oplus \K^{\rm
  t}_1(F)=\F_{p,1-p}$ and the conclusion follows.
\end{proof}

\begin{remark}
For a fibration $F \hookrightarrow X
\rightarrow \S^{2n+1}$ over a generic odd-dimensional sphere, with the
additional assumption that
$\textrm{im}~\dd^{2n+1} : \E^{2n+1}_{2n+1,q-2n} \rightarrow
\E^{2n+1}_{0,q} = 0 = \ker\,\dd^{2n+1} :
\E^{2n+1}_{2n+1,q} \rightarrow \E^{2n+1}_{0,q+2n}$ one can derive an
analogous result to the one of
Proposition~\ref{S1fibre}.~\hfill{$\lozenge$}
\end{remark}
\noindent
Using the relative version of the Leray-Serre spectral sequence
and performing an analysis analogous to those just made allows one to
conclude the following result.

\begin{proposition}
Let $F \hookrightarrow X\rightarrow \S^{n}$, $n \geq 1$ be a
fibration over the $n$-sphere. Then for $i=0,1$ one has
$$\K^{\rm t}_i(X,F) \cong\K^{\rm t}_{e(i+n)}(F) \ . $$
\end{proposition}

\subsection{Hopf Branes}

Intimately related to the four classes of projective D-branes studied
earlier are the Hopf fibrations. Let $r:=\dim_\real(\dalg)$, where
$\dalg$ is one of the four normed division algebras over the field of
real numbers given by $\real$, $\complex$,
$\quat$ or $\oct$. Then the projective plane $\dalg\P^2$ is the
mapping cone of the Hopf fibration
$$\xymatrix{\S^{r-1}\ar[r]&\S^{2r-1}\ar[d] \ . \\ &\S^r}$$ The total
space $X=\S^{2r-1}$ of this fibre bundle is a sphere in $\dalg^2$,
while its base $B=\dalg\P^1\cong\S^r$ is the one-point compactification
$\dalg^\infty$. The Hopf fibrations are the free generators of the
fundamental groups $\pi_{2r-1}(\S^r)$. For the case $r=1$
($\dalg=\real$) the corresponding D-branes are represented as
solitonic kinks. For $r=2$ ($\dalg=\complex$) the D-branes are Dirac
monopoles corresponding to the magnetic monopole bundle over
$\complex\P^1$. For $r=4$ ($\dalg=\quat$) the D-branes are ${\rm
  SU}(2)$ Yang-Mills instantons corresponding to the holomorphic
vector bundle of rank~$2$ over $\complex\P^3$. Finally, the case $r=8$
($\dalg=\oct$) realizes D-branes as ${\rm Spin}(8)$ instantons of the
Hopf bundle over $\real\P^8$. These characterizations~\cite{OS1,16}
are asserted by computing the topological K-homology groups using the
relative version of the Leray-Serre spectral sequence. In fact, they
are special cases of a more general result.

\begin{proposition}\label{Hopf}
For any spherical fibration of the form $\S^{2n+1} \hookrightarrow
X\rightarrow \S^{2m}$ one has $$\K^{\rm t}_i(X,\S^{2n+1})~\cong~
\widetilde{\H}_{2m}(\S^{2m};\zed)~=~\mathbb{Z}$$ for $i=0,1$.
\end{proposition}
\begin{proof}
The second term in the relative Leray-Serre spectral sequence is given
by
\begin{displaymath}
\E^2_{p,q} \cong  \left \{ \begin{array}{ll}
       \widetilde{\H}_{2m}(\S^{2m};\zed) &, \quad
      \textrm{$p=2m$}\\
      \qquad 0 & , \quad\textrm{otherwise} \end{array} \right. \ .
\end{displaymath}
The spectral sequence collapses at the second level so that
 \begin{displaymath}
\E^{\infty}_{p,q} ~\cong~ \E^2_{p,q} ~\cong~  \left \{ \begin{array}{ll}
       \widetilde{\H}_{2m}(\S^{2m};\zed) &, \quad
      \textrm{$p=2m$}\\
      \qquad 0 & , \quad\textrm{otherwise} \end{array} \right. \ .
\end{displaymath}
Since $\E^{\infty}_{p,-p}=0$ unless $p=2m$, one has
$\F_{p,-p}=\F_{2m,-2m}=\widetilde{\H}_{2m}(\S^{2m};\zed)$ and we conclude
that $\K^{\rm
  t}_0(X,\S^{2n+1})\cong\widetilde{\H}_{2m}(\S^{2m};\zed)$. Furthermore,
one has the filtration groups
$\F_{p,1-p}=\F_{2m,1-2m}=\widetilde{\H}_{2m}(\S^{2m};\zed)$ and it follows
that also $\K^{\rm t}_1(X,\S^{2n+1})=\widetilde{\H}_{2m}(\S^{2m};\zed)$.
\end{proof}

\begin{remark}
Proposition~\ref{Hopf} shows that the only stable D-branes (in
addition to the usual point-like D-instantons) in any of the four
non-trivial Hopf fibrations above wrap a spherical submanifold $\S^r$
embedded into $\S^{2r-1}$ as the zero section of the fibre bundle,
even though these submanifolds are homologically trivial. The
worldvolume spheres $\S^r$ are labelled by the classifying
map of the Hopf fibration, which is the generator of $\pi_{r-1}({\rm
  Spin}(r))/\pi_{r-1}({\rm Spin}(r-1))\cong\pi_{r-1}(\S^{r-1})$, and
they are stabilized by the flux given by the $r$-th Chern class
$c_r(\S^{2r-1})\in\H^r(\S^r;\zed)\cong\zed$. For example, for $r=2$
this construction reproduces the well-known result that the stable
branes in $\S^3$ are spherical D2-branes wrapping $\S^2\subset\S^3$
with integral charge labelled by $\pi_1(\S^1)\cong\zed$~\cite{BDS1}. The
stabilizing flux in this case is the magnetic charge
$c_1(\S^3)\in\H^2(\S^2;\zed)\cong\zed$ given by the first Chern class
of the monopole bundle.
\end{remark}


\begin{thebibliography}{99}

\bibitem{UCT3} J.F.~Adams, {\it Stable Homotopy and Generalised
Homology} (The University of Chicago Press, 1974).

\bibitem{KHomM}
T.~Asakawa, S.~Sugimoto and S.~Terashima,
``D-Branes, Matrix Theory and K-Homology'',
J. High Energy Phys. {\bf 0203} (2002) 034
[arXiv:hep-th/0108085].

\bibitem{Aspinwall1}
P.S.~Aspinwall,
``D-Branes on Calabi-Yau Manifolds'',
arXiv:hep-th/0403166.

\bibitem{ABS1} M.F.~Atiyah, R.~Bott and A.~Shapiro, ``Clifford
  Modules'', Topology {\bf 3} (1964) 3--38.

\bibitem{BDS1}
C.~Bachas, M.R.~Douglas and C.~Schweigert,
``Flux Stabilization of D-Branes'',
J. High Energy Phys. {\bf 0005} (2000) 048
[arXiv:hep-th/0003037].

\bibitem{1} P.~Baum and R.G.~Douglas, ``K-Homology and Index Theory'', Proc.
  Symp. Pure Math. {\bf 38} (1982) 117--173.

\bibitem{2} P.~Baum and R.G.~Douglas, ``Index Theory, Bordism and K-Homology'',
  Contemp. Math. {\bf 10} (1982) 1--33.

\bibitem{BDF1} L.G.~Brown, R.G.~Douglas and P.A.~Fillmore,
  ``Extensions of $C^*$-Algebras and K-Homology'', Ann. Math. {\bf
    105} (1977) 265--324.

\bibitem{Berg1}
O.~Bergman, E.G.~Gimon and P.~Ho\v{r}ava,
``Brane Transfer Operations and T-Duality of Non-BPS States'',
J. High Energy Phys. {\bf 9904} (1999) 010
[arXiv:hep-th/9902160].

\bibitem{Berg2}
O.~Bergman, E.G.~Gimon and S.~Sugimoto,
``Orientifolds, RR Torsion and K-Theory'',
J. High Energy Phys. {\bf 0105} (2001) 047
[arXiv:hep-th/0103183].

\bibitem{12} B.~Blackadar, {\it K-theory for Operator Algebras}
(Springer-Verlag, 1986).

\bibitem{Bod} C.F.~B\"odigheimer, ``Splitting the K\"unneth Sequence in K-theory'', Math. Ann.
          \textbf{242} (1979), no. 2, 159-171.

\bibitem{BRS1}
P.~Bordalo, S.~Ribault and C.~Schweigert,
``Flux Stabilization in Compact Groups'',
J. High Energy Phys. {\bf 0110} (2001) 036
[arXiv:hep-th/0108201].

\bibitem{BM1}
P.~Bouwknegt and V.~Mathai,
``D-Branes, $B$-Fields and Twisted K-Theory'',
J. High Energy Phys. {\bf 0003} (2000) 007
[arXiv:hep-th/0002023].

\bibitem{BEM1}
P.~Bouwknegt, J.~Evslin and V.~Mathai,
``T-Duality: Topology Change from $H$-Flux'',
Commun.\ Math.\ Phys.\  {\bf 249} (2004) 383--415
[arXiv:hep-th/0306062].

\bibitem{BEJMS1}
P.~Bouwknegt, J.~Evslin, B.~Jur\v{c}o, V.~Mathai and H.~Sati,
``Flux Compactifications of Projective Spaces and the S-Duality Puzzle'',
arXiv:hep-th/0501110.

\bibitem{18} V.~Braun, ``K-theory Torsion'', arXiv:hep-th/0005103.

\bibitem{4} T.~Br\"ocker and K.~J\"anich, {\it Introduction to Differential
  Topology} (Cambridge University Press, 1982).

\bibitem{19} I.~Brunner and J.~Distler, ``Torsion D-branes in
Nongeometrical Phases'', Adv. Theor. Math. Phys. {\bf 5} (2002)
265--309 [arXiv:hep-th/0102018].

\bibitem{CoFl1} P.E.~Conner, E.E.~Floyd, ``The Relation of Cobordism
to K-theories创, Lecture Notes in Mathematics \textbf{28} (1966),
Springer.

\bibitem{CoFl2} P.E.~Conner, E.E.~Floyd, ``Differentiable Periodic
Maps创, 1964, Springer.

\bibitem{6} J.F.~Davis and P.~Kirk, {\it Lecture Notes in Algebraic
    Topology} (American Mathematical Society, 2001).

\bibitem{Deu} A.~Deutz, ``The Splitting of the K\"unneth sequence in
          K-theory for C$^*$-algebras'', PhD Dissertation, Wayne State
          University, 1981.

\bibitem{DMW1} D.-E.~Diaconescu, G.W.~Moore and E.~Witten, ``$E_8$
  Gauge Theory and a Derivation of K-Theory from M-Theory'',
  Adv. Theor. Math. Phys. {\bf 6} (2003) 1031--1134
  [arXiv:hep-th/0005090].

\bibitem{Douglas2}
M.R.~Douglas,
``Branes within Branes'', in: {\it Strings, Branes and Dualities}
(Kluwer, 1999) 267--275
[arXiv:hep-th/9512077].

\bibitem{Douglas1}
M.R.~Douglas,
``D-Branes, Categories and $\mathcal{N} = 1$ Supersymmetry'',
J.\ Math.\ Phys.\  {\bf 42} (2001) 2818--2843
[arXiv:hep-th/0011017].

\bibitem{DM1}
M.R.~Douglas and G.W.~Moore,
``D-Branes, Quivers and ALE Instantons'',
arXiv:hep-th/9603167.

\bibitem{RDou} R.G.~Douglas, ``C$^*$-algebra Extensions and
K-homology'', Annals of Mathematics Studies, Princeton University
Press, 1980.

\bibitem{FH1}
D.S.~Freed and M.J.~Hopkins,
``On Ramond-Ramond Fields and K-Theory'',
J. High Energy Phys. {\bf 0005} (2000) 044
[arXiv:hep-th/0002027].

\bibitem{FW1}
D.S.~Freed and E.~Witten,
``Anomalies in String Theory with D-Branes'',
Asian J. Math. {\bf 3} (1999) 819
[arXiv:hep-th/9907189].

\bibitem{G-C1}
H.~Garc\'{\i}a-Compe\'an,
``D-Branes in Orbifold Singularities and Equivariant K-Theory'',
Nucl.\ Phys.\ B {\bf 557} (1999) 480--504
[arXiv:hep-th/9812226].

\bibitem{GP1}
B.R.~Greene and M.R.~Plesser,
``Duality in Calabi-Yau Moduli Space'',
Nucl.\ Phys.\ B {\bf 338} (1990) 15--37.

\bibitem{Gukov1}
S.~Gukov,
``K-Theory, Reality and Orientifolds'',
Commun.\ Math.\ Phys.\  {\bf 210} (2000) 621--639
[arXiv:hep-th/9901042].

\bibitem{13} J.A.~Harvey and G.W.~Moore, ``Noncommutative Tachyons and
  K-Theory'', J. Math. Phys. {\bf 42} (2001) 2765--2780
  [hep-th/0009030].

\bibitem{HopHo} M.J.~Hopkins, M.A.Hovey, ``Spin Cobordism Determines
Real K-theory创, Math. Z. \textbf{210} (1992), 181-196.

\bibitem{Horava1}
P.~Ho\v{r}ava,
``Type IIA D-Branes, K-Theory and Matrix Theory'',
Adv.\ Theor.\ Math.\ Phys.\  {\bf 2} (1999) 1373--1404
[arXiv:hep-th/9812135].

\bibitem{Hori1}
K.~Hori,
``D-Branes, T-Duality and Index Theory'',
Adv.\ Theor.\ Math.\ Phys.\  {\bf 3} (1999) 281--342
[arXiv:hep-th/9902102].

\bibitem{3} M.~Jakob, ``A Bordism-Type Description of Homology'',
  Manuscr. Math. {\bf 96} (1998) 67--80.

\bibitem{Johnson1} C.V.~Johnson, {\it D-Branes} (Cambridge University
  Press, 2003).

\bibitem{Kapustin1}
A.~Kapustin,
``D-Branes in a Topologically Non-Trivial $B$-Field'',
Adv.\ Theor.\ Math.\ Phys.\  {\bf 4} (2000) 127--154
[arXiv:hep-th/9909089].

\bibitem{7} M.~Karoubi, {\it K-Theory: An Introduction}
  (Springer-Verlag, 1978).

\bibitem{KamSch} J.~Kaminker, C.~Schochet, ``Topological Obstructions
to Perturbations of Pairs of Operators 创, {\it K-Theory and
Operator Algebras}, Lecture Notes in Mathematics \textbf{575}, 1975,
Springer.

\bibitem{11} O.~Lechtenfeld, A.D.~Popov and R.J.~Szabo,
  ``Noncommutative Instantons in Higher Dimensions, Vortices and
  Topological K-Cycles'', J. High Energy Phys. {\bf 0312}
  (2003) 022 [arXiv:hep-th/0310267].

\bibitem{UCT4} I. Madsen and J. Rosenberg, ``The Universal Coefficient
  Theorem for Equivariant K-Theory of Real and Complex
  $C^*$-Algebras'', Contemp. Math. {\bf 70} (1988) 145--173.

\bibitem{MMS2}
J.M.~Maldacena, G.W.~Moore and N.~Seiberg,
``Geometrical Interpretation of D-Branes in Gauged WZW Models'',
J. High Energy Phys. {\bf 0107} (2001) 046
[arXiv:hep-th/0105038].

\bibitem{MMS1}
J.M.~Maldacena, G.W.~Moore and N.~Seiberg,
``D-Brane Instantons and K-Theory Charges'',
J. High Energy Phys. {\bf 0111} (2001) 062
[arXiv:hep-th/0108100].

\bibitem{Matsuo1}
Y.~Matsuo,
``Topological Charges of Noncommutative Soliton'',
Phys.\ Lett.\ B {\bf 499} (2001) 223--228
[arXiv:hep-th/0009002].

\bibitem{17} M.~Matthey, ``Mapping the Homology of a Group to the
  K-Theory of its $C^*$-Algebra'', Illinois Math. J. {\bf 46} (2002)
  953--977.

\bibitem{MM1}
R.~Minasian and G.W.~Moore,
``K-Theory and Ramond-Ramond Charge'',
J. High Energy Phys. {\bf 9711} (1997) 002
[arXiv:hep-th/9710230].

\bibitem{MW1}
G.W.~Moore and E.~Witten,
``Self-Duality, Ramond-Ramond Fields and K-Theory'',
J. High Energy Phys. {\bf 0005} (2000) 032
[arXiv:hep-th/9912279].

\bibitem{Myers1}
R.C.~Myers,
``Dielectric-Branes'',
J. High Energy Phys. {\bf 9912} (1999) 022
[arXiv:hep-th/9910053].

\bibitem{OS1}
K.~Olsen and R.J.~Szabo,
``Brane Descent Relations in K-Theory'',
Nucl.\ Phys.\ B {\bf 566} (2000) 562--598
[arXiv:hep-th/9904153].

\bibitem{16} K.~Olsen and R.J.~Szabo, ``Constructing D-Branes from
  K-Theory'', Adv. Theor. Math. Phys. {\bf 4} (2000)
  889--1025 [arXiv:hep-th/9907140].

\bibitem{Periwal1}
V.~Periwal,
``D-Brane Charges and K-Homology'',
J. High Energy Phys. {\bf 0007} (2000) 041
[arXiv:hep-th/0006223].

\bibitem{Polchinski1} J.~Polchinski, {\it String Theory, Vol. 2}
  (Cambridge University Press, 1998).

\bibitem{RosSch} J.~Rosenberg, C.~Schochet, ``The K\"unneth Theorem
    and the Universal Coefficient Theorem for Kasparov's Generalized
    K-functor'', Duke Math. J. \textbf{55}, no 2 (1987), 431-474.

\bibitem{Sen1}
A.~Sen,
``Tachyon Condensation on the Brane-Antibrane System'',
J. High Energy Phys. {\bf 9808} (1998) 012
[arXiv:hep-th/9805170].

\bibitem{8} R.M.~Switzer, {\it Algebraic Topology: An Introduction}
  (Springer-Verlag, 1978).

\bibitem{Sz1}
R.J.~Szabo,
``Superconnections, Anomalies and Non-BPS Brane Charges'',
J.\ Geom.\ Phys.\  {\bf 43} (2002) 241--292
[arXiv:hep-th/0108043].

\bibitem{Sz2}
R.J.~Szabo,
``D-Branes, Tachyons and K-Homology'',
Mod.\ Phys.\ Lett.\ A {\bf 17} (2002) 2297--2316
[arXiv:hep-th/0209210].

\bibitem{Witten3}
E.~Witten,
``Chern-Simons Gauge Theory as a String Theory'',
Prog.\ Math.\  {\bf 133} (1995) 637--678
[arXiv:hep-th/9207094].

\bibitem{Witten2a}
E.~Witten,
``Phases of $\mathcal{N} = 2$ Theories in Two Dimensions'',
Nucl.\ Phys.\ B {\bf 403} (1993) 159--222
[arXiv:hep-th/9301042].

\bibitem{Witten2}
E.~Witten,
``Bound States of Strings and $p$-Branes'',
Nucl.\ Phys.\ B {\bf 460} (1996) 335--350
[arXiv:hep-th/9510135].

\bibitem{Witten1}
E.~Witten,
``D-Branes and K-Theory'',
J. High Energy Phys. {\bf 9812} (1998) 019
[arXiv:hep-th/9810188].

\bibitem{14} U.~W\"urgler, ``Riemann-Roch Transformationen und
  Kobordismen'', Comm. Math. Helv. {\bf 46} (1971) 414--424.

\bibitem{UCT1} Z.~Yosimura, ``Universal Coefficient Sequences for
          Cohomology Theories of CW-Spectra'', Osaka J. Math. {\bf 16}
          (1979) 201--217.


\end{thebibliography}
\end{document}